\documentclass[12pt,preprint]{aastex}
\usepackage{epsfig}
\usepackage{natbib}
\usepackage{graphicx}
\usepackage{slashbox}
\usepackage{multirow}
\usepackage{lscape}
\usepackage{mathrsfs,amssymb}
\usepackage{amsmath}
\usepackage{subfigure}
\usepackage{amssymb}
\usepackage{multirow}
\usepackage{tabularx}
\usepackage{rotating}
\usepackage{amsmath}

\newcommand       \cm           {\,{\rm cm}}

\newcommand       \K            {\,{\rm K}}

\newcommand       \simlt        {\lesssim}
\newcommand       \simgt        {\gtrsim}

\newcommand       \mum          {\,{\rm \mu m}}

\newcommand       \simali       {\sim\,}

\newcommand       \Iratio         {I_{3.4}/I_{3.3}}
\newcommand       \Aratio        {A_{3.4}/A_{3.3}}
\newcommand       \Aali           {A_{3.4}}
\newcommand       \Aaro          {A_{3.3}}
\newcommand       \Iratiodfa         {I_{6.85}/I_{6.2}}
\newcommand       \km        {\,{\rm km}}
\newcommand       \mol       {\,{\rm mol}}
\newcommand       \cals       {\,{\rm cal}}
\newcommand       \kcal       {\,{\rm kcal}}

\newcommand       \Etot      {E_{\rm tot}}

\newcommand       \Adfa       {A_{6.85}}
\newcommand       \Idfa       {I_{6.85}}

\newcommand       \Acc        {A_{6.2}}
\newcommand       \Icc        {I_{6.2}}
\countdef\decade=200
\advance\decade by \year
\countdef\hours=201
\advance\hours by \time
\divide\hours by 60
\countdef\mins=202
\advance\mins by \hours
\multiply\mins by 60
\multiply\hours by 100

\countdef\miltime=203
\advance\miltime by \hours
\advance\miltime by \time
\advance\miltime by -\mins

\shorttitle{C--H Stretches of Superhydrogenated PAHs}
\title{
\vspace*{-2.0em}
{\normalsize\rm Accepted for publication in {\it The Astrophysical
    Journal Supplement Series} (12-31-2019)}\\
\vspace*{1.0em}
Superhydrogenated Polycyclic Aromatic Hydrocarbon Molecules:
Vibrational Spectra in the Infrared
}
\author{X.J.~Yang\altaffilmark{1,2},
            Aigen Li\altaffilmark{2}, and
            R.~Glaser\altaffilmark{3}}
\altaffiltext{1}{Department of Physics,
                      Xiangtan University,
                      411105 Xiangtan, Hunan Province, China;
                      {\sf xjyang@xtu.edu.cn}}
\altaffiltext{2} {Department of Physics and Astronomy,
                  University of Missouri,
                  Columbia, MO 65211, USA;
                  {\sf lia@missouri.edu}}
\altaffiltext{3} {Department of Chemistry,
                  University of Missouri,
                  Columbia, MO 65211, USA;
                  {\sf glaserr@missouri.edu}}

\begin{document}

\begin{abstract}
Superhydrogenated polycyclic aromatic hydrocarbons (PAHs) may be
present in H-rich and ultraviolet-poor benign regions. The addition of
excess H atoms to PAHs converts the aromatic bonds into aliphatic
bonds, the strongest of which falls near 3.4$\mum$. Therefore,
superhydrogenated PAHs are often hypothesized as a carrier of the
3.4$\mum$ emission feature which typically accompanies the stronger
3.3$\mum$ aromatic C--H stretching feature. To assess this hypothesis,
we use density function theory to compute the infrared (IR)
vibrational spectra of superhydrogenated PAHs and their ions of
various sizes (ranging from benzene, naphthalene to perylene and
coronene) and of various degrees of hydrogenation. For each molecule,
we derive the intrinsic oscillator strengths of the 3.3$\mum$ aromatic
C--H stretch ($\Aaro$) and the 3.4$\mum$ aliphatic C--H stretch ($\Aali$). By comparing the computationally-derived mean ratio of $\langle\Aali/\Aaro\rangle\approx 1.98$ with the mean ratio of the observed intensities
$\langle\Iratio\rangle\approx 0.12$, we find that the degree of
superhydrogenation --- the fraction of carbon atoms attached with
extra hydrogen atoms --- is only $\simali$2.2\% for neutral PAHs which
predominantly emit the 3.3 and 3.4$\mum$ features. We also determine
for each molecule the intrinsic band strengths of the 6.2$\mum$
aromatic C--C stretch ($\Acc$) and the 6.85$\mum$ aliphatic C--H
deformation ($\Adfa$). We derive the degree of superhydrogenation from
the mean ratio of the observed intensities
$\langle\Iratiodfa\rangle\simlt 0.10$ and
$\langle\Adfa/\Acc\rangle\approx 1.53$ for neutrals and
$\langle\Adfa/\Acc\rangle\approx 0.56$ for cations to be $\simlt$3.1\%
for neutrals and $\simlt$8.6\% for cations. We conclude that
astrophysical PAHs are primarily aromatic 
and are only marginally superhydrogenated.
\end{abstract}



\section{Introduction\label{sec:intro}}
The so-called ``unidentified'' infrared (IR) emission (UIE) bands,
which are composed of a distinctive set of broad emisison features
at 3.3, 6.2, 7.8, 8.6 and 11.3$\mum$,
are ubiquitously detected in a wide range of
Galactic and extragalactic environments (see Tielens 2008).
The hypothesis of polycyclic aromatic hydrocarbon (PAH)
molecules as a viable carrier of the UIE bands,
originally proposed by L\'eger \& Puget (1984)
and Allamandola et al.\ (1985), has been widely accepted.
The PAH hypothesis attributes the UIE bands
to the vibrational modes of PAHs,
with the 3.3$\mum$ feature assigned to
      C--H stretching modes,
the 6.2$\mum$ and 7.7$\mum$ features to
      C--C stretching modes,
the 8.6$\mum$ feature to
      C--H in-plane bending modes,
and the 11.3$\mum$ feature to
      C--H out-of-plane bending modes.
%
According to this hypothesis, PAHs are present
in the interstellar medium (ISM) in various sizes,
geometries, and charging states
(Allamandola et al.\ 1989, Peeters 2004).
%
The relative strengths of these bands depend on
the size, charge and molecular structure
of the PAH molecule (Allamandola et al.\ 1999,
Draine \& Li 2001) and the physical conditions
(e.g., the intensity and hardness of
the starlight illuminating the molecule,
the electron density and gas temperature;
see Bakes \& Tielens 1994, Weingartner \& Draine 2001).

In the diffuse ISM and photodissociated regions (PDRs)
where hydrogen (H) atoms are abundant,
astronomical PAHs are exposed to the continuous
bombardment of H atoms and may likely have excess
peripheral H atoms
(see Andrews et al.\ 2016 and references therein).
In the following, we term superhydrogenated PAHs
as those PAHs whose edges contain excess H atoms.
In the literature, superhydrogenated PAHs are often
also called hydrogenated PAHs.
In this work we will use the term
``superhydrogenated PAHs''
(or ``superhydrogenation'')
interchangeably used with
the term ``hydrogenated PAHs''
(or ``hydrogenation'').
The interaction between PAHs and H atoms has been
studied both theoretically (Cassam-Chena\"i et al.\ 1994,
Bauschlicher 1998, Le Page et al.\ 2009)
and experimentally (Ricks et al.\ 2009, Boschman et al.\ 2012,
Kl$\ae$rke et al.\ 2013, Cazaux et al.\ 2016).
These studies have demonstrated that it is possible
to superhydrogenate PAH cations,
particularly in regions rich in ultraviolet (UV) photons
(e.g., the surface of PDRs).
Theoretical studies have also shown that superhydrogenation
of neutral PAHs in H-rich, UV-poor benign regions
(e.g., protoplanetary nebulae) is possible
(e.g., see Rauls \& Hornek$\ae$r 2008, Rasmussen et al.\ 2011).
Experimentally, it has been demonstrated that
coronene (C$_{24}$H$_{12}$) could be fully
superhydrogenated to form perhydrocoronene
(C$_{24}$H$_{36}$) in low UV flux regions
(see Thrower et al.\ 2012, 2014).
Wolf et al.\ (2016) explored experimentally
the photo-stability of cationic pyrene
(C$_{16}$H$_{10}$$^{+}$)
with six (C$_{16}$H$_{16}$$^{+}$) or
16 extra H atoms (C$_{16}$H$_{26}$$^{+}$)
and found superhydrogenated pryene cations
would undergo backbone fragmentation upon
absorption of two (for C$_{16}$H$_{16}$$^{+}$)
or one  (for C$_{16}$H$_{26}$$^{+}$) photons of
energy just below 3\,eV. On the other hand,
by combining thermal desorption mass
spectrometry measurements
and density functional theory (DFT) calculations,
Jensen et al.\ (2019) have shown
the existence of stable configurations
of superhydrogenated neutral coronene.
Halasinski et al.\ (2005) and Hammonds et al.\ (2009)
obtained the electronic spectra of hydrogenated PAHs
and their ions, respectively through the matrix isolation
spectroscopy experiments and the time-dependent DFT
computations. They argued that hydrogenated PAHs might
be responsible for some of the diffuse interstellar bands.

Superhydrogenated PAHs have been suggested
to be (at least partly) responsible for
the 3.4$\mum$ emission feature
detected in many UIE sources
which always accompanies the (often much stronger)
3.3$\mum$ feature
(e.g., see Geballe et al.\ 1985, 1989,
Jourdain de Muizon et al.\ 1986, 1990,
Nagata et al.\ 1988,
Allamandola et al.\ 1989,
Sandford et al.\ 1991,
Joblin et al.\ 1996,
Sloan et al.\ 1997,
Goto et al.\ 2003,
Smith et al.\ 2004, 
Kondo et al.\ 2012,
Yamagishi et al.\ 2012,
Seok \& Li 2017,
Quiti\'{a}n-Lara et al.\ 2018).
The exact carrier of the 3.4$\mum$ emission feature
remains unidentified, although it is often thought to
arise from the aliphatic side chains
attached as functional groups to PAHs
(see Yang et al.\ 2017a  
and references therein).
However, Wagner et al.\ (2000) obtained
the IR emission spectra of five gas-phase
UV laser-excited PAHs,
two of which are methylated and
three of which are peripherally hydrogenated.
They found that hydrogenated PAHs produce
a better match to astrophysical data
than methylated PAHs.
  The 3.4$\mum$ emission feature could
  also be due to the anharmonicity of
  the aromatic C--H stretching vibration
  (see Barker et al.\ 1987, Maltseva et al.\ 2016).
  Let $v$ be the vibrational quantum number.
  In a harmonic oscillator, the spacing between all adjacent
  vibrational energy levels is constant, hence the $\Delta v=1$
  vibrational transitions between high $v$ levels result in
  the same spectral line as that of the $v =1\rightarrow0$ transition.
  In contrast, anharmonicity would continuously decrease the spacing
  between the adjacent vibrational states for higher values of $v$,
  and therefore the $\Delta v=1$ transitions between higher $v$ levels
  occur at increasingly longer wavelengths. The anharmonicity model
  interprets the weaker feature at 3.4$\mum$
  as the $v=2\rightarrow1$ ``hot band''
  of the 3.3$\mum$ fundamental $v =1\rightarrow0$ aromatic
  C--H stretching mode (see Barker et al.\ 1987).\footnote{%
   The anharmonicity model also predicts a weak band 
   at $\simali$1.6--1.8$\mum$, the overtone of the aromatic
   C--H stretch and/or combination bands
   (Brenner \& Barker 1992,  Geballe et al. 1994, Chen et al.\ 2019).
   }

In superhydrogenated PAHs, some peripheral C atoms
have two H atoms and the extra H atom converts
the originally aromatic ring into an aliphatic ring.
This creates two aliphatic C--H stretching bands:
one due to the symmetric and the other to
the asymmetric C--H stretching modes.
These bands would fall near 3.4$\mum$ and could
(at least partly) account for the 3.4$\mum$ emission
(Schutte et al.\ 1993, Bernstein et al.\ 1996,
Sandford et al.\ 2013, Steglich et al.\ 2013).
Pauzat \& Ellinger (2001) suggested that
hydrogenated PAHs also produce series of bands
that may be at the origin of the broad plateau
observed below the 3.4$\mum$ feature.

Superhydrogenated PAHs also exhibit two
aliphatic C--H deformation bands at
$\simali$6.85 and 7.25$\mum$
(e.g., see Sandford et al.\ 2013).
Their low intensities put them at the limit
of modern observational techniques.
Observationally, these two bands have been
detected both in the Milky Way
and in the Large and Small Magellanic Clouds
(e.g., see Acke et al.\ 2010,
Sloan et al.\ 2014, Materese et al.\ 2017),
but only in a limited number of objects
(see Table~3 of Yang et al.\ 2016a
for a summary).
This will change with the launch of
the {\it James Webb Space Telescope} (JWST).
The Mid-IR Instrument (MIRI) on {\it JWST}
will cover the wavelength range of
the aliphatic C--H deformation bands
with a medium spectral resolution
of $\simali$1550--3250
and unprecedented sensitivity.
On the other hand,
the Near-IR Spectrograph (NIRSpec)
on {\it JWST} with a spectral resolution
up to $\simali$2700 will allow one to probe
the aromatic and aliphatic C--H stretches
at 3.3 and 3.4$\mum$ in depth.
{\it JWST}'s unique high sensitivity and
near- and mid-IR medium spectral resolution
capabilities will open up an IR window unexplored
by the {\it Spitzer Space Telescope} and unmatched
by the {\it Infrared Space Observatory} (ISO)
and thus will probably place the detection of
superhydrogenated PAHs on firm ground
and enable far more detailed band
analysis than previously possible.

The opportunity to thoroughly probe
superhydorgenated PAHs in various
astrophysical regions using {\it JWST}
motivates us to employ DFT to
compute the IR spectra of a series of
superhydrogenated PAH molecules
and their cations,
with special attention paid to the intrinsic
strengths of the aliphatic C--H bands
at 3.4, 6.85 and 7.25$\mum$.
In \S\ref{sec:Method} we briefly describe
the computational methods
and the structures of our target molecules.
\S\ref{sec:Results} presents the computed
IR spectra as well as the intrinsic oscillator
strengths of the aromatic and aliphatic C--H bands.
The astrophysical implications are discussed
in \S\ref{sec:Discussion}.
Finally, we summarize our major results
in \S\ref{sec:Summary}.


\section{Computational Methods
         and Target Molecules
         \label{sec:Method}
         }
We use the Gaussian09 and Gaussian16 softwares
(Frisch et al.\ 2009)
to calculate the IR vibrational spectra
of a range of superhydrogenated PAHs
and their cations. 
We employ the hybrid DFT method 
B3LYP (Frisch et al.\ 2009)
in conjunction with the 6-311+G$^{\ast\ast}$ 
basis set, i.e., 
triple $\zeta$ functions are included to 
describe the valence orbitals,
diffuse functions are applied 
to the heavy (i.e., carbon) atoms, 
and polarization functions are applied to 
both heavy atoms and hydrogen atoms.
%
The neutral hydrocarbons are closed-shell systems
and they will be computed with restricted wave 
functions (RB3LYP).
The cationic hydrocarbons are the result of
single electron oxidation and these radical cations
will be computed with unrestricted wave 
functions (UB3LYP).
We optimize the molecule structures 
and then calculate the harmonic 
vibrational spectra 
(see Yang et al.\ 2017b and references therein). 
%
%
%

\begin{figure*}
\centering{
\includegraphics[scale=0.8,clip]{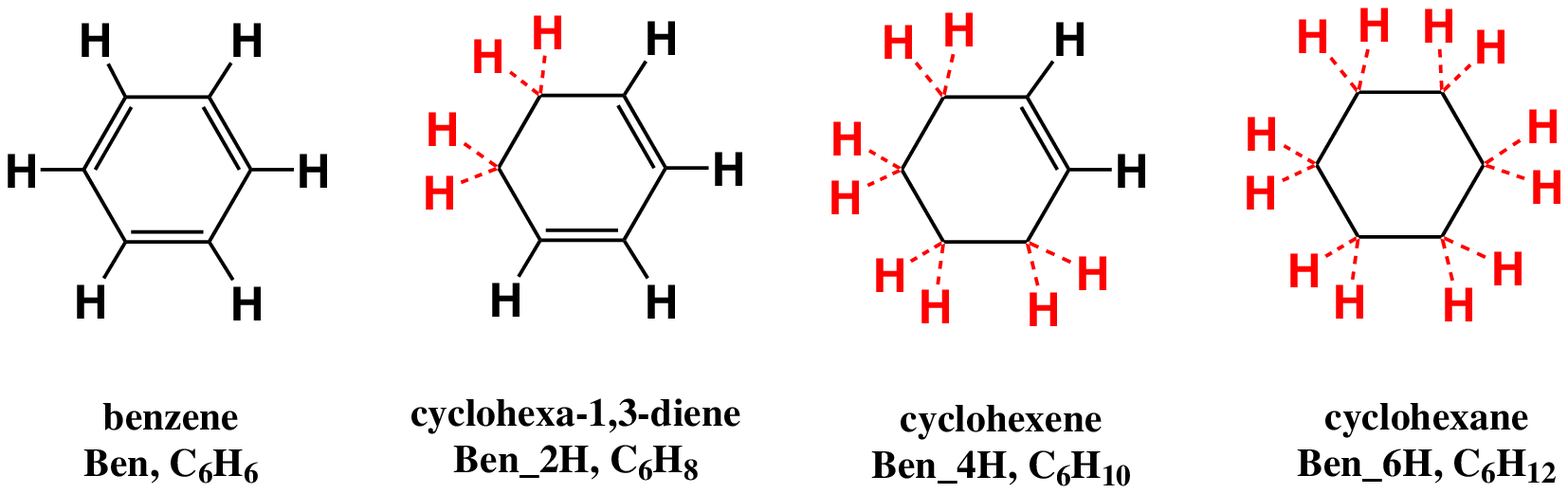}
}
\caption{\footnotesize
         \label{fig:HBenzene_structure}
         Structures of benzene and hydrogenated benzenes.
         }
\end{figure*}

Our target molecules include the derivatives
of benzene (Figure~\ref{fig:HBenzene_structure}),
of naphthalene (Figure~\ref{fig:HNaph_structure}),
of perylene (Figure~\ref{fig:HPery_structure})
and of those experimentally investigated by
Sandford et al.\ (2013;
see Figure~\ref{fig:HPAH_Sandford_structure}).
For all our target molecules, we consider hydrogenation
products that result from the addition of an even number
of H atoms. The radical species resulting from the addition
of an odd number of H atoms
are likely to have short lifetimes.
%
%
We will refer to hydrogenated species
by the abbreviation of the first three or four letters
of their parent PAH name followed by the number
of extra H atoms (e.g., Pery$\_$2H refers to perylene
attached with 2 extra H atoms). But for those molecules
studied by Sandford et al.\ (2013),
we shall adopt the abbreviations
given by them (see Table~1 of
Sandford et al.\ 2013).
More descriptive names shall also be
used if such names are common.
Many of the structures allow for structural isomers
(i.e., H atoms are attached at different positions)
and some of the structural isomers may adopt
several conformations of the same connectivity
but different spacial arrangement.
Isomers and conformers will be distinguished
by the addition of a letter.

In general, we will focus on the most likely
and/or most stable structure.
For benzene, for example, we will only consider
cyclo--1,3--hexadiene
(Ben$\_$2H; see Figure~\ref{fig:HBenzene_structure})
and ignore the less stable isomer cyclo--1,4--hexadiene.
Also, we will consider only cyclohexene
(Ben$\_$4H; see Figure~\ref{fig:HBenzene_structure})
and ignore all derivatives
in which the CH groups are not geminal.
Moreover, we will consider only the best conformation
for all these systems such as the chair conformation
of cyclohexane
(Ben$\_$6H; see Figure~\ref{fig:HBenzene_structure})
while ignoring the less stable boat conformations.
For dihydroperylene
(Pery$\_$2H; see Figure~\ref{fig:HPery_structure})
we will consider three structural isomers
(Pery$\_$2H$\_$RamII,
Pery$\_$2H$\_$RamIII,
and Pery$\_$2H$\_$RamIV;
see Figure~\ref{fig:HPery_structure}).
We will describe these isomers and their conformations
in more detail below along with the structures of
the other superhydrogenated perylenes.

To verify our computations,
we compare our computational results
with experimental measurements.
Figure~\ref{fig:CHSpec_Naph_ExpAndCal} shows
the computed IR spectra (color lines)
along with the experimental results (black line)
for the hydrogenated species of naphthalene,
i.e., Series~A marked by Sandford et al.\ (2013).\footnote{
  We note that here the band strengths of
  the experimental spectra are scaled
  to that of the calculated spectra
  since Sandford et al.\ (2013) did not
  report the absolute band strengths
  of these molecules.
  }
In Figure~\ref{fig:CHSpec_Naph_ExpAndCal},
the red dashed lines represent the computational spectra
applied with the frequency scale factor ($\gamma$)
of $\simali$0.9688 given by Borowski (2012).
As we can see that the scaled computational spectra
are systematically blue-shifted
with respect to the experimental spectra.
Hence, we determine an optimized scale factor
of $\gamma\approx0.963$. With this scale factor,
the agreement between computations (purple solid lines)
and experiments (black solid lines) is remarkably improved
for bands which correspond to pure fundamental vibrations,
and this fact attests to the quality of our computations.

In Figure~\ref{fig:CHSpec_HHP_ExpAndCal},
we further compare our computational spectra
of HHP (i.e., Pyre$\_$6H, C$_{16}$H$_{16}$)
and THB[a]p (C$_{20}$H$_{16}$;
see Figure~\ref{fig:HPAH_Sandford_structure})
with the experimental spectra of Sandford et al.\ (2013).
As we can see, with a line width of 10$\cm^{-1}$
and a scale factor of 0.963 for the frequencies,
our calculations agree quite well with the experiments.
Therefore, we believe that our calculations are reliable,
and the optimized scaling factor of $\gamma$\,=\,0.963
for frequencies will be applied in the following.

The intensity scaling is much more complicated
than the frequency scaling
since the experimental data for
the band intensities of hydrogenated PAHs
are rare and, 
in band assignment, it is often difficult to 
obtain a one-to-one correspondence 
between the experimental and computational spectra. 
Therefore,  a common way is to derive the relative strength,
the strength of one band (e.g., the 3.4$\mum$ band)
relative to another band (e.g., the 3.3$\mum$ band),
and then compare the relative band strengths
of the computational data with that of experimental data,
with the band intensity obtained 
by integrating the intensity profile of the band
which contains several neighboring peaks.
This will be discussed in detail in \S\ref{subsec:SandfordSet}. 
In the following, we will present the calculated intensities 
without any scaling.

\section{Results\label{sec:Results}}

\subsection{Benzene Derivatives \label{subsec:Benzenes}}

We start from the smallest PAH molecule benzene
and its hydrogenated derivatives.
The computed total energies and thermochemical
parameters are given in Table~\ref{tab:E_ThermPara_Benzene}.
The calculated spectra for neutral Ben$\_$2H, Ben$\_$4H
and Ben$\_$6H are shown in Figure~\ref{fig:BenNISTSpec}
along with the corresponding experimental spectra from
the {\it National Institute of Standards and Technology} (NIST).
The calculated frequencies are scaled with our optimized factor
of $\gamma=0.963$.
For each spectral feature,
we assign a line width of 20$\cm^{-1}$.
Since NIST only provides information about the absorbance
but not the condensations, we cannot derive the absolute
intensities for the features of the NIST experimental spectra.
Therefore, we just scale the intensity of the NIST spectra
with an appropriate factor to match our calculated spectra.
Figure~\ref{fig:BenNISTSpec} shows that for Ben$\_$2H,
the calculation is in good agreement
with the NIST experimental spectrum
in the range of $\simali$2700--3300$\cm^{-1}$.
In the range of $\simali$500--1500$\cm^{-1}$,
a smaller scaling factor would lead to a better match.
For Ben$\_$4H and Ben$\_$6H, the current scaling factor
for frequencies already gives a satisfactory agreement
with the experiment.

The upper panel of Figure~\ref{fig:BenSpec}
shows the spectra of neutral benzene,
toluene (i.e., methyl-benzene)
and all the hydrogenated sepcies of benzene.
It is quite clear that with the presence of
aliphatic C--H bonds, the 3.4$\mum$ feature
(at $\simali$2900$\cm^{-1}$) arising from
the aliphatic C--H stretch shows up.\footnote{%
  For benzene, all the C atoms are aromatic,
  while for Ben$\_$6H, all the C atoms are aliphatic.
  }
Meanwhile, all the benzene derivatives
except benzene also exhibit
a feature at $\simali$450--500$\cm^{-1}$
which arises from the out-of-plane bending
of aliphatic C--H.
%
%
The aromatic C--C stretch at $\simali$6.8$\mum$
(around 1470$\cm^{-1}$) of benzene
becomes weaker in toluene,
Ben$\_$2H and Ben$\_$4H, and is absent
in Ben$\_$6H in which all the C atoms are aliphatic.

The lower panel of Figure~\ref{fig:BenSpec}
shows the spectra of the cations of benzene
and its hydrogenated derivatives as well as toluene.
It is apparent that highly hydrogenated cations
exhibit several strong features that are not
seen in their neutral counterparts.
For Ben$\_$4H+,
a feature at $\simali$1300$\cm^{-1}$
originating from the aliphatic C--H
out-of-plane bending is quite prominent.
Ben$\_$6H+ shows strong features
at $\simali$380$\cm^{-1}$ from
the aliphatic C--C stretch,
at $\simali$700$\cm^{-1}$ from
the aliphatic C--H out-of-plane bending,
and at $\simali$840$\cm^{-1}$
from the aliphatic C--H in-plane bending.
Furthermore, the aliphatic C--H stretch features
of cations occur at longer wavelengths
with respect to the neutrals,
especially for Ben$\_$6H+.
Moreover, all the C--H stretch features are
significantly suppressed for cations,
while their features in the range of
$\simali$1200--1600$\cm^{-1}$ are
significantly enhanced.

The calculated intensities for the major aliphatic
vibrational modes as well as the 3.3 and 6.2$\mum$
aromatic modes are accumulated in
Table~\ref{tab:AValueARatio_Benzene}.
Unless otherwise noted,
$\Aali$, the band strength or intensity of
the 3.4$\mum$ aliphatic C--H stretch,
and $\Adfa$, the band strength of
the 6.85$\mum$ aliphatic C--H deformation,
are given on a per aliphatic C--H bond basis
all through this paper.
While $\Aaro$, the band strength of
the 3.3$\mum$ aromatic C--H stretch 
is given on a per aromatic C--H bond basis,
$\Acc$,  the band strength of
the 6.2$\mum$ C--C stretch,
is given on a per aromatic C atom basis. 
The mean intensity for each feature 
obtained by averaging over
all the hydrogenated derivatives
of benzene (i.e.,  Ben$\_$2H+,
Ben$\_$4H+, and Ben$\_$6H+)
is also tabulated in Table~\ref{tab:AValueARatio_Benzene}.
%

\subsection{Naphthalene Derivatives \label{subsec:Naphthalenes}}
\begin{figure*}
\centering{
\includegraphics[scale=0.7,clip]{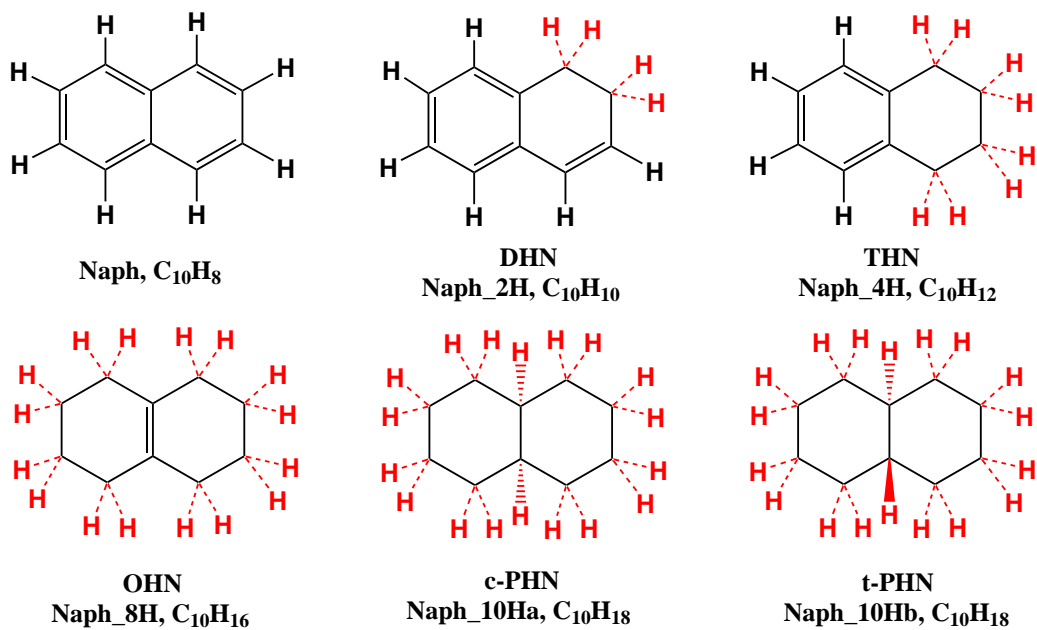}
}
\caption{\footnotesize
         \label{fig:HNaph_structure}
         Structures of naphthalene and hydrogenated naphthalenes.
         The hydrogenated naphthalenes are labelled
         ``Series A'' in Sandford et al.\ (2013).
         }
\end{figure*}

\begin{figure*}
\centering{
\includegraphics[scale=0.32,clip]{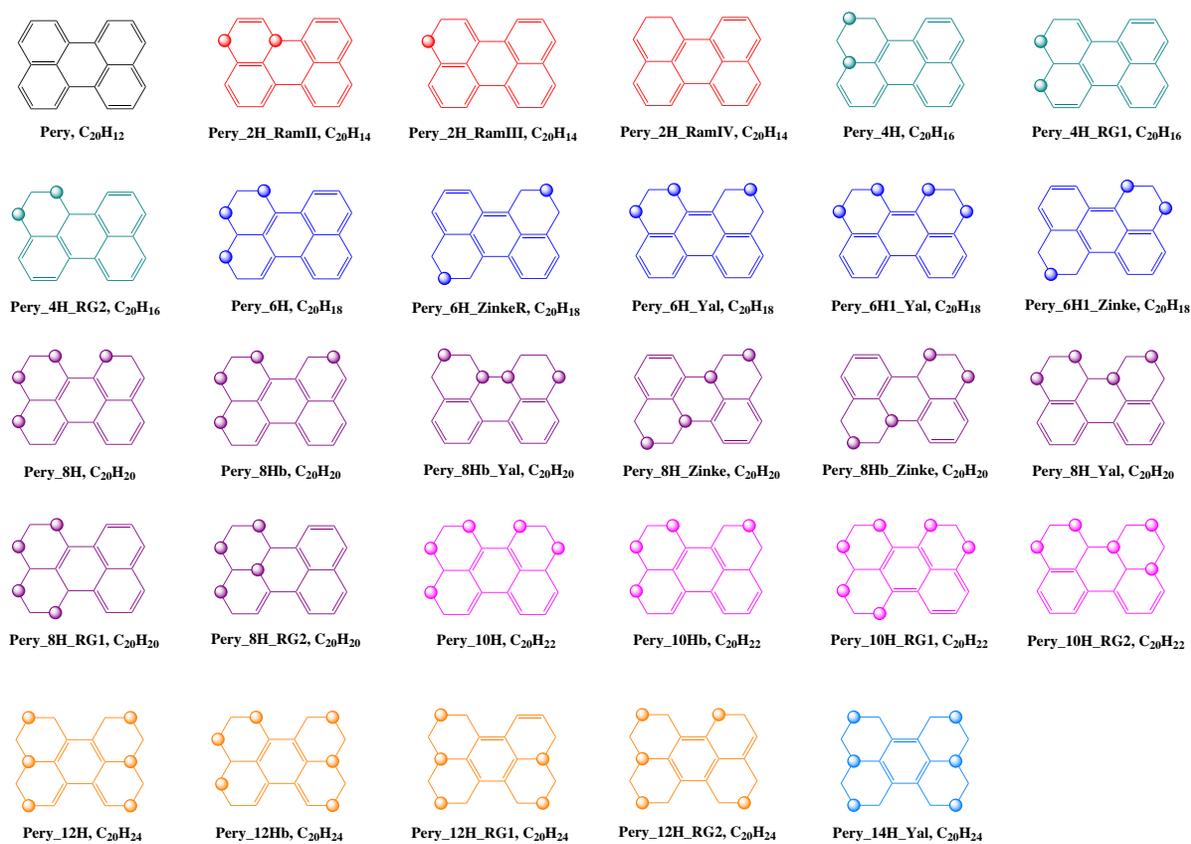}
}
\caption{\footnotesize
         \label{fig:HPery_structure}
         Structures of perylene and hydrogenated perylenes.
         All the molecules are named as Pery plus
         the number of extra H atoms. The hydrogenated
         molecules are shown in different colors
         with each color for PAHs having the same
         number of extra H atoms.
         }
\end{figure*}

\begin{figure*}
\centering{
\includegraphics[scale=0.45,clip]{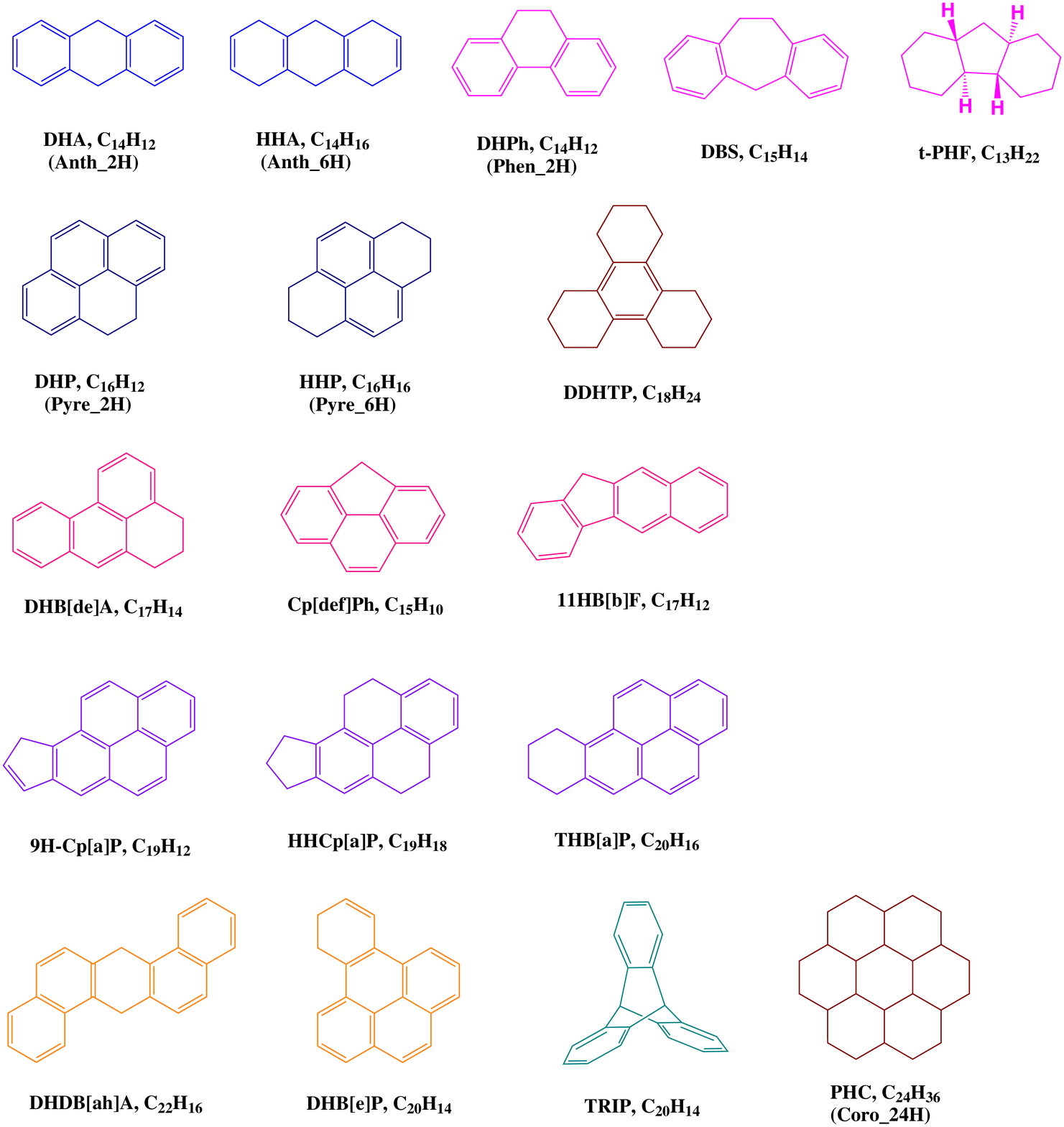}
}
\caption{\footnotesize
          \label{fig:HPAH_Sandford_structure}
          Structures of all the hydrogenated PAHs
          (except hydrogenated naphthalenes)
          experimentally studied by Sandford et al.\ (2013).
          These molecules were marked
          Series B, Series C, ..., and Series H
          in Sandford et al.\ (2013) and are shown here
          in different colors with each color for one series.
          }
\end{figure*}

\begin{figure*}
\centering
{
\includegraphics[width=1.0\textwidth,angle=0]{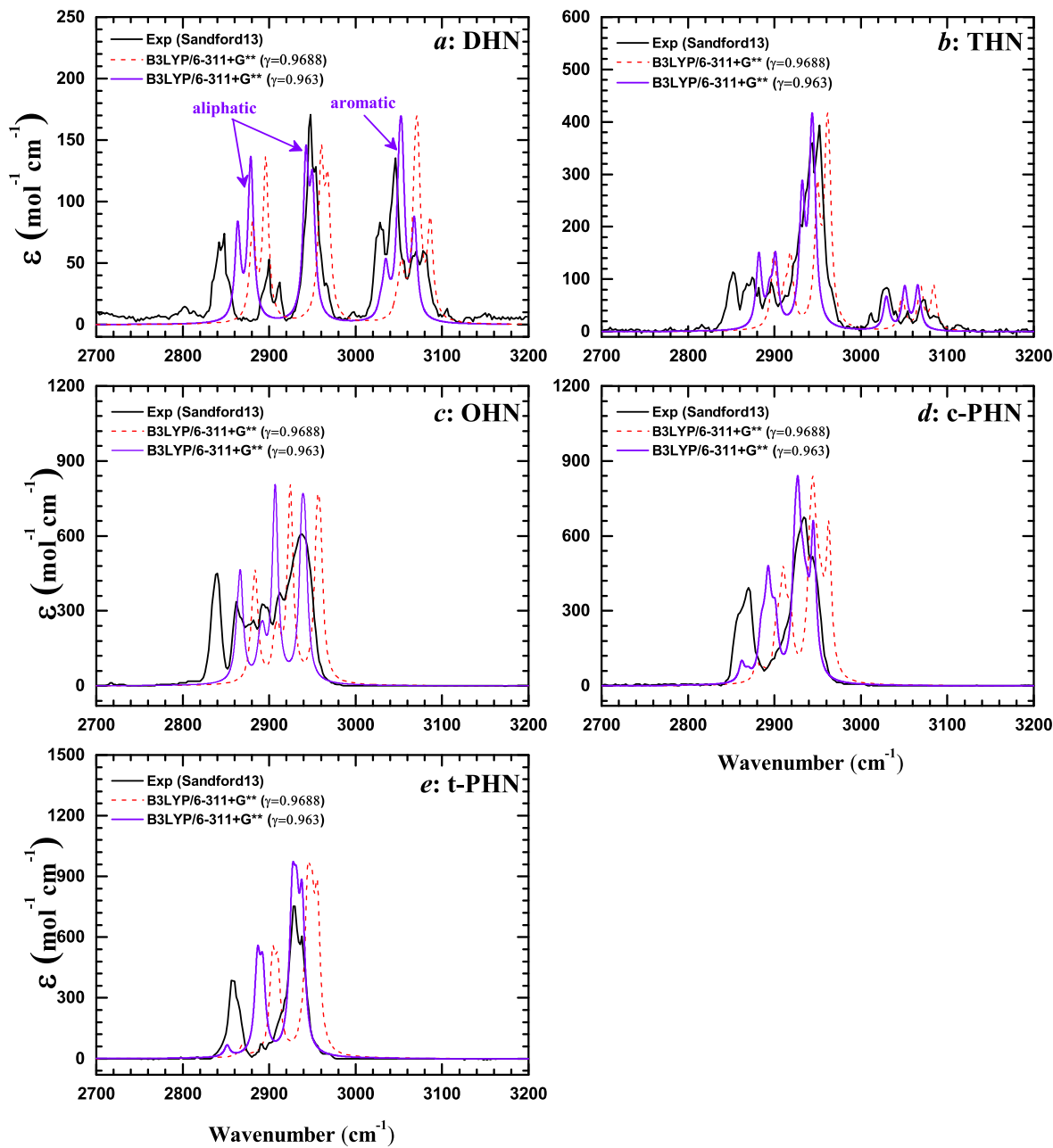}
}
\vspace{-10mm}
\caption{\footnotesize
         \label{fig:CHSpec_Naph_ExpAndCal}
         Comparison of the absorption spectra of
         hydrogenated naphthalenes computed at level
         {\rm B3LYP/6-311+G$^{\ast\ast}$}
         with the experimental spectra
         of Sandford et al. (2013)
         [marked with ``Exp\,(Sandford13)''].
         We assign a line width of 4$\cm^{-1}$
         for the computed spectra.
         The frequencies of the computed spectra
         are multiplied by a scaling factor ($\gamma$)
         to be comparable with
         the experimental spectra (black solid lines).
         The red dashed lines are the computed spectra
         applied with $\gamma=0.9688$ to the frequencies,
         and the purple solid lines are those applied with
         an optimized $\gamma$ of 0.963.
         The band intensities --- expressed as
         the molar absorption coefficient
         ($\varepsilon$) in units of $\mol^{-1}\cm^{-1}$
         --- of the experimental spectra are scaled
         to that of the computed spectra
         since Sandford et al.\ (2013) did not
         measure the absolute band intensities.
         Note that in panel (a) for DHN,
         the 3.4$\mum$ aliphatic C--H stretch
         consists of two complexes at $\simali$2870$\cm^{-1}$
         and $\simali$2950$\cm^{-1}$,
         while the 3.3$\mum$ aromatic C--H stretch
         has only one complex at $\simali$3050$\cm^{-1}$.
         }
\end{figure*}

We then consider PAH molecules with two benzene rings,
i.e., naphthalene and its hydrogenated derivatives.
For this group, we calculate the vibrational spectra
of five molecules (see Figure~\ref{fig:HNaph_structure}).
The vibrational spectra of these molecules,
marked as ``Series A'' in Sandford et al.\ (2013),
were experimentally obtained by Sandford et al.\ (2013).
This allows us to confront our computational spectra
with the experimental spectra.
The computed total energies and thermochemical
parameters are given in Table~\ref{tab:E_ThermPara_Naph}.

The calculated spectra are shown
in Figure~\ref{fig:Naph_Spec_NC}
which clearly shows that for the aliphatic
C--H stretch features, the neutrals have
much higher intensities
and peak at shorter wavelengths
with respect to cations.
For highly hydrogenated cations,
e.g., Naph$\_$10H, the aliphatic C--H stretch
peaks at $\simali$3.5$\mum$.
Meanwhile, the cations have much stronger features
at $\simali$1300--1500$\cm^{-1}$ than the neutrals,
just as the benzene derivatives.
For methylated naphthalene,
as shown in Figure~\ref{fig:Naph_Spec_NC},
the aliphatic C--H stretch
also shows up in the computed spectra,
but not as pronounced as hydrogenated naphthalene.
Also, compared with hydrogenated naphthalene,
the aliphatic C--H stretch of methylated naphthalene
occurs at somewhat shorter wavelengths.

The intensities of the major aliphatic vibrational modes
for the hydrogenated cations are shown
in Table~\ref{tab:AValueARatio_Naph}.
Also tabulated are the mean intensities of
individual features obtained by
averaging over all five hydrogenated cations.
For c-PHN  (i.e., Naph$\_$10Ha;
see Figure~\ref{fig:HNaph_structure})
and t-PHN (i.e., Naph$\_$10Hb;
see Figure~\ref{fig:HNaph_structure}),
they are fully hydrogenated
and thus have no aromatic features.\footnote{%
   The fully hydrogenated species c-PHN 
   and t-PHN are isomers.
   For c-PHN, the hydrogen atoms attached 
   to the two carbon atoms at the junction of the 
   two benzene rings are on the same side of 
   the PAH plane, while for t-PHN the hydrogen atoms
   are on the opposite side.
   }
They are both fully hydrogenated
and thus have no aromatic features.
For OHN (i.e., Naph$\_$8H;
see Figure~\ref{fig:HNaph_structure}),
although it is not fully hydrogenated,
it has no H attached to aromatic C atoms and thus
the 3.3$\mum$ aromatic C--H stretch is also absent.
The aromatic C--C stretch features are seriously
suppressed and essentially have negligible intensities
for both neutrals and cations.

\subsection{Perylene Derivatives \label{subsec:Perylenes}}

A larger PAH molecule, perylene (C$_{20}$H$_{12}$),
which has as many edge C atoms as possible to be
hydrogenated for PAHs of five six-membered rings,
is selected as our next sample.
A series of extra H atoms has been considered,
from two all the way up to 14
which corresponds to a complete hydrogenation
of all the edge C atoms.
For each situation, we consider several isomers
of which the extra H atoms are attached
at different positions.
Table~\ref{tab:E_ThermPara_Pery_Neutral}
present the computed total energies
and thermochemical parameters for
the neutrals and cations, respectively.

We show in
Figure~\ref{fig:Pery2HSpec}
the calculated spectra for each hydrogenation
along with that of perylene
and methyl-perylene.\footnote{%
  For Pery$\_$14H, only one isomer is calculated,
  so its spectrum is shown with the Pery$\_$12H isomers
  in Figure~\ref{fig:Pery2HSpec}.
  }
In all these figures, the neutrals are displayed
in the upper panels, and the cations in the lower ones.
For the neutrals, the spectra do not seem to
exhibit significant differences among different isomers.
Again, the most prominent aliphatic feature
is the aliphatic C--H stretch at
3.4$\mum$ for all the neutrals
which becomes stronger with the degree
of hydrogenations and shows a strong peak
around 2950$\cm^{-1}$ accompanied by
a series of satellite features
at longer wavelengths.
%
As the number of extra H atoms increases,
the aromatic features becomes weaker as expected,
and intend to shift to the red side,
especially for heavy hydrogenations
(e.g., superhydrogenated perylene
with eight or more extra H atoms).
On the other hand, the aliphatic C--H
deformation bands at 6.85$\mum$ (1470$\cm^{-1}$)
and 7.45$\mum$ (1310$\cm^{-1}$)
become stronger as the hydrogenation increases.
Moreover,  the aliphatic C--H stretch
of methyl-perylene appears to occur
at a shorter wavelength than perylene.
Compared with that of the neutral molecules,
both the 3.3$\mum$ aromatic and 3.4$\mum$ aliphatic
C--H stretch features are significantly suppressed for
all the cations, while the cationic C--C stretch
and C--H bending bands
at $\simali$1200--1600$\cm^{-1}$ are
considerably enhanced.
This is similar to benzene and naphthalene
and their derivatives.
Unfortunately, we cannot compare these computed
spectra with the experimental spectra
since, to our knowledge, there are no experimental
data available for the perylene derivatives.

The intensities computed for the aliphatic bands
at 3.4 and 6.85$\mum$ and the aromatic bands
at 3.3 and 6.2$\mum$ are shown
in Table~\ref{tab:AValueARatio_NeuPery}.
Also tabulated in Tables~\ref{tab:AValueARatio_NeuPery}
are the mean intensities for each band,
obtained by averaging over all the derivaties
of perylene.

\subsection{The Sandford et al.\ (2013) Molecules
                     \label{subsec:SandfordSet}}
Sandford et al.\ (2013) obtained the 2.5--20$\mum$
(i.e., 500--4000$\cm^{-1}$) absorption spectra
of 23 hydrogenated PAHs and related molecules
isolated in the argon matrix at 15$\K$.
We now consider all the hydrogenated PAH species
experimentally studied by Sandford et al.\ (2013)
except the derivatives of naphthalene,
i.e., Series~B to Series~H as marked
by Sandford et al.\ (2013),
which have already been discussed in detail in \S\ref{subsec:Naphthalenes}.
To highlight the vibrational features originated from
the addition of extra H atoms, we also calculate
the vibrational spectra of the parental molecules,
e.g., the parental molecule of the Series~B molecules
of Sandford et al.\ (2013) is anthracene (C$_{14}$H$_{10}$).
Again, we consider both neutrals and cations
and tabulate in Table~\ref{tab:E_ThermPara_LongChain}
the computed total energies and
thermochemical parameters for these molecules.
%

In Figure~\ref{fig:Anth_Spec_NC}
we present the calculated spectra,
with each subfigure for one group
of which the molecules share the same parent
(e.g., the 1st subfigure is
for the derivatives of anthracene,
and the 2nd subfigure is
for the derivatives of phenanthrene).
Again, we also show in each figure the spectra
computed for the mono-methylated molecules.
It is apparent that, as shown in the upper panel
of each figure, all hydrogenated neutral molecules
exhibit the aliphatic C--H stretching feature
around 3.4$\mum$. This feature is also seen
in methylated neutral molecules
but at a slightly shorter wavelength.
The aliphatic C--H deformation feature
at $\simali$6.85$\mum$ is also prominent
in the spectra of all the molecules
in the range of $\simali$1400--1450$\cm^{-1}$.
However, the 7.25$\mum$ aliphatic C--H deformation feature
at $\simali$1350--1400$\cm^{-1}$ is much less pronounced.
In Table~\ref{tab:AValueARatio_NeuSandford}
we present the intensities of the 3.4 and 6.85$\mum$
aliphatic C--H bands as well as the 3.3 and 6.2$\mum$
aromatic bands. Also tabulated are the corresponding
aliphatic-to-aromatic intensity ratios
$A_{3.4}/A_{3.3}$ and $A_{6.85}/A_{6.2}$.
Similarly, we show in the lower panel of each figure
the spectra of the cationic counterparts of those
presented in the upper panel. Clearly, the C--H stretch
at $\simali$3.4$\mum$ is considerably reduced
while the C--C stretching and C--H bending features
in $\simali$1200--1600$\cm^{-1}$
are remarkably enhanced.
The intensities of the 3.3, 3.4, 6.2 and 6.85$\mum$
bands calculated for the cationic species are also
tabulated in Table~\ref{tab:AValueARatio_NeuSandford}.
For both neutrals and cations, we also derive
the mean intensity for each band by averaging
over all the derivatives (see
Table~\ref{tab:AValueARatio_NeuSandford}).

\begin{figure*}
\centering
{
\includegraphics[width=1.5\textwidth,angle=0]{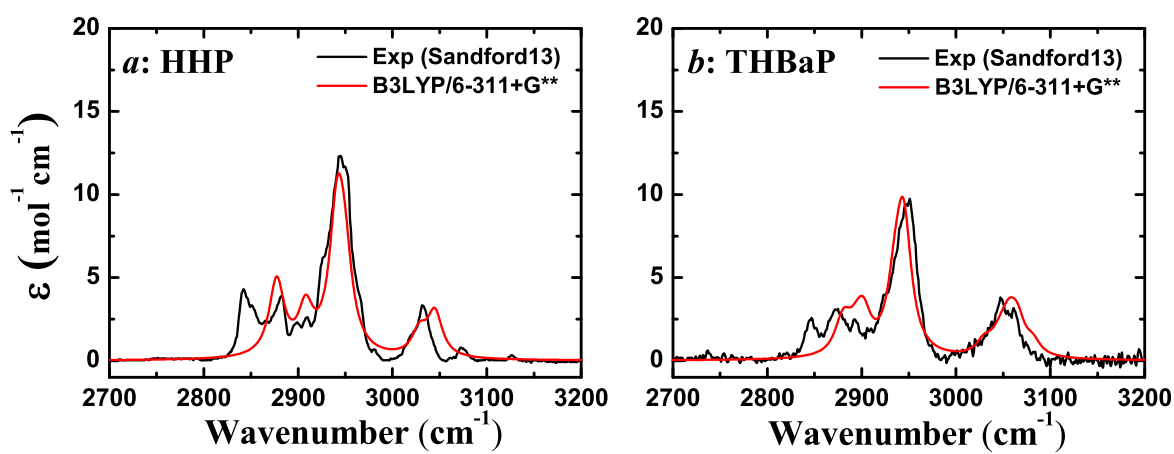}\vspace{-3.5cm}\hspace{-6.5cm}
}
\caption{\footnotesize
         \label{fig:CHSpec_HHP_ExpAndCal}
         Comparison of the absorption spectra of
         HHP (i.e., Pyre$\_$6H, C$_{16}$H$_{16}$;
         see Figure~\ref{fig:HPAH_Sandford_structure})
         and THB[a]P  (i.e., C$_{20}$H$_{16}$;
         see Figure~\ref{fig:HPAH_Sandford_structure})
         computed at level
         {\rm B3LYP/6-311+G$^{\ast\ast}$}
         with the experimental spectra
         of Sandford et al.\ (2013)
         [marked with ``Exp\,(Sandford13)''].
         To be comparable with the experimental
         spectra (black lines),
         the frequencies of the computed spectra
         (red lines) are multiplied by
         an optimized scaling factor of 0.963
         and a line width of 10$\cm^{-1}$ is assigned.
         The molar absorption coefficients $\varepsilon$
         of the experimental spectra are scaled
         to that of the computed spectra
         since Sandford et al.\ (2013) did not
         measure the absolute band intensities.
         }
\end{figure*}

\begin{figure*}
\begin{center}
\begin{minipage}[t]{1.0\textwidth}
\resizebox{16.8cm}{6cm}{\includegraphics[clip]{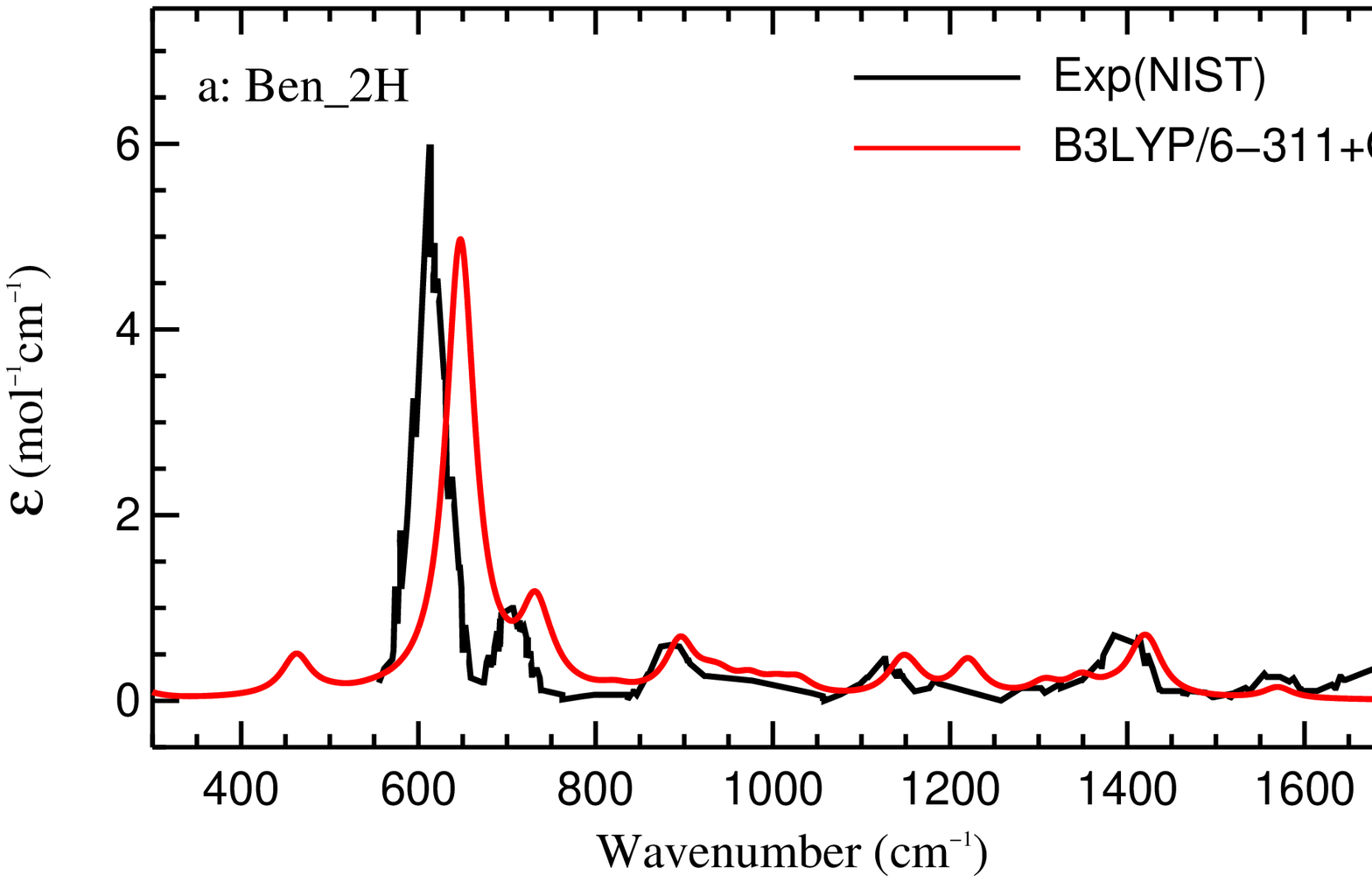}}\vspace{0.2cm}
\resizebox{16.8cm}{6cm}{\includegraphics[clip]{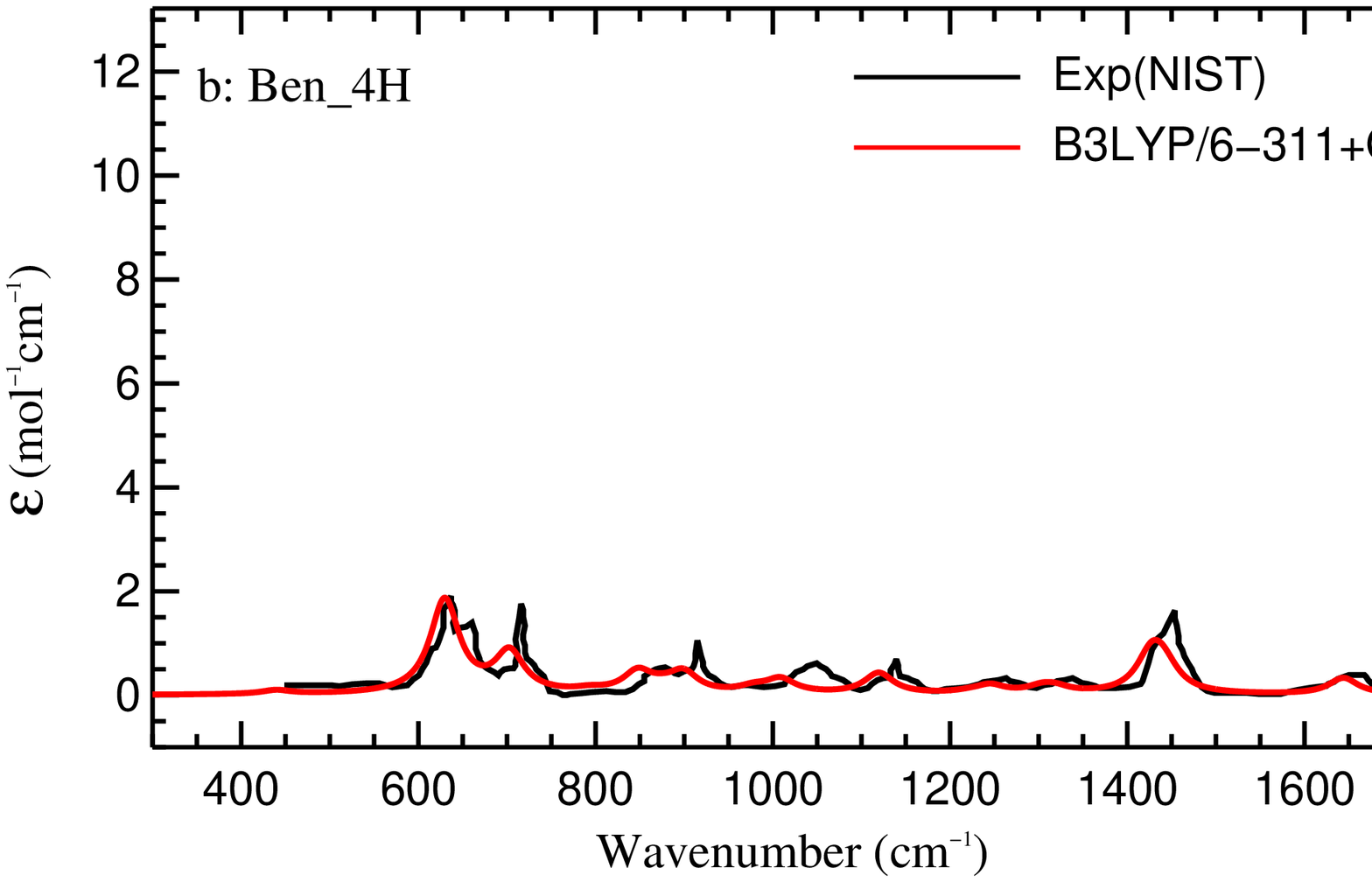}}\vspace{0.2cm}
\resizebox{16.8cm}{6cm}{\includegraphics[clip]{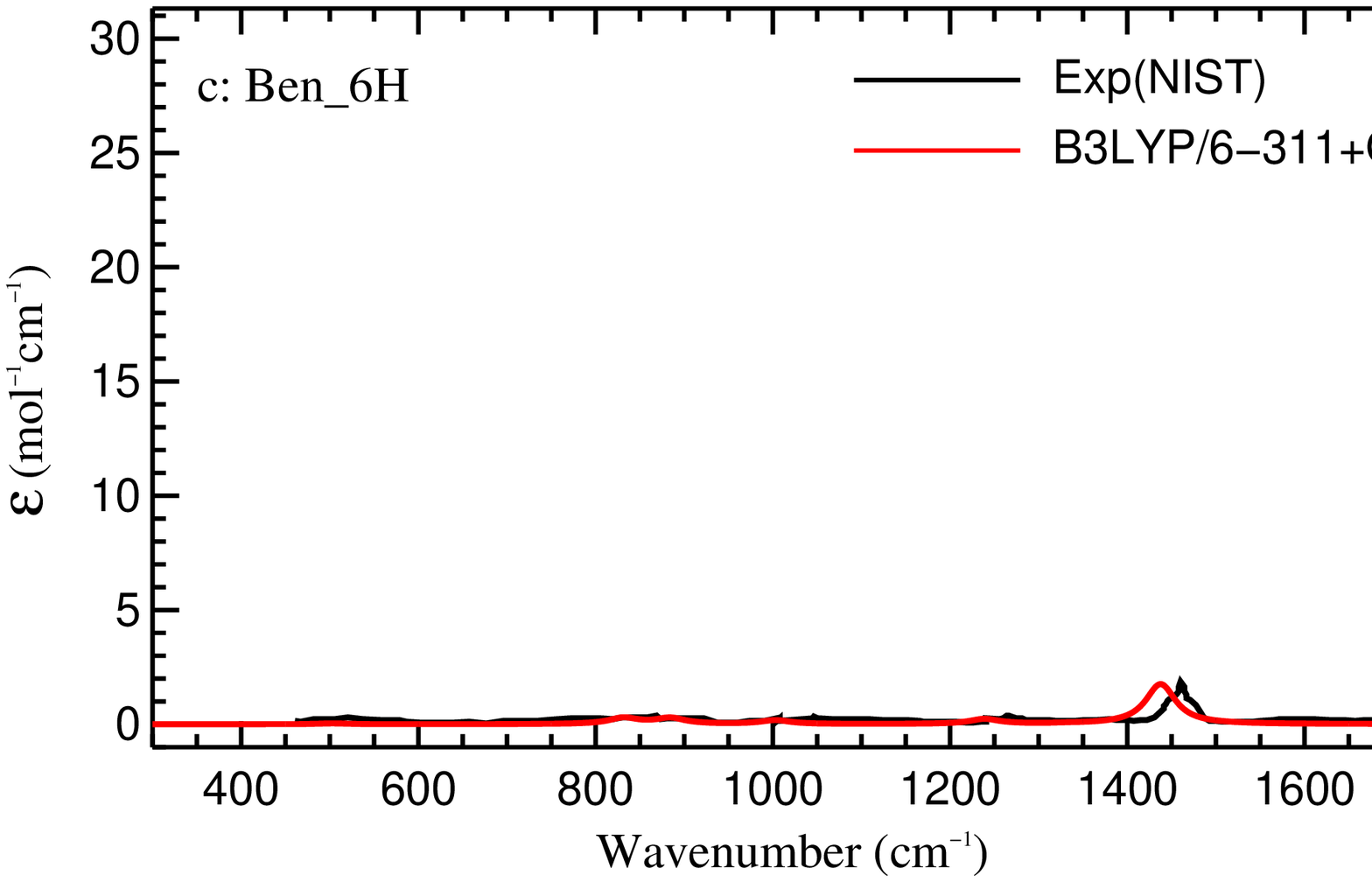}}\vspace{0.2cm}
\end{minipage}\hspace{-1.5cm}
\end{center}
\caption{\footnotesize
         \label{fig:BenNISTSpec}
         Comparison of the gas-phase absorption spectra
         experimentally measured by NIST
         (labelled with ``Exp\,(NIST)''; red lines)
         with the computed, frequency-scaled spectra
         (black lines) of hydrogenated benzenes.
         The molar absorption coefficients $\varepsilon$
         for the NIST experimental data
         are scaled to be comparable
         to the computed spectra
         by multiplying the NIST absorbance
         with an artificial factor,
         as NIST only gives the absorbance and
         does not provide information
         on the concentration to derive $\varepsilon$.
         }
\end{figure*}

\begin{figure*}
\begin{center}
\begin{minipage}[t]{\textwidth}
\resizebox{16.8cm}{8cm}{\includegraphics[clip]{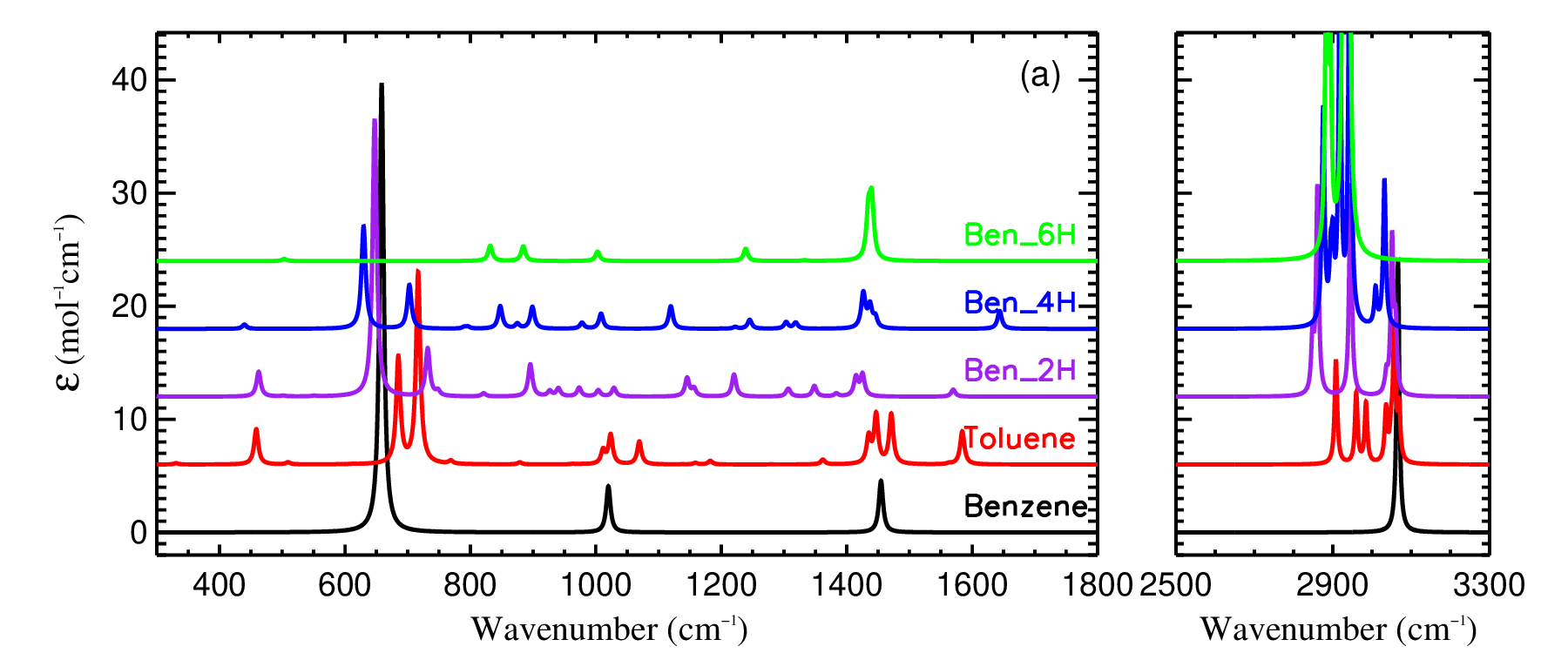}}\vspace{0.2cm}
\resizebox{16.8cm}{8cm}{\includegraphics[clip]{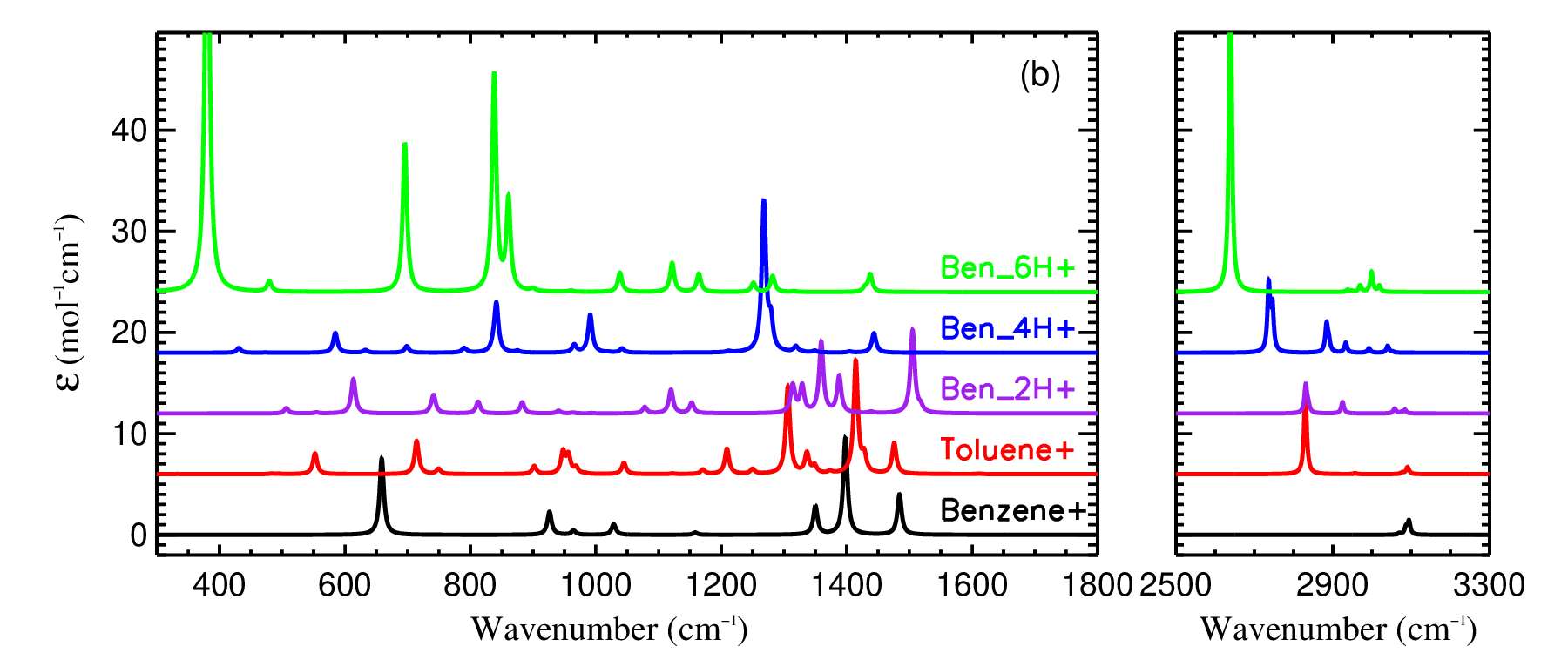}}\vspace{0.2cm}
\end{minipage}\hspace{-1.5cm}
\end{center}
\caption{\footnotesize
         \label{fig:BenSpec}
         Comparison of the calculated spectra
         of hydrogenated benzenes
         with that of benzene
         and methyl-benzene (i.e., toluene).
         The upper panels are for neutrals
         and the lower ones are for cations.
         The frequencies are scaled with
         a factor of 0.963, and a line width
         of 4$\cm^{-1}$ is assigned.
         }
\end{figure*}

\begin{figure*}
\begin{center}
\begin{minipage}[t]{1.0\textwidth}
\resizebox{16.8cm}{8cm}{\includegraphics[clip]{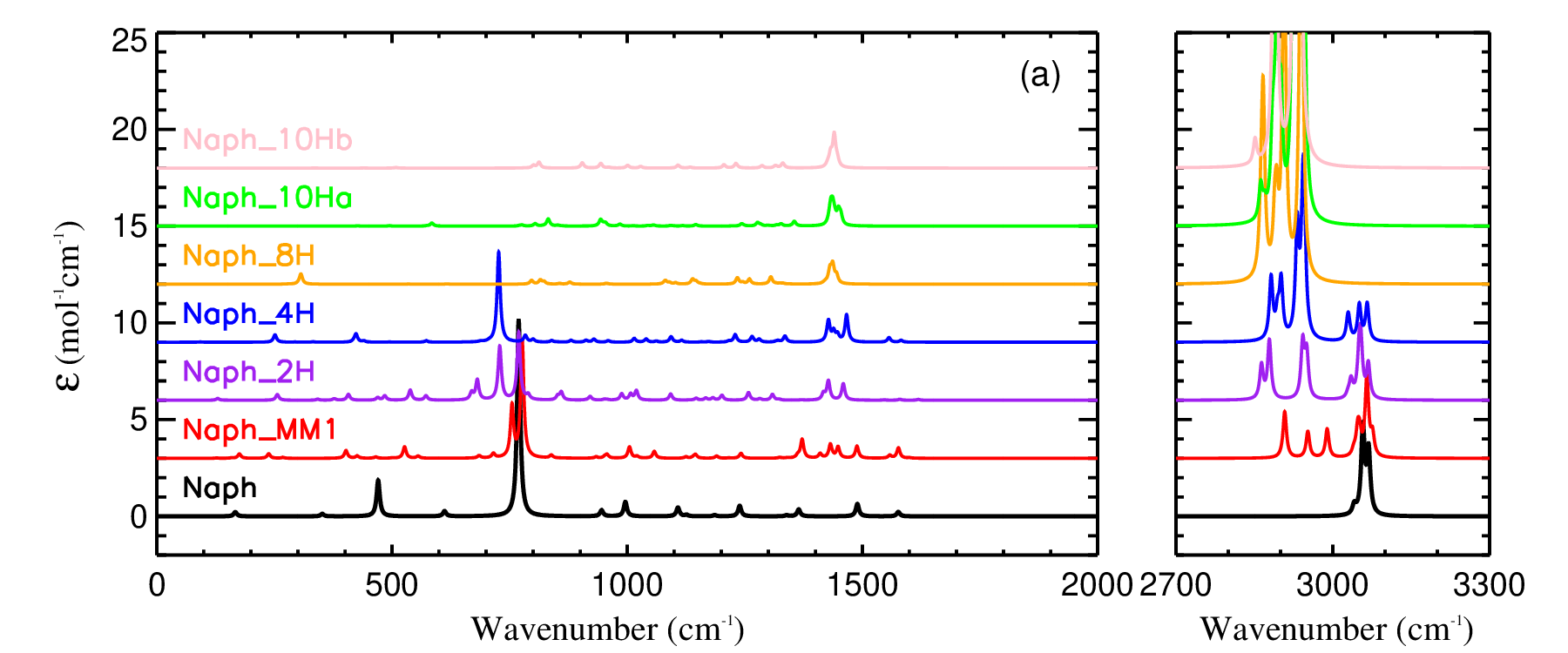}}\vspace{0.2cm}
\resizebox{16.8cm}{8cm}{\includegraphics[clip]{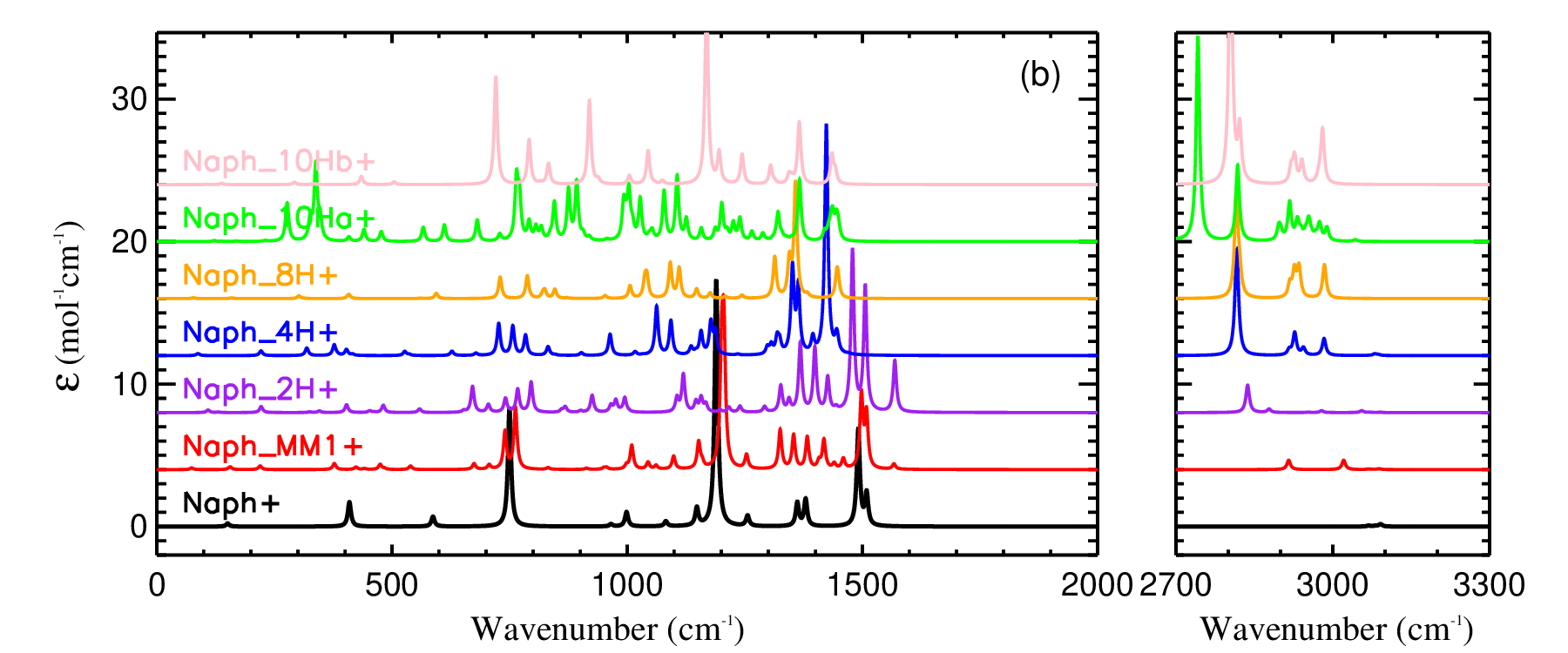}}\vspace{0.2cm}
\end{minipage}\hspace{-1.5cm}
\end{center}
\caption{\footnotesize
         \label{fig:Naph_Spec_NC}
         Same as Figure~\ref{fig:BenSpec}
         but for naphthalene and its hydrogenated
         and mono-methylated derivatives, 
         where ``MM1'' refers to
         mono-methylated species in which 
         the methyl group is attached 
         at position ``1'' of a PAH molecule 
         according to standard 
         {\it International Union Pure and Applied Chemistry} 
         (IUPAC) numbering. 
         }
\end{figure*}

\begin{figure*}
\begin{center}
\begin{minipage}[t]{1.0\textwidth}
\resizebox{16.8cm}{8cm}{\includegraphics[clip]{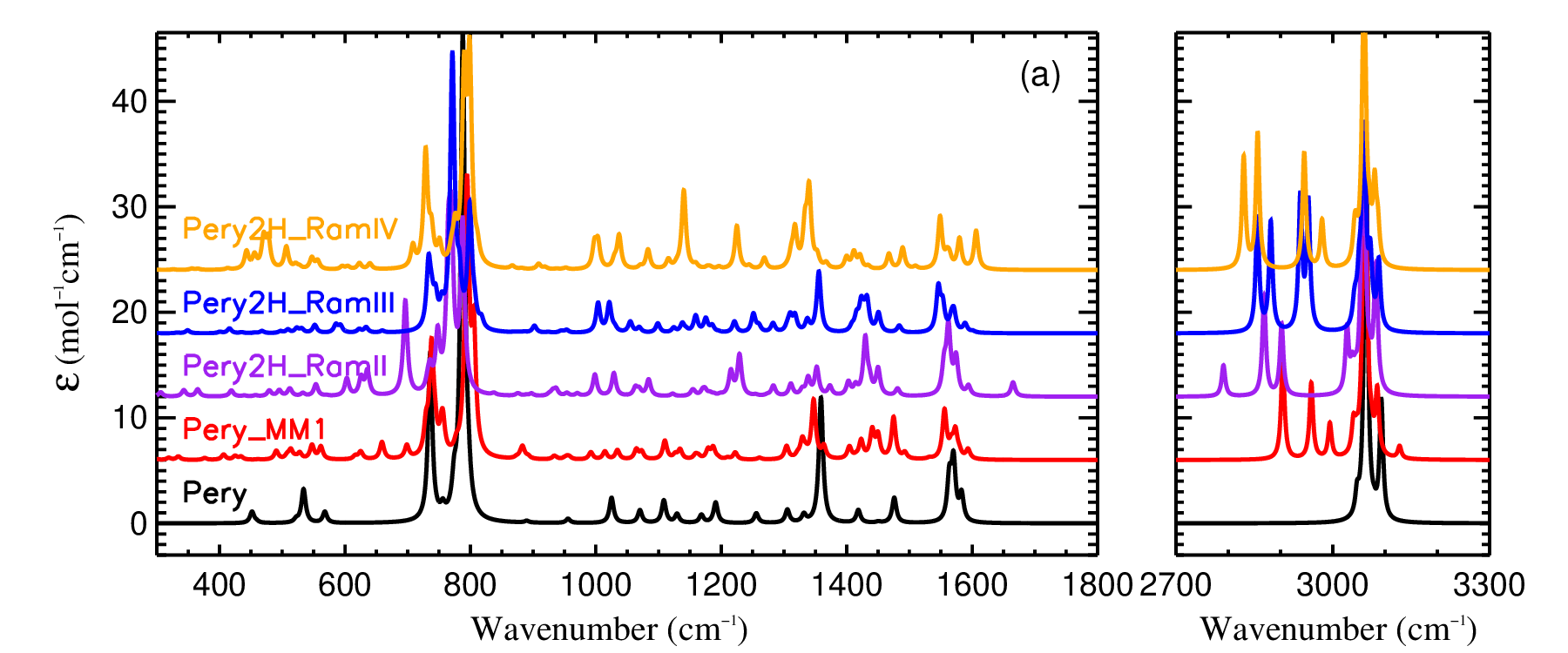}}\vspace{0.2cm}
\resizebox{16.8cm}{8cm}{\includegraphics[clip]{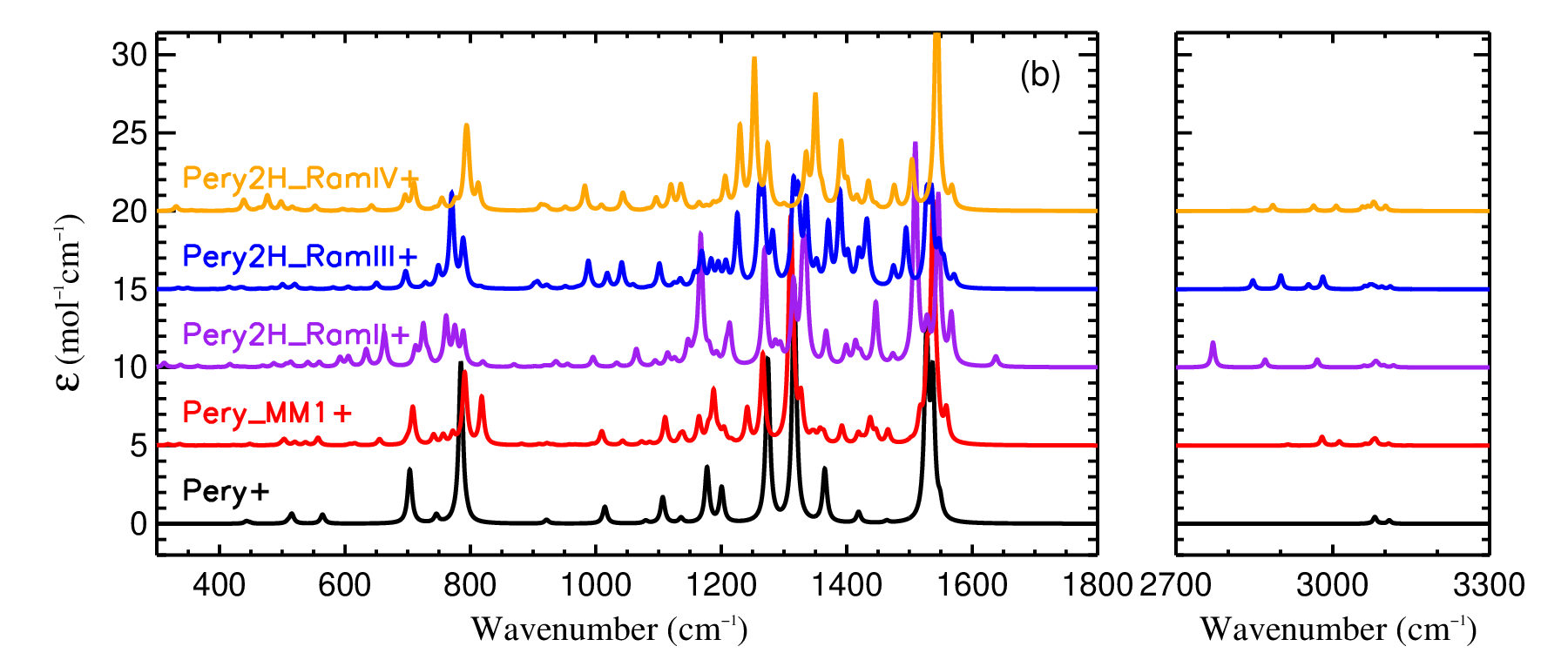}}\vspace{0.2cm}
\end{minipage}\hspace{-1.5cm}
\end{center}
\caption{\footnotesize
         \label{fig:Pery2HSpec}
         Comparison of the calculated spectra
         of Pery$\_$2H (perylene with two excess H atoms)
         with that of perylene and methyl-perylene.
         The upper panels are for neutrals
         and the lower ones are for cations.
         The frequencies are scaled with
         a factor of 0.963, and a line width
         of 4$\cm^{-1}$ is assigned.
         }
\end{figure*}

In Figure~\ref{fig:NHaliOverNHaro} we show
$A_{\rm CH,ali}/A_{\rm CH,aro}$
as a function of $N_{\rm H,ali}/N_{\rm H,aro}$,
where $A_{\rm CH,ali}$ is the intensity of
the 3.4$\mum$ aliphatic C--H stretch
of a given molecule,
$A_{\rm CH,aro}$ is the intensity of
the 3.3$\mum$ aromatic C--H stretch
of the given molecule, and
$N_{\rm H,\,ali}$ and $N_{\rm H,\,aro}$ are
respectively the total number of
aliphatic and aromatic C--H bonds
of the given molecule.
The slope of the line fit to the data
in Figure~\ref{fig:NHaliOverNHaro} is
$d\left(A_{\rm CH,ali}/A_{\rm CH,aro}\right)/
d\left(N_{\rm H,ali}/N_{\rm H,aro}\right)\approx1.85$,
which is lower than that of Sandford et al.\ (2013)
by $\simali$33\%. 
This difference appears to be mainly caused
by heavily hydrogenated species 
HHP (Pyre$_{-}$6H, C$_{16}$H$_{16}$)
and THB[a]P (C$_{20}$H$_{16}$), 
and slightly hydrogenated species
9H-Cp[a]P (C$_{19}$H$_{12}$)
and 11HB[b]F (C$_{17}$H$_{12}$).
The difference would be reduced to
within 10\% if these species are excluded.
Compared with the spectra of Sandford et al.\ (2013) 
obtained from the matrix isolation spectroscopy, 
the $A_{\rm CH,ali}/A_{\rm CH,aro}$ intensity ratios
computed here for HHP and THB[a]P are appreciably lower.
As illustrated in Figure~\ref{fig:CHSpec_HHP_ExpAndCal},
the computed spectra for both molecules are
in close agreement with the experimental spectra
of Sandford et al.\ (2013), except that the experimental
spectra exhibit an extra feature at $\simali$2840$\cm^{-1}$.
This would raise $A_{\rm CH,ali}$
and therefore result in a larger slope
for the experimental data. 
Moreover, Pauzat \& Ellinger (2001) 
performed DFT calculations on 
hydrogenated naphthalene, anthracene,
and pyrene and obtained 
$d\left(A_{\rm CH,ali}/A_{\rm CH,aro}\right)/
d\left(N_{\rm H,ali}/N_{\rm H,aro}\right)\approx1.74$,
which is close to that derived here.

Maltseva et al.\ (2018) applied
advanced laser spectroscopic techniques
combined with mass spectrometry
to obtain the gas-phase absorption spectra of
four supersonically cooled superhydrogenated
PAH species in the 3.175--3.636$\mum$
wavelength region, including slightly
hydrogenated DHA and DHPh 
(see Figure~\ref{fig:HPAH_Sandford_structure})
and heavily hydrogenated 
THN (see Figure~\ref{fig:HNaph_structure}) and
HHP (see Figure~\ref{fig:HPAH_Sandford_structure}).
They obtained a slope of
$d\left(A_{\rm CH,ali}/A_{\rm CH,aro}\right)/
d\left(N_{\rm H,ali}/N_{\rm H,aro}\right)\approx1.57$
which is closer to that derived here 
but considerably lower 
than that of Sandford et al.\ (2013).\footnote{%
   Maltseva et al.\ (2018) also obtained 
   the experimental spectra of methylated PAHs
   in the C--H stretch wavelength region.
   The aliphatic to aromatic C--H stretch
   band ratio of methylated PAHs 
   experimentally derived by Maltseva et al.\ (2018) 
   is close to that determined from quantum-chemical
   computations (see Yang et al.\ 2013, 2016b).
   }
Maltseva et al.\ (2018) argued that the difference
might be traced back to the influence of the environment
on band intensities. It is known that the incorporation of
PAHs into rare gas matrices could cause a suppression
on the intensities of the vibrational bands compared to
that of isolated molecules (Joblin et al.\ 1994).
Maltseva et al.\ (2018) speculated that it is possible
that aromatic C--H stretch bands might be suppressed
to a larger extent under rare gas matrix conditions
than aliphatic C--H stretch bands. 
However, as shown in Figure~\ref{fig:CHSpec_HHP_ExpAndCal},
the relative strengths of the aliphatic and aromatic C--H
stretches of HHP and THB[a]P of Sandford et al.\ (2013)
measured with the matrix isolation spectroscopy method
agree closely with that computed here,
except that the experimental spectra show
an extra feature at  $\simali$2840$\cm^{-1}$
which is not seen in the computed spectra. 

In future work, a more precise assignment 
of the experimental and computational
spectral bands of hydrogenated PAHs 
and accurate intensity scaling 
would be necessary for accurately determining
their band strengths. 
Based on what are presently available,
we argue that the relative band strengths
derived here are generally reliable.

\section{Astrophysical Implications}\label{sec:Discussion}
\subsection{Average Spectra \label{subsec:AverageSpec}}
To highlight the features originated from hydrogenation,
we obtain the mean spectra of hydrogenated PAHs,
methylated PAHs and their bare parental compounds
as well as their cationic counterparts
(see Figure~\ref{fig:AverageSpec_Compare}).
The mean spectra of hydrogenated PAHs
are derived by averaging the computed spectra,
on a per aliphatic C--H bond basis,
over all the hydrogenated species shown in
Figures~\ref{fig:HBenzene_structure}--\ref{fig:HPAH_Sandford_structure},
including hydrogenated benzenes, 
hydrogenated naphthalenes, 
hydrogenated perylenes, 
and the hydrogenated molecules 
of Sandford et al.\ (2013).
For methylated PAHs, we average over
all the molecules listed in Figure~2
of Yang et al.\ (2013).
For bare PAHs, we average over benzene,
naphthalene, anthracene, phenanthrene,
pyrene, perylene and coronene,
the parental molecules of
the hydrogenated species shown in
Figures~\ref{fig:HBenzene_structure}--\ref{fig:HPAH_Sandford_structure}.
For both bare PAHs and methylated PAHs,
their mean spectra are obtained
on a per C atom basis.

As shown in Figure~\ref{fig:AverageSpec_Compare},
the 3.4$\mum$ feature is clearly seen in
the mean spectra of neutral hydrogenated PAHs
and of neutral methylated PAHs.
Meanwhile, a bump at $\simali$1430$\cm^{-1}$
(which is close to 6.85$\mum$) is also seen in
neutral hydrogenated and methylated PAHs.
For hydrogenated neutral PAHs, 
the average band strengths (per aliphatic C--H bond)
are $\langle \Aali\rangle \approx 33.6\pm8.8\km\mol^{-1}$
and $\langle \Adfa\rangle \approx 2.59\pm1.46\km\mol^{-1}$
(see Table~\ref{tab:AValueARatio_All}).
%
In contrast, neither the 3.4$\mum$ feature
nor the 6.85$\mum$ feature is seen
in the spectra of bare PAHs.
The mean spectra of hydrogenated PAH ions
and of methylated PAH ions also exhibit
the aliphatic C--H stretch at 3.4$\mum$,
but this feature is severely suppressed
with respect to neutrals.
In comparison, cations have much
stronger bands than neutrals
at $\simali$1200--1600$\cm^{-1}$.
For hydrogenated PAH ions, 
the average band strengths (per aliphatic C--H bond)
are $\langle \Aali\rangle \approx 13.6\pm8.7\km\mol^{-1}$
and $\langle \Adfa\rangle \approx 13.2\pm13.1\km\mol^{-1}$
(see Table~\ref{tab:AValueARatio_All}).
%
%
In the following, we shall focus on
the 3.4 and 6.85$\mum$ aliphatic C--H features.

\subsection{$\Aratio$ \label{subsec:C--HSpec}}
Figure~\ref{fig:A33A34_all} shows $\Aratio$,
the band-strength ratio of
the 3.4$\mum$ aliphatic C--H stretch
to the 3.3$\mum$ aromatic C--H stretch
computed for all the hydrogenated PAHs
and their ions listed in
Figures~\ref{fig:HBenzene_structure}--\ref{fig:HPAH_Sandford_structure}.
For neutral hydrogenated PAHs,
the band-strength ratios $\Aratio$,
with an average value of
$\langle\Aratio\rangle\approx1.98$
and a standard deviation of $\simali$0.60
(see Table~\ref{tab:AValueARatio_All}),
do not vary much from one molecule to another.
In contrast, $\Aratio$ varies more appreciably
among ions (with $\langle\Aratio\rangle\approx7.73$
and a standard deviation of $\simali$6.56;
see Table~\ref{tab:AValueARatio_All}),
not only for different molecules
but also for different isomers of the same molecule.
Nevertheless, $\Aratio$ basically exhibits
a low-end of $\simali$1.0
for all the cationic hydrogenated PAHs.

\subsection{$\Adfa/\Acc$\label{subsec:C--CSpec}}
Figure~\ref{fig:A685A62_all} shows $\Adfa/\Acc$,
the band-strength ratio of
the 6.85$\mum$ aliphatic C--H deformation
to the 6.2$\mum$ aromatic C--C stretch
computed for all the hydrogenated PAHs
and their ions listed in
Figures~\ref{fig:HBenzene_structure}--\ref{fig:HPAH_Sandford_structure}.
For the Sandford et al.\ (2013) molecules,
the $\Adfa/\Acc$ ratio varies considerably
from one molecule to another.
Nevertheless, for $\simali$70\% of
the Sandford et al.\ (2013) molecules
$\Adfa/\Acc$ does not exceed $\simali$5.0.
For the hydrogenated derivatives of
neutral benzene and perylene,
the $\Adfa/\Acc$ ratios are generally
in the range of $\simali$1.0--3.0,
with an average value of
$\langle\Adfa/\Acc\rangle\sim1.11$
and a standard deviation of $\simali$0.73.
For the hydrogenated cations,
the $\Adfa/\Acc$ ratios are more scattered
than their neutral counterparts,
ranging from $\simali$0.1 to $\simali$140.
However, $\Adfa/\Acc$ does not exceed
$\simali$1.0 for $\simali$82\% of the cations.
Note that the cations of THN, DHB[de]A,
Ben$\_$4H, and Pery$\_8$H$\_$RG2
have a much larger $\Adfa/\Acc$ ratio
than others since their C--C stretch modes
are significantly suppressed.
Finally, if we exclude those molecules with
extreme $\Adfa/\Acc$ ratios, we derive
$\langle\Adfa/\Acc\rangle\sim1.53$
and $\simali$0.56 for neutrals and cations,
respectively, with a standard deviation
of $\simali$1.23 and $\simali$0.50
(see Table~\ref{tab:AValueARatio_All}).

%

\subsection{Degrees of Superhydrogenation}\label{subsec:DoH}
With the computed intrinsic band strength $\Aratio$,
we can estimate the hydrogenation of the UIE carrier.
We first make an assumption that the 3.4$\mum$ feature
comes exclusively from hydrogenated PAHs.
This will place an upper limit on the hydrogenation
of the UIE carriers since those PAHs with aliphatic
sidegroups (e.g., methylated PAHs) and anharmonicity
of the aromatic C--H stretch could also contribute
to the 3.4$\mum$ feature, sometimes prominently
(see Li \& Draine 2012).

Let $f_{\rm H} \equiv N_{\rm C,super}/\left[ N_{\rm C, super}+ N_{\rm C, arom}\right]$
be the degree of superhydrogenation, where $N_{\rm C,super}$
is the number of ``superhydrogenated'' C atoms
and $N_{\rm C, arom}$ is the number of aromatic C atoms.
Let $I_{3.3}$ and $I_{3.4}$ respectively be the observed intensities
of the 3.3 and 3.4$\mum$ emission features.
If we assume that one ``superhydrogenated'' C atom corresponds to
2 aliphatic C--H bonds\footnote{%
  Here we only consider the more normal situation that
  the extra H is attached to the C atom on the edge of
  a benzene ring. If the extra H is attached to the C atom
  in the middle (e.g., the two H atoms shown
  in the middle of the structure of Naph$\_$10Ha,
  i.e., c-PHN, in Figure~\ref{fig:HNaph_structure}),
  one hydrogenated H atom corresponds
  to one aliphatic C--H bond.
  }
and one aromatic C atom
corresponds to 3/4 aromatic C--H bonds 
(intermediate between benzene C$_6$H$_6$ 
and coronene C$_{24}$H$_{12}$), 
then 
$I_{3.4}/I_{3.3}
\approx \left(2/0.75\right)\,\times\,\left(N_{\rm C,super}/N_{\rm C,arom}\right)\,
\times\,\left(\Aali/\Aaro\right)$, 
i.e., 
$N_{\rm C,super}/N_{\rm C,arom}\approx
2.67\times\,\left(I_{3.4}/I_{3.3}\right)
\times\,\left(\Aaro/\Aali\right)$. 
The degree of superhydrogenation is
\begin{equation}
f_{\rm H} \equiv N_{\rm C,super}/\left[ N_{\rm C, super}+ N_{\rm C, arom}\right]
\approx \left[1+
2.67\times\frac{I_{3.3}}{I_{3.4}}
\times\frac{A_{3.4}}{A_{3.3}}\right]^{-1} ~~,
\end{equation}
where $A_{3.3}$ and $A_{3.4}$ are measured
on per unit C--H bond basis.
Yang et al.\ (2013) have compiled and analyzed
the UIE spectra of 35 sources
available in the literature
which exhibit both the 3.3$\mum$
and 3.4$\mum$ C--H features.
They derived a median ratio of
$\langle\Iratio\rangle\approx 0.12$,
with the majority (31/35) of
these sources having $\Iratio < 0.25$
(see Figure~1 of Yang et al.\ 2013).
By taking $\langle\Iratio\rangle\approx 0.12$
and $\langle\Aali/\Aaro\rangle\approx 1.98$
for the neutrals and
$\langle\Aali/\Aaro\rangle\approx 7.73$
for the cations (see Table~\ref{tab:AValueARatio_All}
and \S\ref{subsec:C--HSpec}),
we obtain $f_{\rm H}\approx2.2\%$
and $\approx0.57\%$, respectively.
This suggests that the hydrogenation of the UIE
emitters is quite small. 
Note that, as the 3.3$\mum$ feature is predominantly
emitted by neutral PAHs, we conclude that, even if
the 3.4$\mum$ feature exclusively arises from
superhydrogenated PAHs, the degree of superhydrogenation
of the UIE carriers would not exceed $\simali$2.2\%.
%


Similarly, if we assume that one ``superhydrogenated'' 
C atom corresponds to 2 aliphatic C--H bonds, 
the degree of superhydrogenation
could also be derived from the 6.85$\mum$
aliphatic C--H deformation band and
the 6.2$\mum$ C=C stretch band as follows:
\begin{equation}
f_{\rm H} \approx \left[1+
2\times\frac{I_{6.2}}{I_{6.85}}
\times\frac{A_{6.85}}{A_{6.2}}
\times\frac{B_{6.85}(T)}{B_{6.2}(T)}\right]^{-1} ~~,
\end{equation}
where $A_{6.2}$ is measured on
a per aromatic C atom basis,
$A_{6.85}$ is measured on
a per unit C--H bond basis,
$B_{6.2}(T)$ and $B_{6.85}(T)$ are
the Planck functions of temperature $T$
at 6.2 and 6.85$\mum$, respectively.
Observationally, the detection of 
the 6.85$\mum$ emission band
in the ISM of the Milky Way is much rarer 
than the 3.4$\mum$ emission band.
Yang et al.\ (2016a) have compiled
the UIE spectra of Galactic sources 
which exhibit the 6.85$\mum$ band
and found that,
except for several Galactic protoplanetary nebulae,
the 6.85$\mum$ band is weaker 
than the 6.2$\mum$ band 
by a factor of $\simgt$\,10.\footnote{%
    For several Galactic protoplanetary nebulae,
    the 6.85$\mum$ feature is much stronger,
    with $\Idfa/\Icc \simgt 1$ for some of these
    sources (see Yang et al.\ 2016a, Materese et al.\ 2017).
    Such a high $\Idfa/\Icc$ ratio is also seen
    in some protoplanetary nebulae in 
    the Small and Large Magellanic Clouds 
    (see Sloan et al.\ 2014, Matsuura et al.\ 2014).
    }
With $\langle I_{6.85}/I_{6.2} \rangle\simlt0.10$
(Yang et al.\ 2016a),
the mean ratio of the observed intensities
of the 6.85$\mum$ band
to the 6.2$\mum$ band,
$B_{6.85}/B_{6.2}\approx1.04\pm0.24$
for $200\simlt T\simlt 800\K$
(see Yang et al.\ 2016a),
and $\langle\Adfa/\Acc\rangle\approx 1.53$
and $\langle\Adfa/\Acc\rangle\approx 1.23$
respectively for the neutrals and cations
(see Table~\ref{tab:AValueARatio_All}),
we obtain $f_{\rm H}\approx3.1\%$
and $\approx8.6\%$ for the UIE carriers.
This also supports the results obtained from
the 3.4$\mum$ feature that the superhydrogenation
of the UIE carriers is insignificant.

Thanks in large part to the fact that the 3$\mum$ region
is accessible to ground-based telescopes,
the 3.3 and 3.4$\mum$ bands have been
the subject of extensive scrutiny. 
ISO and AKARI have also provided a wealth of data 
on these bands. Operating at 5--38$\mum$, 
{\it Spitzer}/IRS unfortunately missed 
the PAH C--H stretch at 3.3$\mum$
and the accompanying satellite features
at 3.4--3.6$\mum$. 
Compared with {\it Spitzer}, JWST will have more 
than an order of magnitude increase in sensitivity 
and spatial resolution as well as a broader wavelength
coverage in the near-IR.
It is expected that JWST/NIRSpec,
operating at 0.6--5$\mum$,
will be able to examine these bands
so as to better contrain the degree of 
superhydrogenation of PAHs
and its environmental dependence. 
The MIRI instrument on board JWST 
which covers the wavelength 
range of 5 to 28$\mum$ will allow one to 
extend the mid-IR spectroscopy into new regimes 
that ISO and {\it Spitzer} could not probe,
including the 6.85 and 7.25$\mum$ aliphatic
C--H deformation bands 
in objects which were too faint for ISO and {\it Spitzer}.
Objects of particular interest for exploring 
the aromatic and aliphatic C--H emission bands
include carbon star outflows, protoplanetary nebulae,
planetary nebulae, protoplanetary disks around young stars,
reflection nebulae, HII regions, photodissociated regions,
as well as extragalactic objects 
(e.g., protoplanetary and planetary nebulae in 
the Small and Large Magellanic Clouds,
the starburst ring of the barred
spiral galaxy NGC~1097,
and the superwind halo of 
the prototypical starburst galaxy M82).
%
One would imagine that the 3.4$\mum$ band
is more likely to be seen in benign regions. 
It is puzzling that the 3.4$\mum$ emission 
is detected in the harsh superwind of M82
and exhibits appreciable enhancements
with distance from the galactic plane
(see Yamagishi et al.\ 2012).
With the upcoming JWST, smaller spatial scales 
can be probed and spectral mapping in these bands
would be valuable for exploring their nature
and environmental dependence.
%


 \section{Summary}\label{sec:Summary}
We have used the hybrid DFT method B3LYP
in conjunction with the 6-311+G$^{\ast\ast}$ basis set
to compute the IR vibrational spectra of
superhydrogenated PAHs and their cations
of various sizes (ranging from benzene,
naphthalene to perylene and coronene)
and of various degrees of hydrogenation
(ranging from minimally hydrogenated PAHs
to heavily hydrogenated PAHs).
For comparison, we have also computed
the spectra of mono-methylated PAHs
as well as their bare parental PAHs.
The principal results are as follows:
\begin{enumerate}
\item The 3.4$\mum$ aliphatic C--H stretch
           and the 6.85$\mum$ aliphatic C--H deformation
           are seen in all these superhydrogenated species,
           more pronouncedly than in methyl PAHs.
\item For all these superhydrogenated molecules,
          we have derived from the computed spectra
          the intrinsic band strengths of
          the 3.3$\mum$ aromatic C--H stretch ($\Aaro$),
          the 3.4$\mum$ aliphatic C--H stretch ($\Aali$),
          the 6.2$\mum$ aromatic C--C stretch ($\Acc$), and
          the 6.85$\mum$ aliphatic C--H deformation ($\Adfa$).
          By averaging over all these molecules,
          for hydrogenated neutral PAHs 
          we have determined 
          the mean band strengths 
          (per aliphatic C--H bond) of
          $\langle \Aali\rangle \approx 33.6\km\mol^{-1}$
          and $\langle \Adfa\rangle \approx 2.59\km\mol^{-1}$,
          and the mean band-strength ratios of
          $\langle \Aali/\Aaro\rangle\approx 1.98$ and
          $\langle\Adfa/\Acc\rangle\approx1.53$.
          For hydrogenated PAH cations,
          the corresponding band strengths
          and band-strength ratios are
          $\langle \Aali\rangle \approx 13.6\km\mol^{-1}$
          and $\langle \Adfa\rangle \approx 13.2\km\mol^{-1}$,
          and $\langle\Aali/\Aaro\rangle\approx 7.73$ and
          $\langle\Adfa/\Acc\rangle\approx0.56$.
\item By comparing the computationally-derived mean ratio
          of $\langle\Aali/\Aaro\rangle\approx 1.98$
          with the mean ratio of the observed intensities
          $\langle\Iratio\rangle\approx 0.12$,
          we have estimated the degree of superhydrogenation
          to be only $\simali$2.2\% for neutral PAHs
          which predominantly emit the 3.3 and 3.4$\mum$ features.
          We have also derived the degree of superhydrogenation
          from the mean ratio of the observed intensities
          $\langle\Iratiodfa\rangle\simlt 0.10$ and
          $\langle\Adfa/\Acc\rangle\approx 1.53$ for neutrals
          and $\langle\Adfa/\Acc\rangle\approx 0.56$ for cations
          to be $\simlt$3.1\% for neutrals
          and $\simlt$8.6\% for cations.
          The actual degrees of superhydrogenation
          could be even lower
          since methylated PAHs and the anharmonicity of PAHs
          could also contribute to the observed 3.4 and 6.85$\mum$
          aliphatic C--H bands.
          Therefore, we conclude that astrophysical PAHs
          are primarily aromatic and are not significantly superhydrogenated.
\end{enumerate}

\acknowledgments{%
We thank the anonymous referee for his/her
very helpful comments and suggestions.
XJY is supported in part by NSFC 11873041
and the NSFC-CAS Joint Research Funds
in Astronomy (U1731106, U1731107).
AL is supported in part by NASA grant 80NSSC19K0572.
RG is supported in part by NSF-PRISM grant
Mathematics and Life Sciences (0928053).
Computations were performed using the high-performance computer
resources of the University of Missouri Bioinformatics Consortium.
}


\clearpage

\clearpage

\begin{figure*}
\figurenum{\ref{fig:Pery2HSpec}}
\leavevmode
\begin{center}
\begin{minipage}[t]{1.0\textwidth}
\resizebox{16.8cm}{8cm}{\includegraphics[clip]{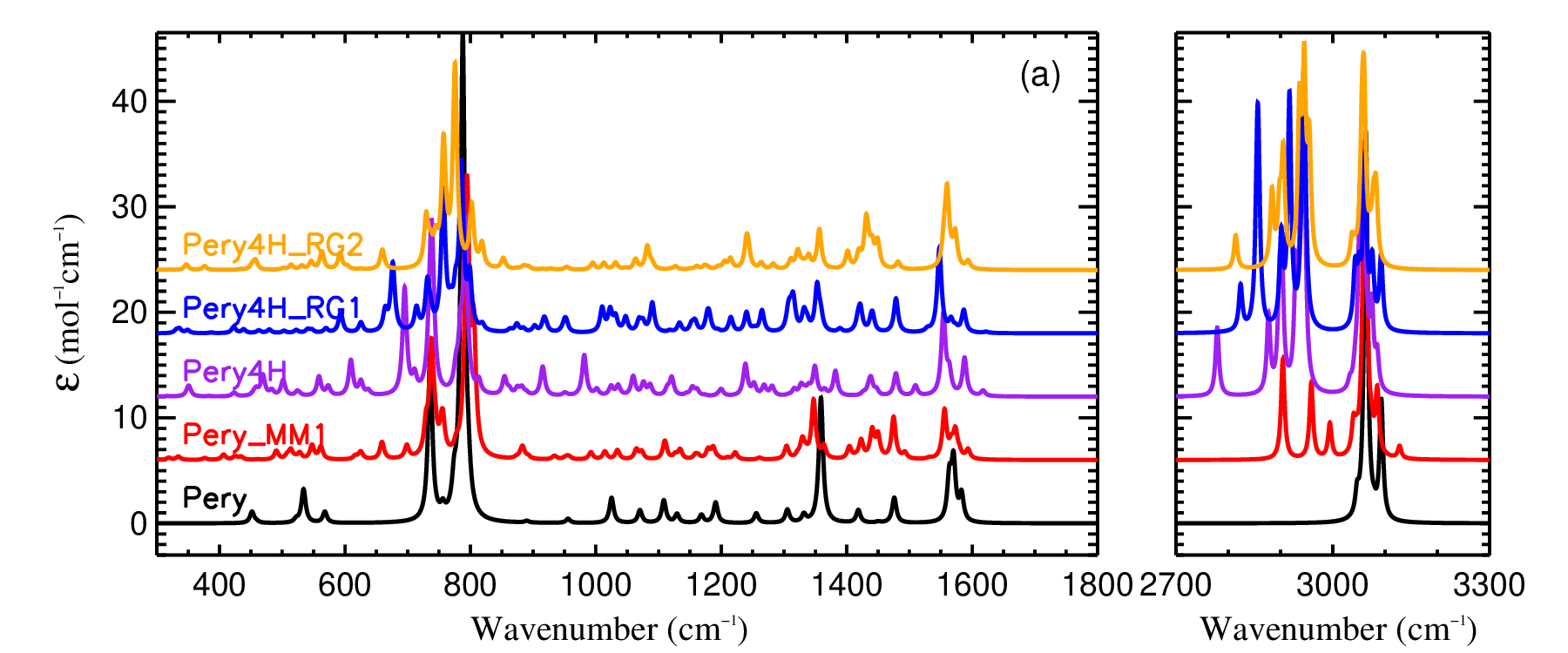}}\vspace{0.2cm}
\resizebox{16.8cm}{8cm}{\includegraphics[clip]{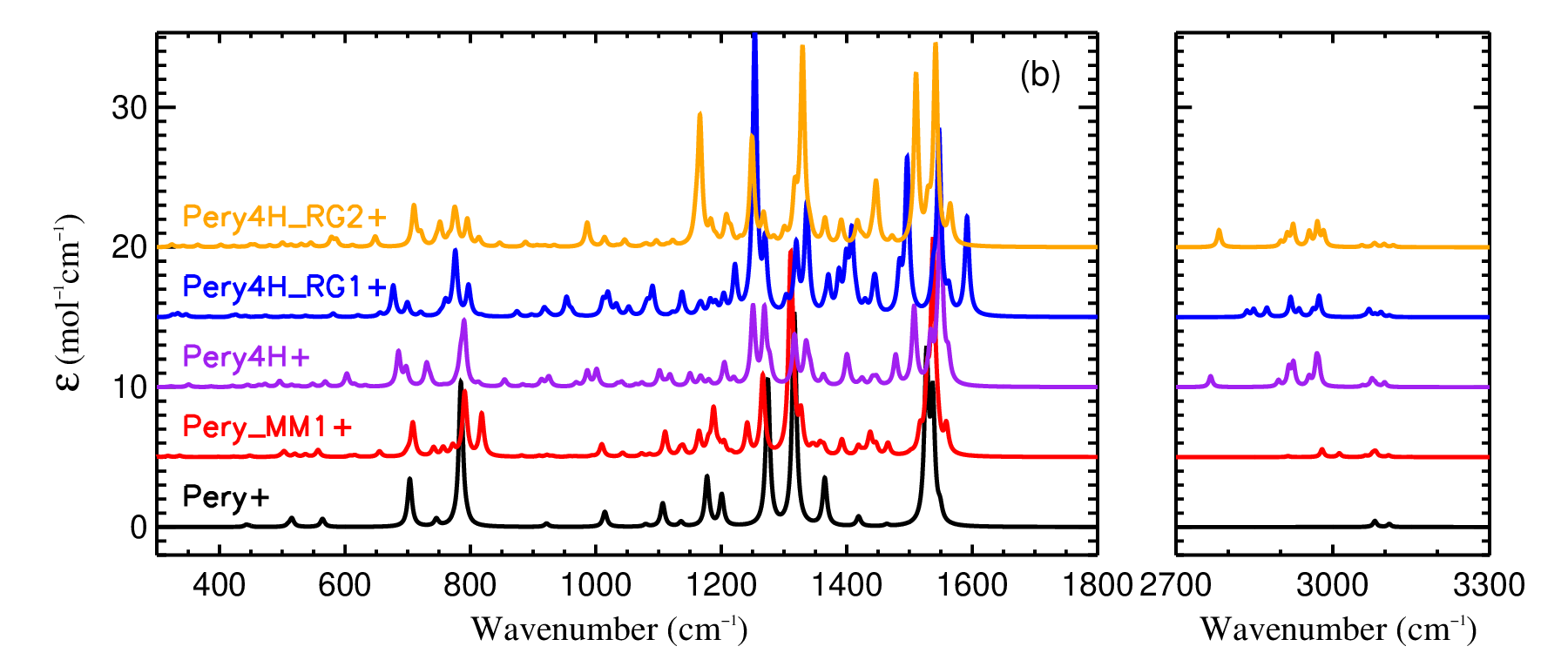}}\vspace{0.2cm}
\end{minipage}\hspace{-1.5cm}
\end{center}
\caption{\footnotesize
         Continued, but for Pery$\_$4H
         (perylene with four excess H atoms).
         }
\end{figure*}

\begin{figure*}
\figurenum{\ref{fig:Pery2HSpec}}
\leavevmode
\begin{center}
\begin{minipage}[t]{1.0\textwidth}
\resizebox{16.8cm}{8cm}{\includegraphics[clip]{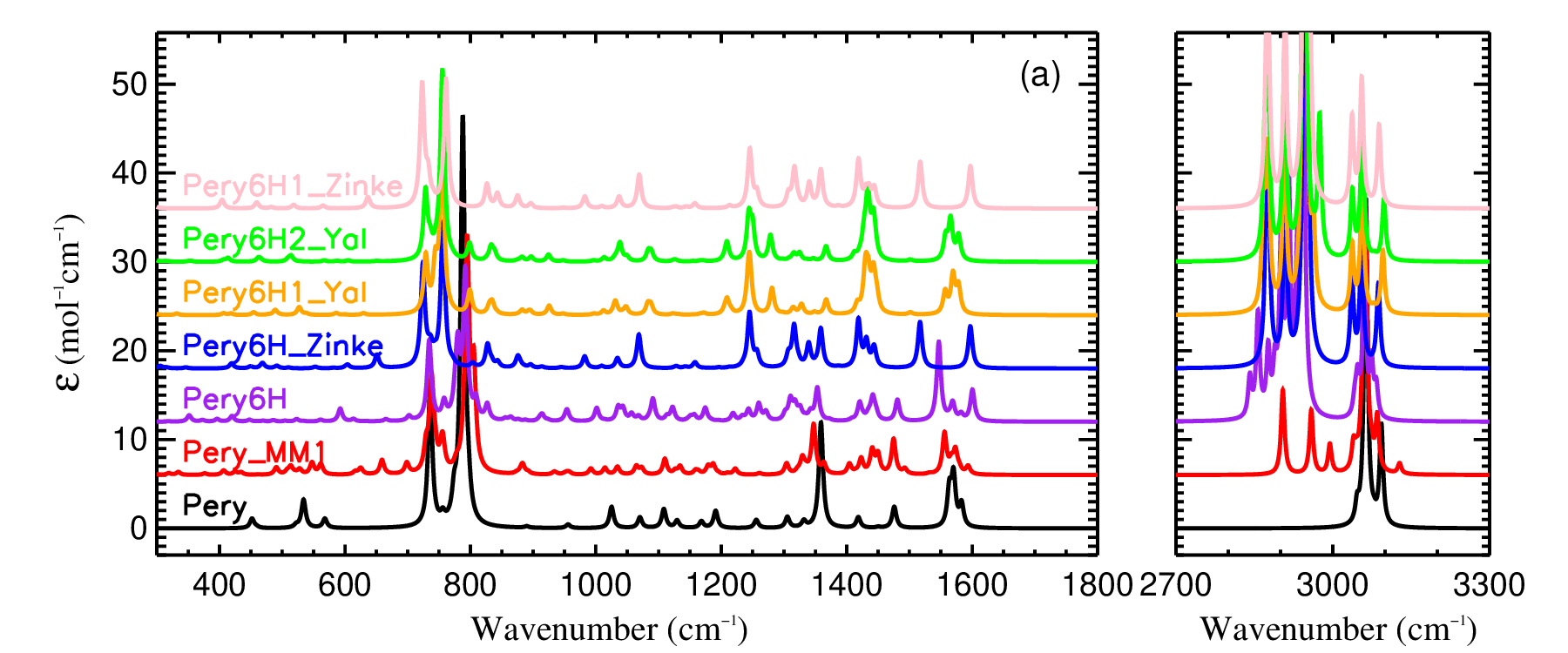}}\vspace{0.2cm}
\resizebox{16.8cm}{8cm}{\includegraphics[clip]{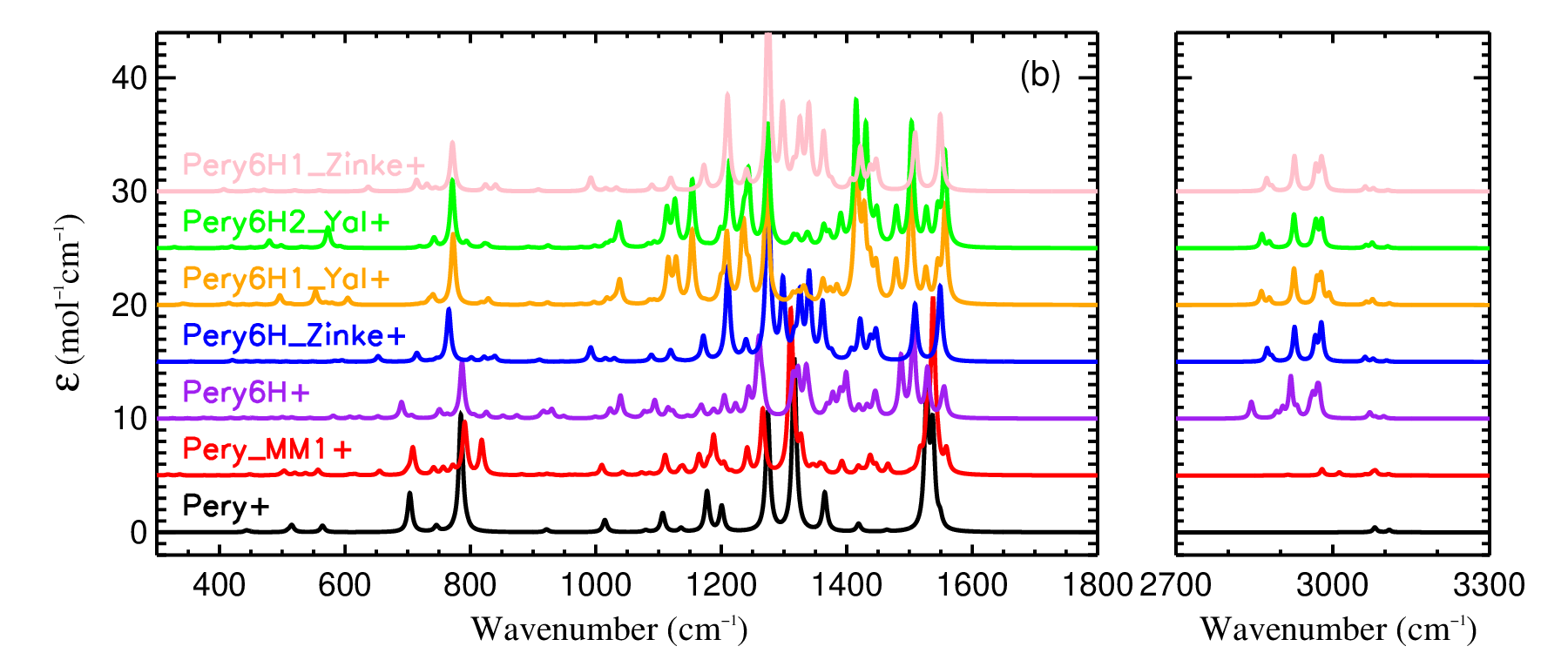}}\vspace{0.2cm}
\end{minipage}\hspace{-1.5cm}
\end{center}
\caption{\footnotesize
         Continued, but for Pery$\_$6H
         (perylene with six excess H atoms).
         }
\end{figure*}

\begin{figure*}
\figurenum{\ref{fig:Pery2HSpec}}
\leavevmode
\begin{center}
\begin{minipage}[t]{1.0\textwidth}
\resizebox{16.8cm}{8cm}{\includegraphics[clip]{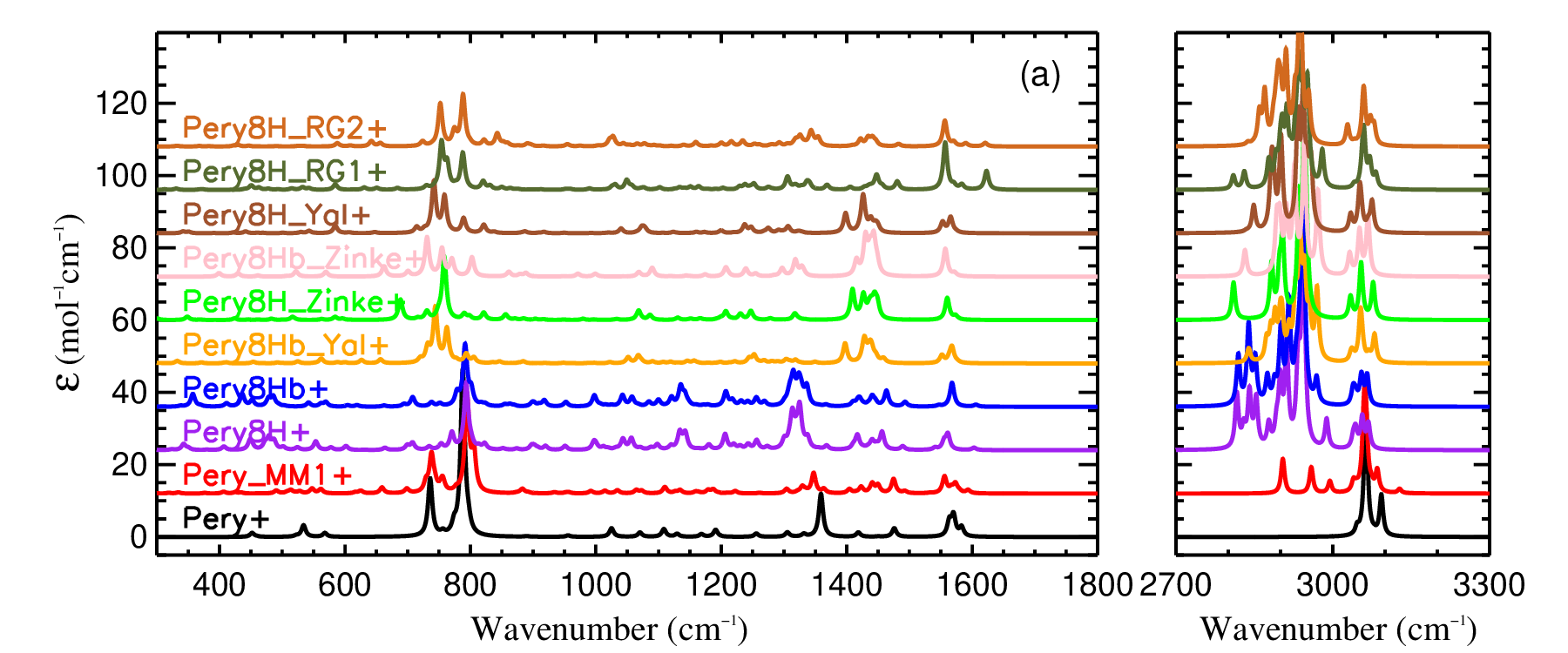}}\vspace{0.2cm}
\resizebox{16.8cm}{8cm}{\includegraphics[clip]{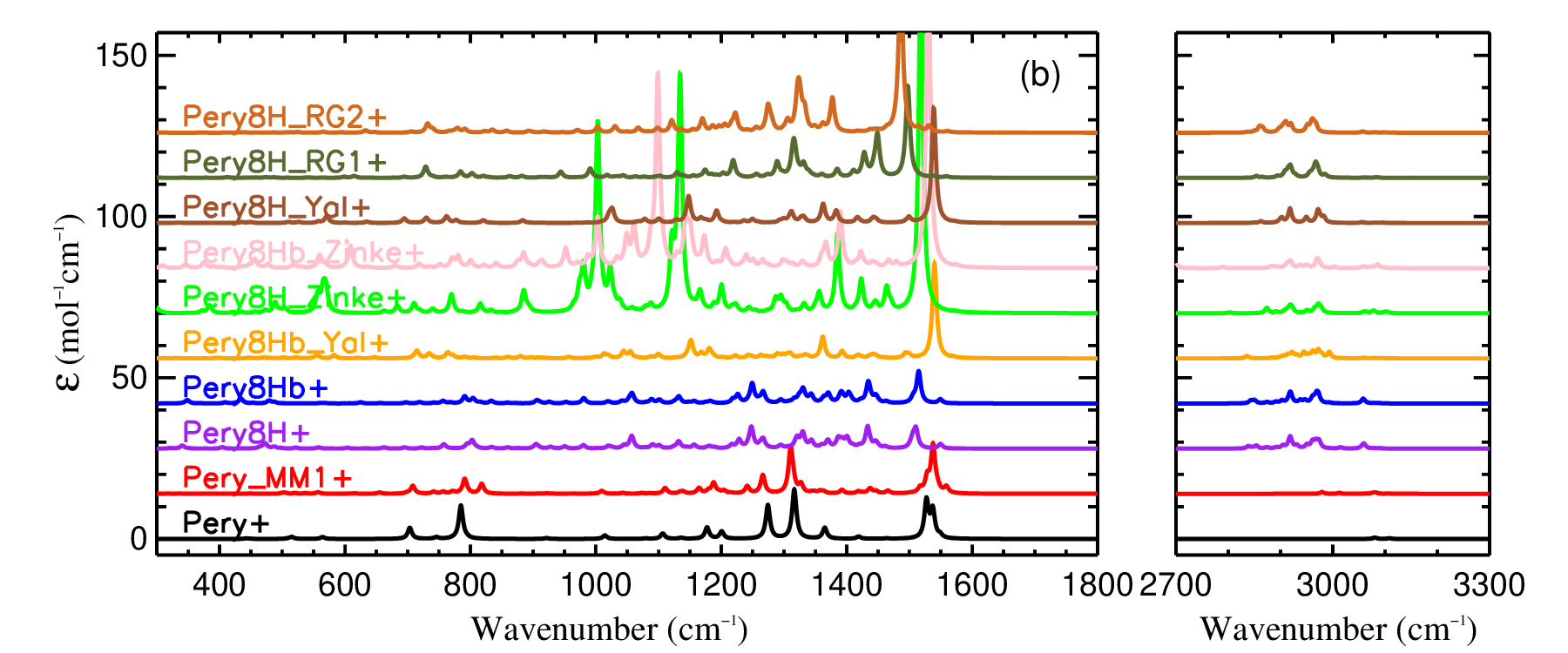}}\vspace{0.2cm}
\end{minipage}\hspace{-1.5cm}
\end{center}
\caption{\footnotesize
         Continued, but for Pery$\_$8H
         (perylene with eight excess H atoms).
         }
\end{figure*}

\begin{figure*}
\figurenum{\ref{fig:Pery2HSpec}}
\leavevmode
\begin{center}
\begin{minipage}[t]{1.0\textwidth}
\resizebox{16.8cm}{8cm}{\includegraphics[clip]{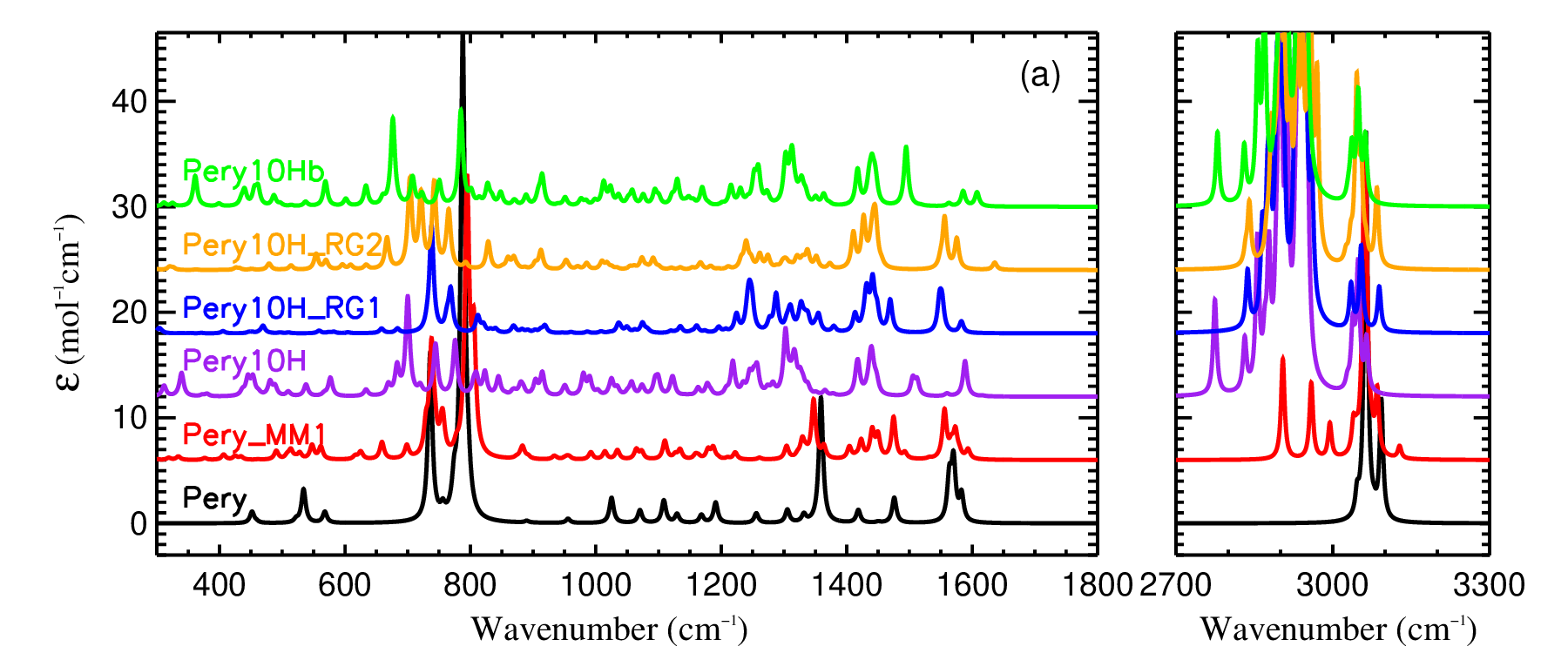}}\vspace{0.2cm}
\resizebox{16.8cm}{8cm}{\includegraphics[clip]{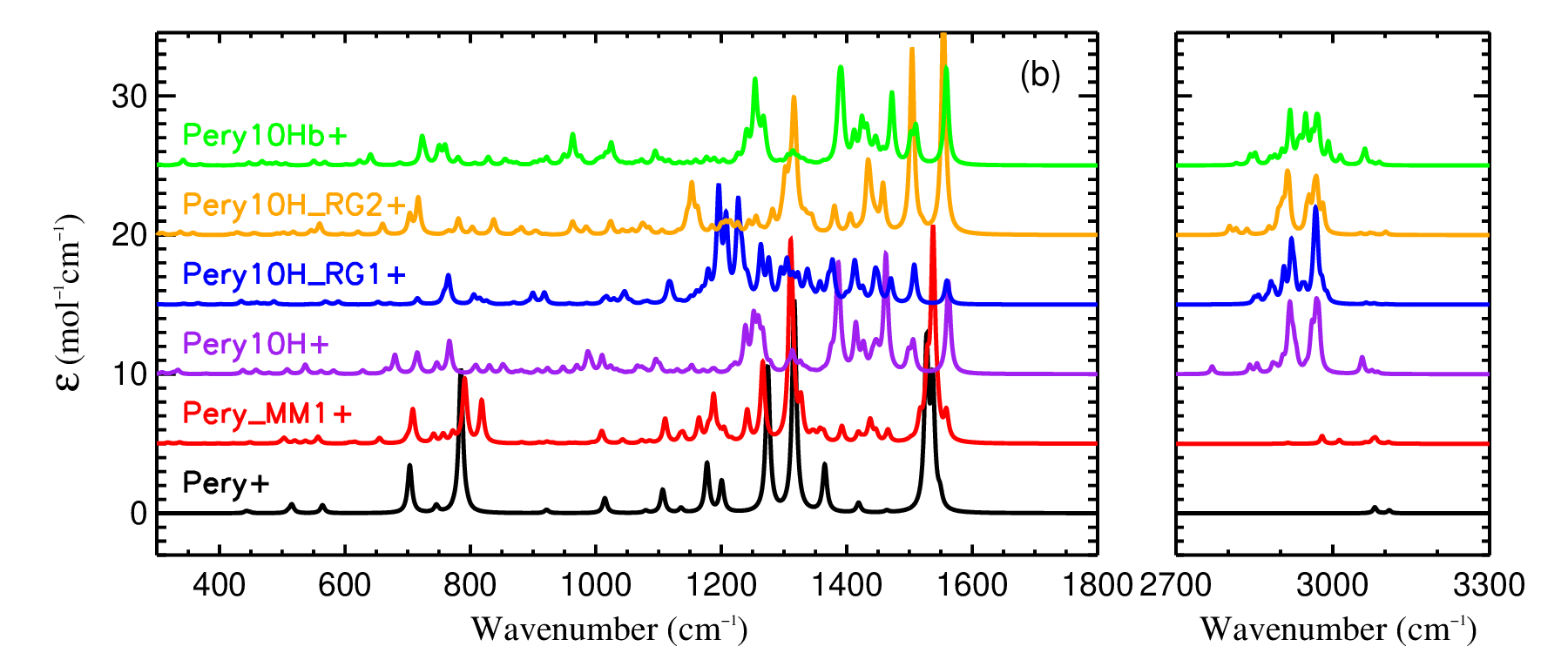}}\vspace{0.2cm}
\end{minipage}\hspace{-1.5cm}
\end{center}
\caption{\footnotesize
         Continued, but for Pery$\_$10H
         (perylene with ten excess H atoms).
         }
\end{figure*}

\begin{figure*}
\figurenum{\ref{fig:Pery2HSpec}}
\leavevmode
\begin{center}
\begin{minipage}[t]{1.0\textwidth}
\resizebox{16.8cm}{8cm}{\includegraphics[clip]{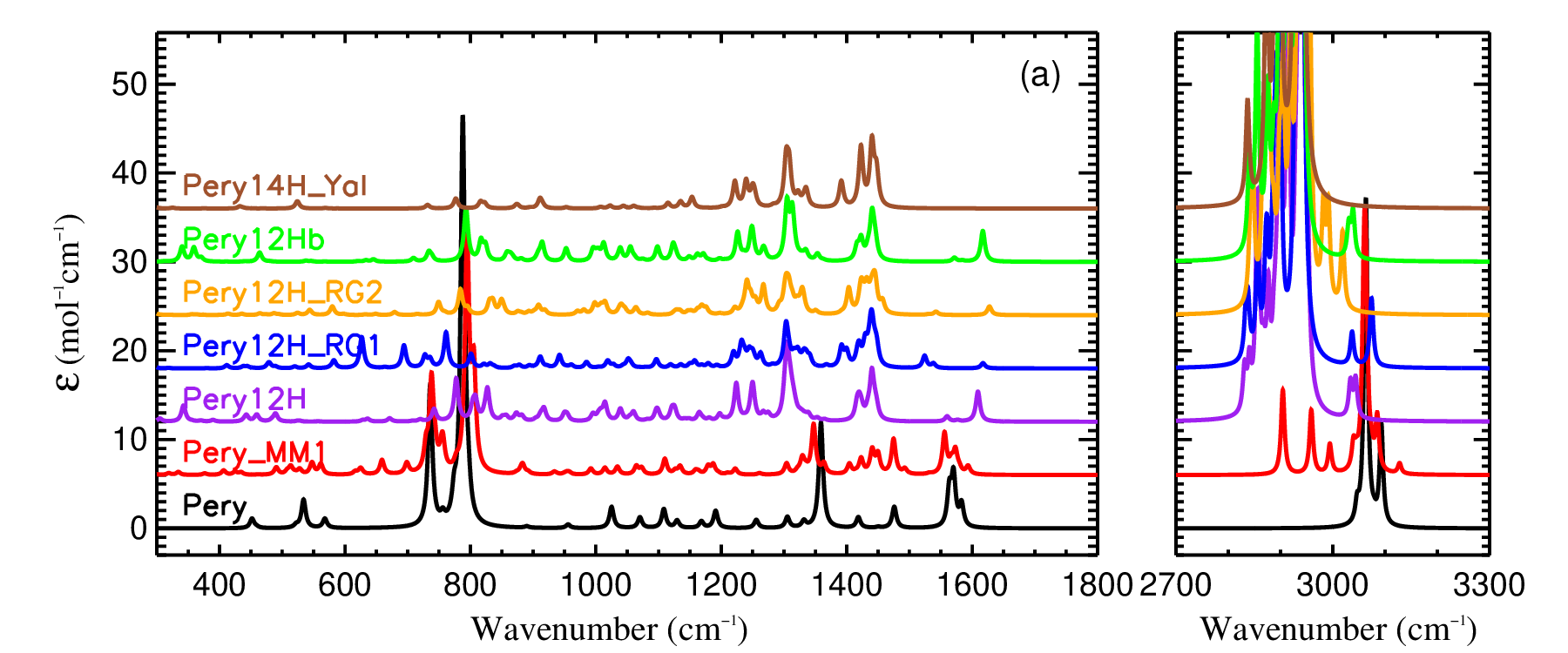}}\vspace{0.2cm}
\resizebox{16.8cm}{8cm}{\includegraphics[clip]{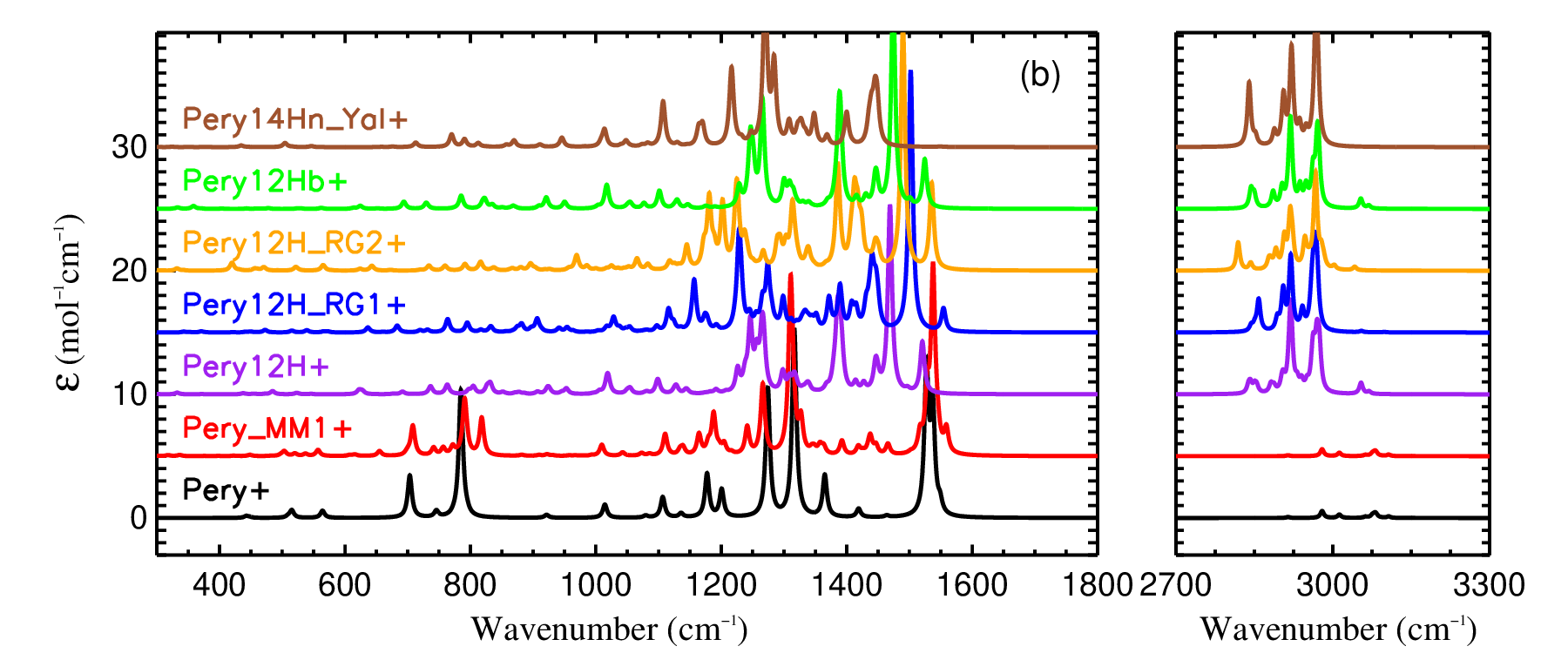}}\vspace{0.2cm}
\end{minipage}\hspace{-1.5cm}
\end{center}
\caption{\footnotesize
         Continued, but for Pery$\_$12H
         (perylene with 12 excess H atoms)
         and Pery$\_$14H
         (perylene with 14 excess H atoms).
         }
\end{figure*}

\clearpage

\begin{figure*}
\begin{center}
\begin{minipage}[t]{1.0\textwidth}
\resizebox{16.8cm}{8cm}{\includegraphics[clip]{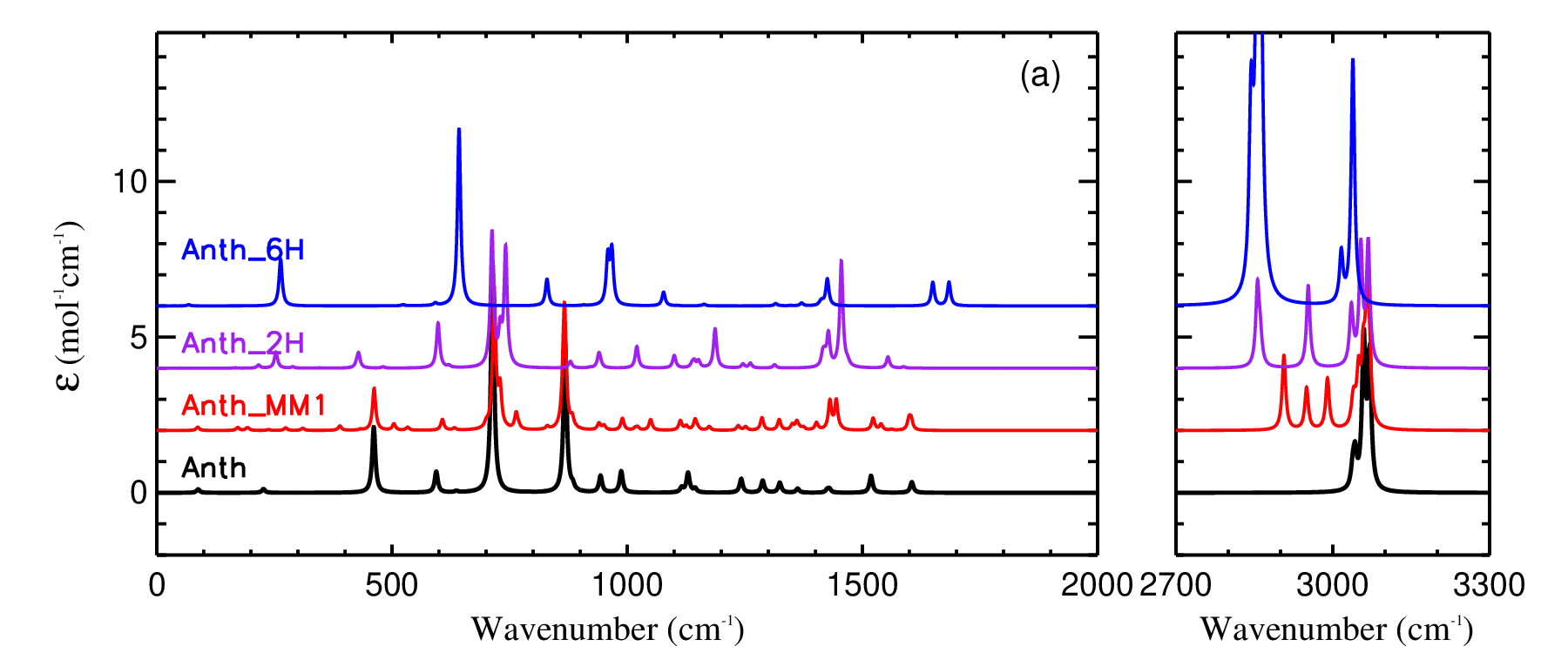}}\vspace{0.2cm}
\resizebox{16.8cm}{8cm}{\includegraphics[clip]{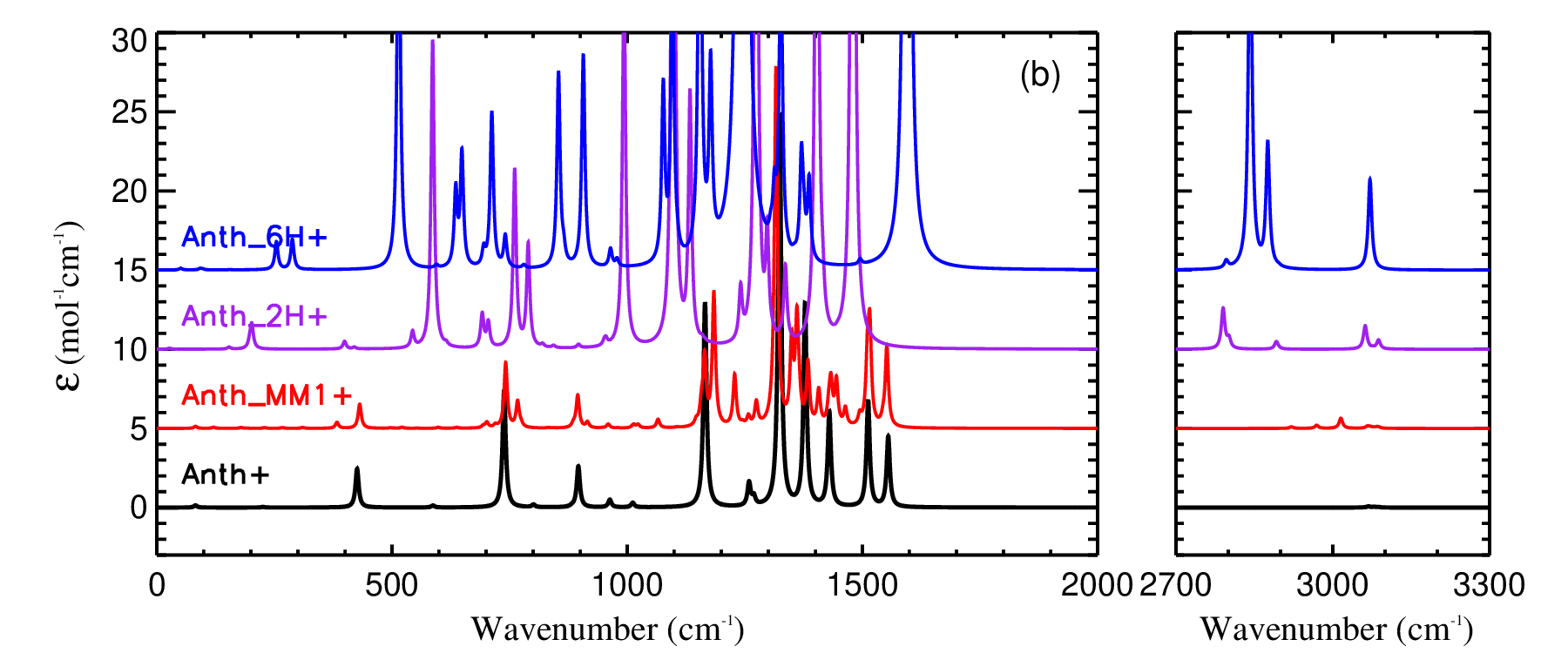}}\vspace{0.2cm}
\end{minipage}\hspace{-1.5cm}
\end{center}
\caption{\footnotesize
         \label{fig:Anth_Spec_NC}
         Calculated vibrational spectra of
         neutral (upper panels) and
         cationic (lower panels) anthracene derivatives
         (``Series B'' of Sandford et al.\ 2013)
         compared with anthracene and
         mono-methylated-anthracene.
         The frequencies are scaled with
         a factor of 0.963, and a line width
         of 4$\cm^{-1}$ is assigned.
         }
\end{figure*}

\begin{figure*}
\figurenum{\ref{fig:Anth_Spec_NC}}
\leavevmode
\begin{center}
\begin{minipage}[t]{1.0\textwidth}
\resizebox{16.8cm}{8cm}{\includegraphics[clip]{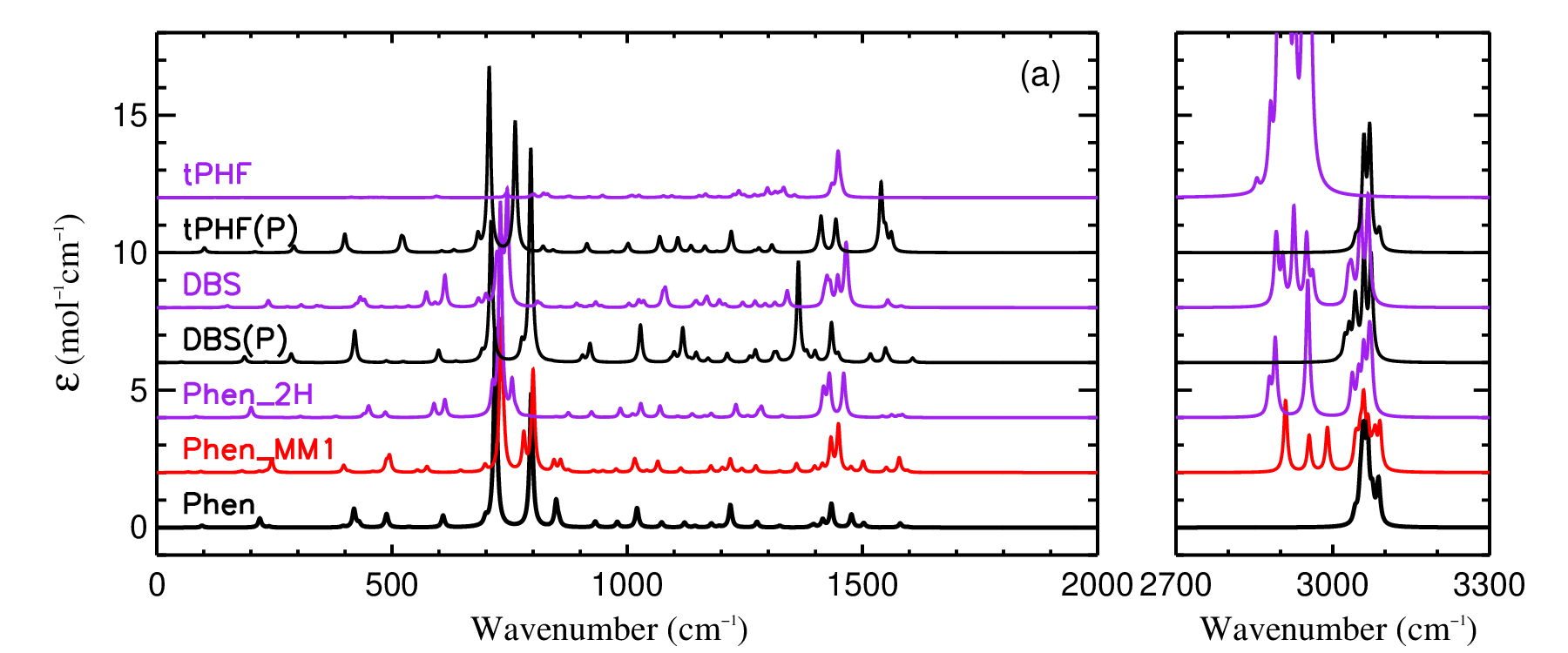}}\vspace{0.2cm}
\resizebox{16.8cm}{8cm}{\includegraphics[clip]{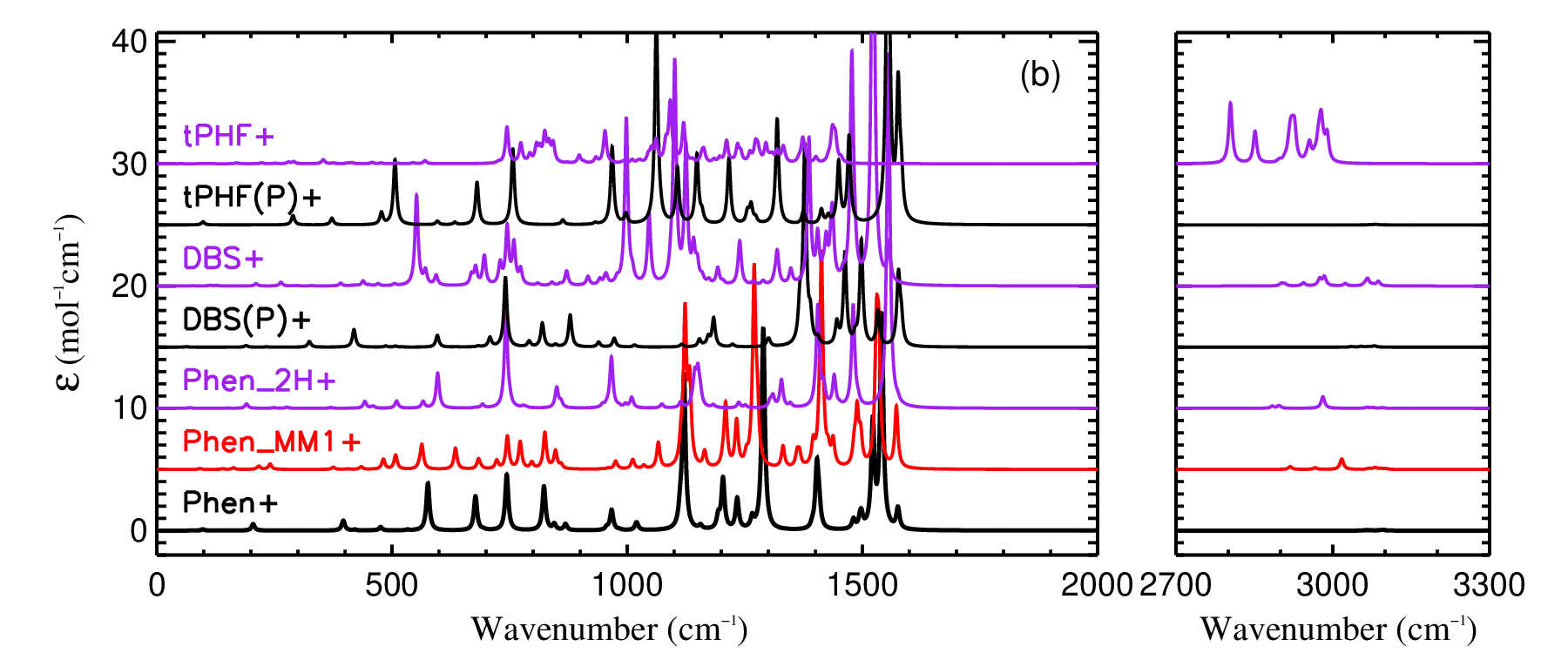}}\vspace{0.2cm}
\end{minipage}\hspace{-1.5cm}
\end{center}
\caption{\footnotesize
         Continued, but for the derivatives of
         phenanthrene (``Series C'' of Sandford et al.\ 2013).
         }
\end{figure*}

\begin{figure*}
\figurenum{\ref{fig:Anth_Spec_NC}}
\leavevmode
\begin{center}
\begin{minipage}[t]{1.0\textwidth}
\resizebox{16.8cm}{8cm}{\includegraphics[clip]{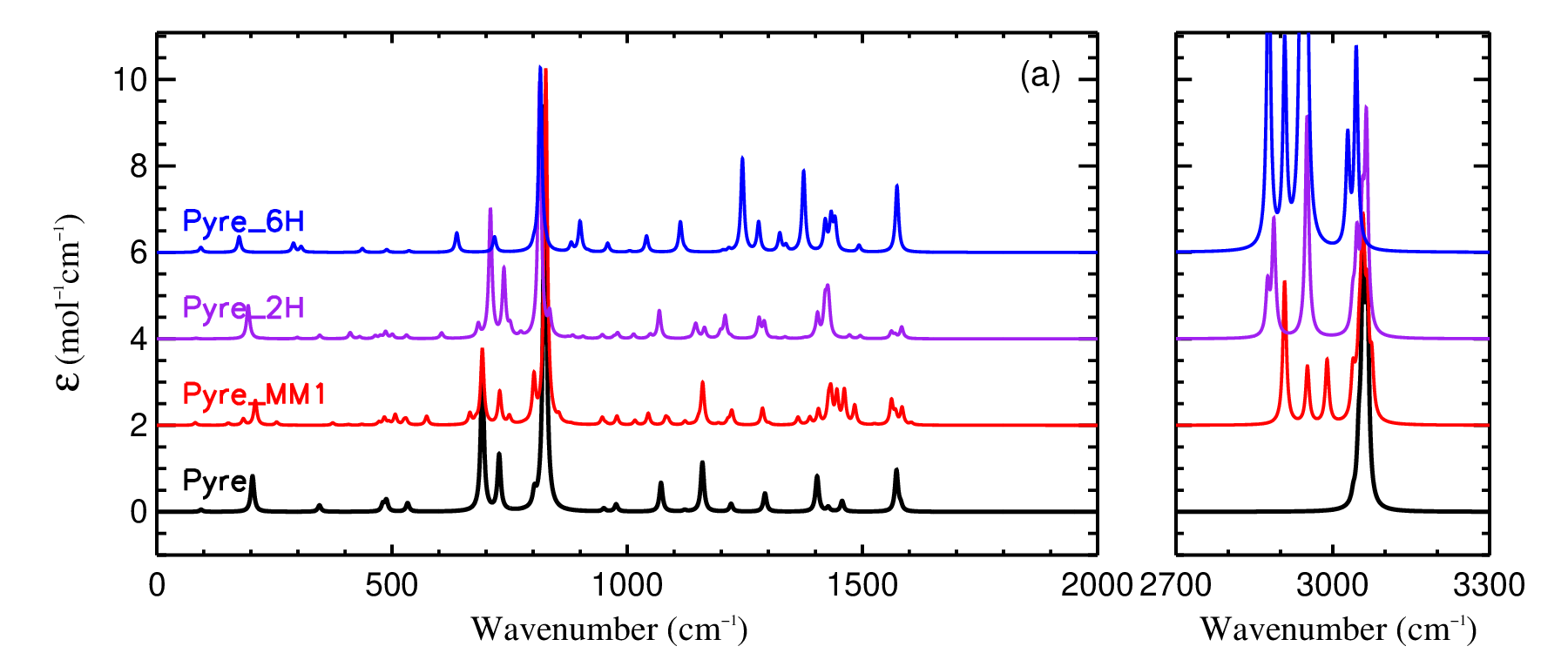}}\vspace{0.2cm}
\resizebox{16.8cm}{8cm}{\includegraphics[clip]{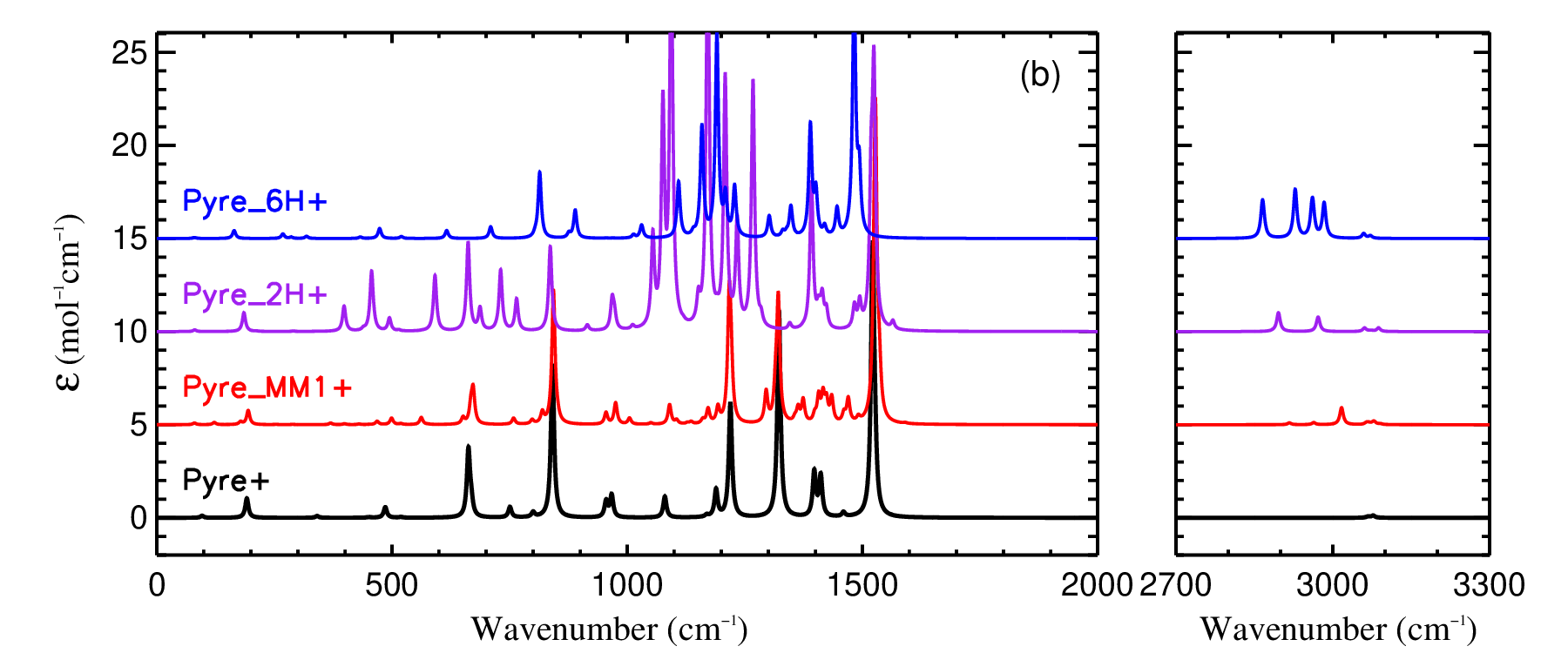}}\vspace{0.2cm}
\end{minipage}\hspace{-1.5cm}
\end{center}
\caption{\footnotesize
         Continued, but for the derivatives of
         pyrene (``Series D'' of Sandford et al.\ 2013).
         }
\end{figure*}

\begin{figure*}
\figurenum{\ref{fig:Anth_Spec_NC}}
\leavevmode
\begin{center}
\begin{minipage}[t]{1.0\textwidth}
\resizebox{16.8cm}{8cm}{\includegraphics[clip]{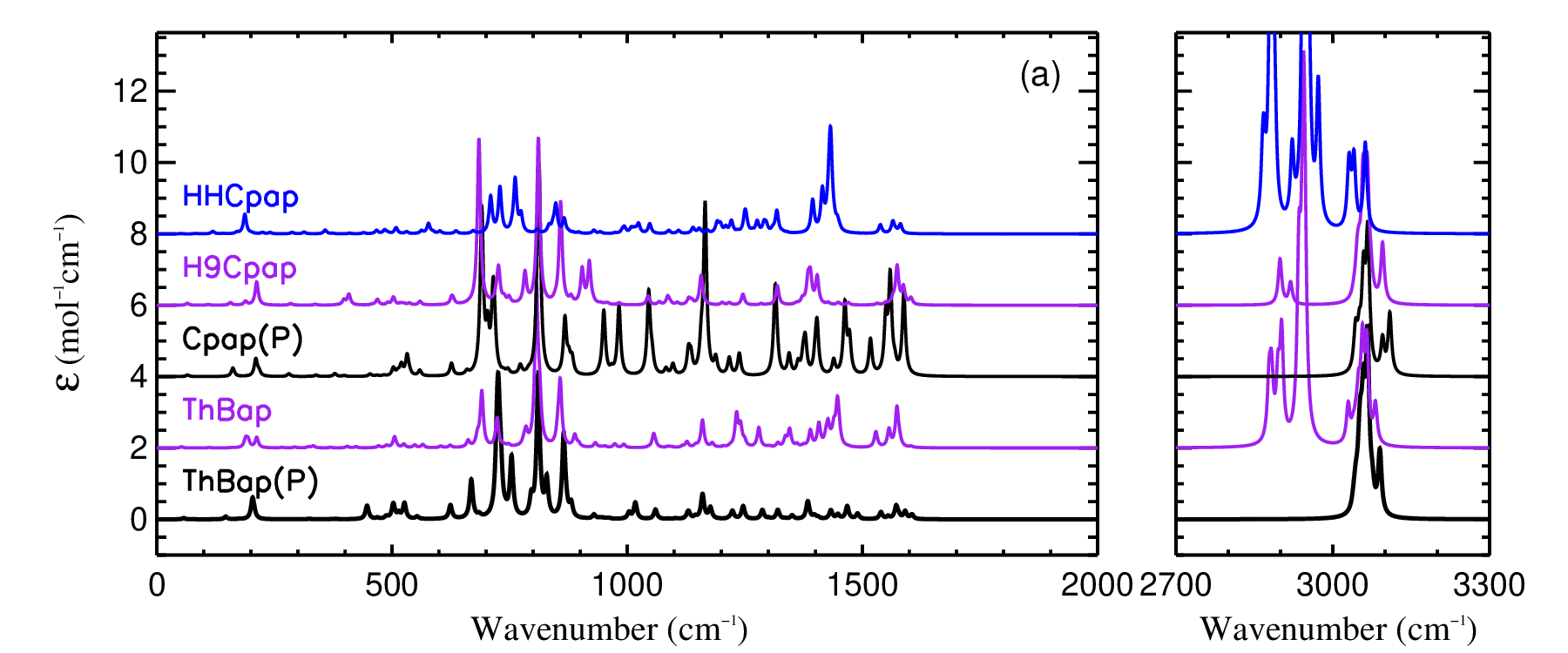}}\vspace{0.2cm}
\resizebox{16.8cm}{8cm}{\includegraphics[clip]{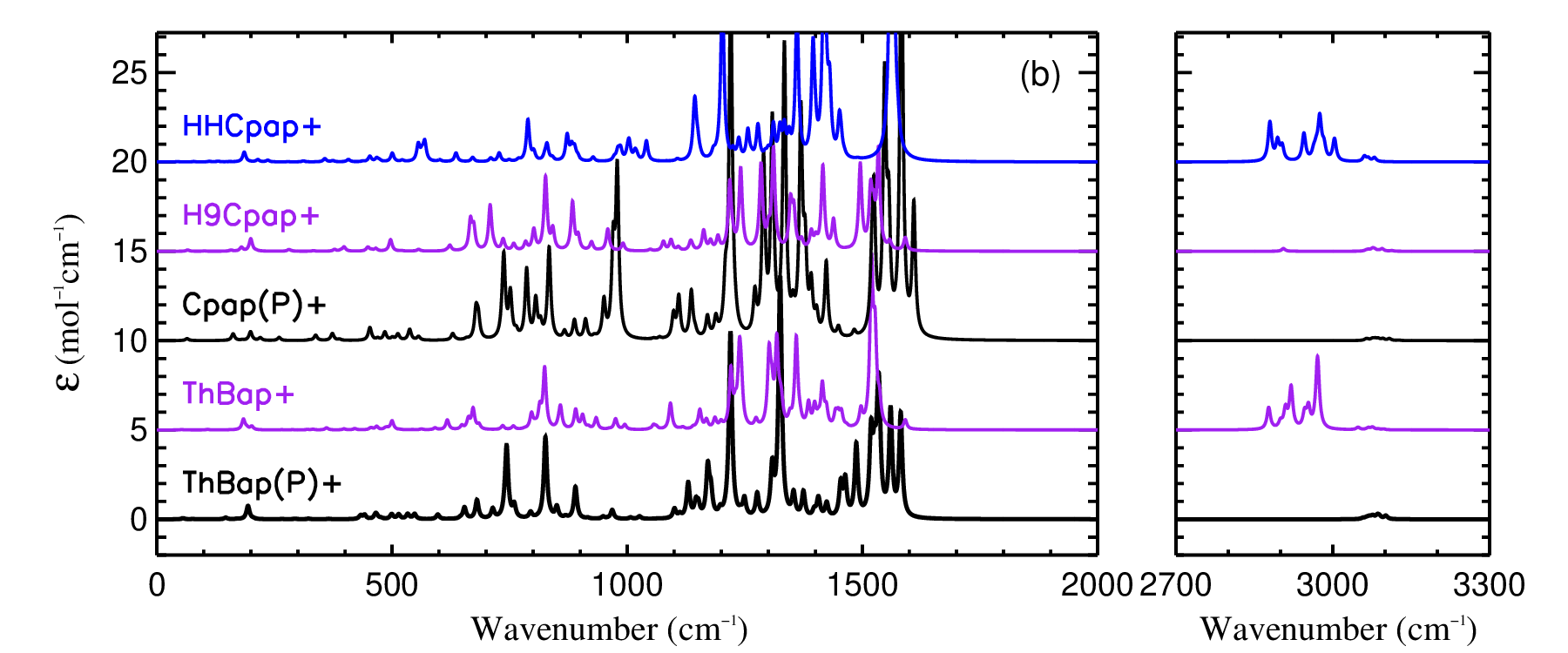}}\vspace{0.2cm}
\end{minipage}\hspace{-1.5cm}
\end{center}
\caption{\footnotesize
         Continued, but for  the ``Series E''
         molecules of Sandford et al.\ (2013).
         }
\end{figure*}

\begin{figure*}
\figurenum{\ref{fig:Anth_Spec_NC}}
\leavevmode
\begin{center}
\begin{minipage}[t]{1.0\textwidth}
\resizebox{16.8cm}{8cm}{\includegraphics[clip]{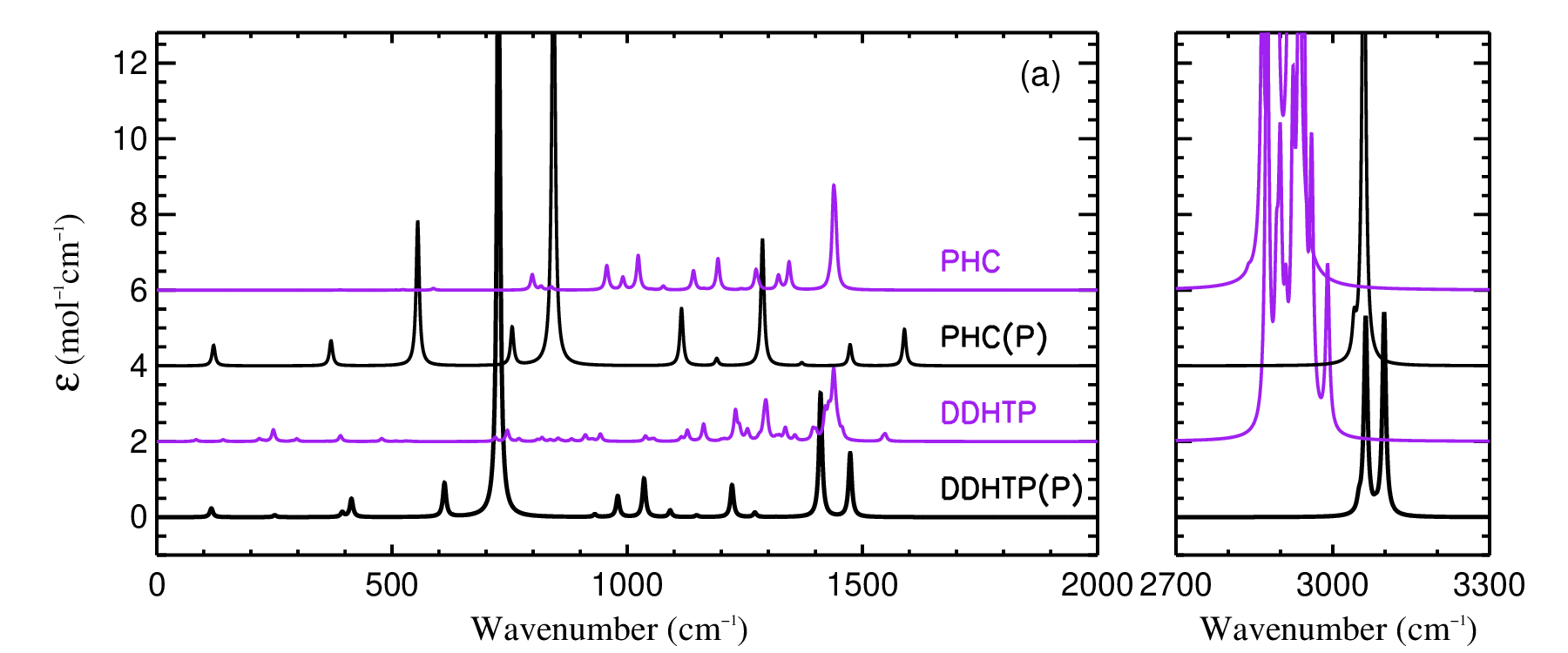}}\vspace{0.2cm}
\resizebox{16.8cm}{8cm}{\includegraphics[clip]{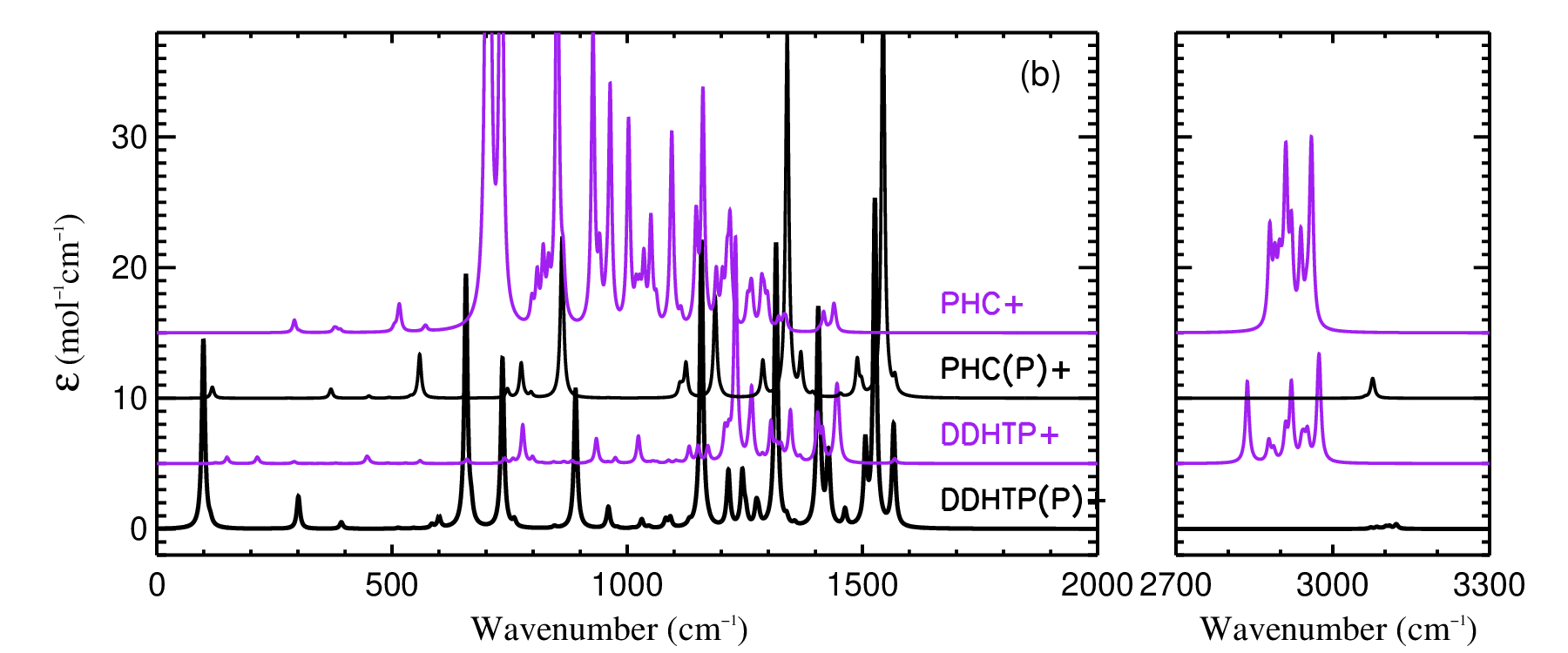}}\vspace{0.2cm}
\end{minipage}\hspace{-1.5cm}
\end{center}
\caption{\footnotesize
         Continued, but for  the ``Series F''
         molecules of Sandford et al.\ (2013).
         }
\end{figure*}

\begin{figure*}
\figurenum{\ref{fig:Anth_Spec_NC}}
\leavevmode
\begin{center}
\begin{minipage}[t]{1.0\textwidth}
\resizebox{16.8cm}{8cm}{\includegraphics[clip]{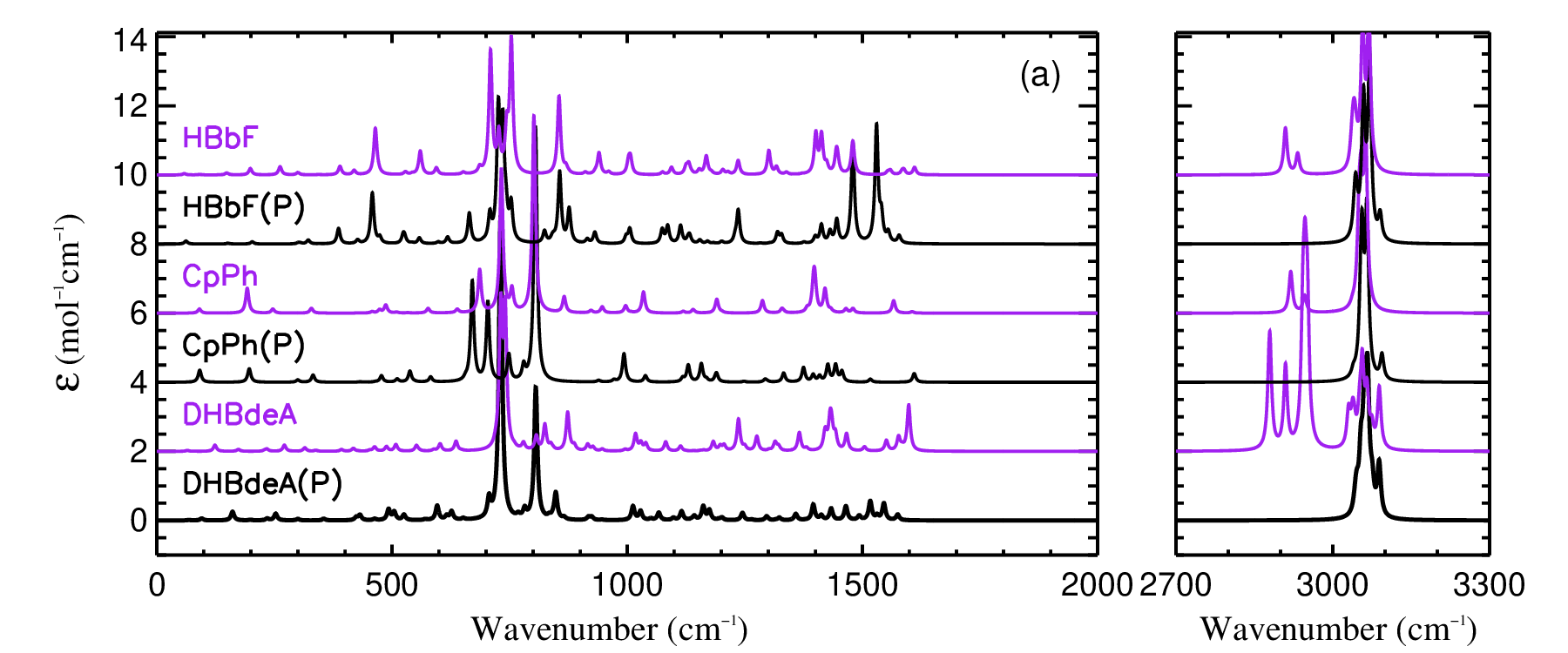}}\vspace{0.2cm}
\resizebox{16.8cm}{8cm}{\includegraphics[clip]{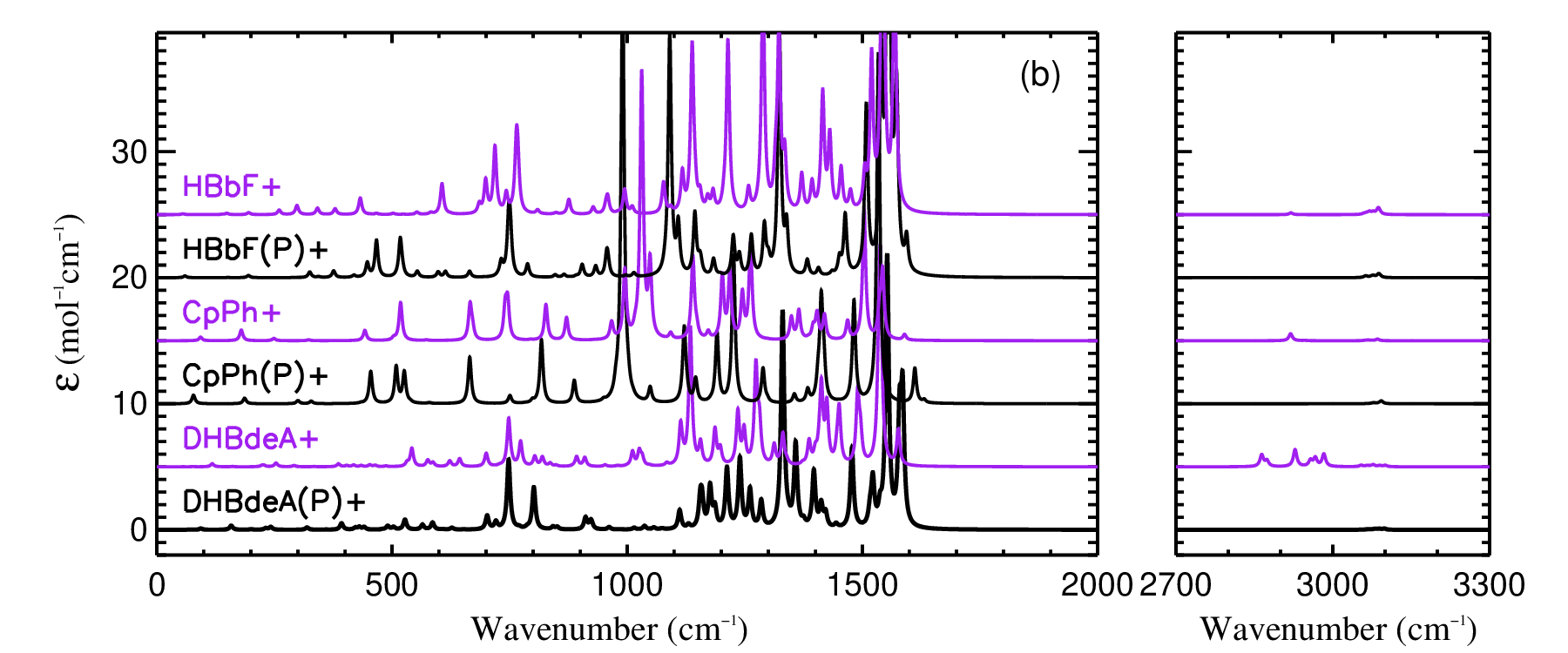}}\vspace{0.2cm}
\end{minipage}\hspace{-1.5cm}
\end{center}
\caption{\footnotesize
         Continued, but for  the ``Series G''
         molecules of Sandford et al.\ (2013).
         }
\end{figure*}

\begin{figure*}
\figurenum{\ref{fig:Anth_Spec_NC}}
\leavevmode
\begin{center}
\begin{minipage}[t]{1.0\textwidth}
\resizebox{16.8cm}{8cm}{\includegraphics[clip]{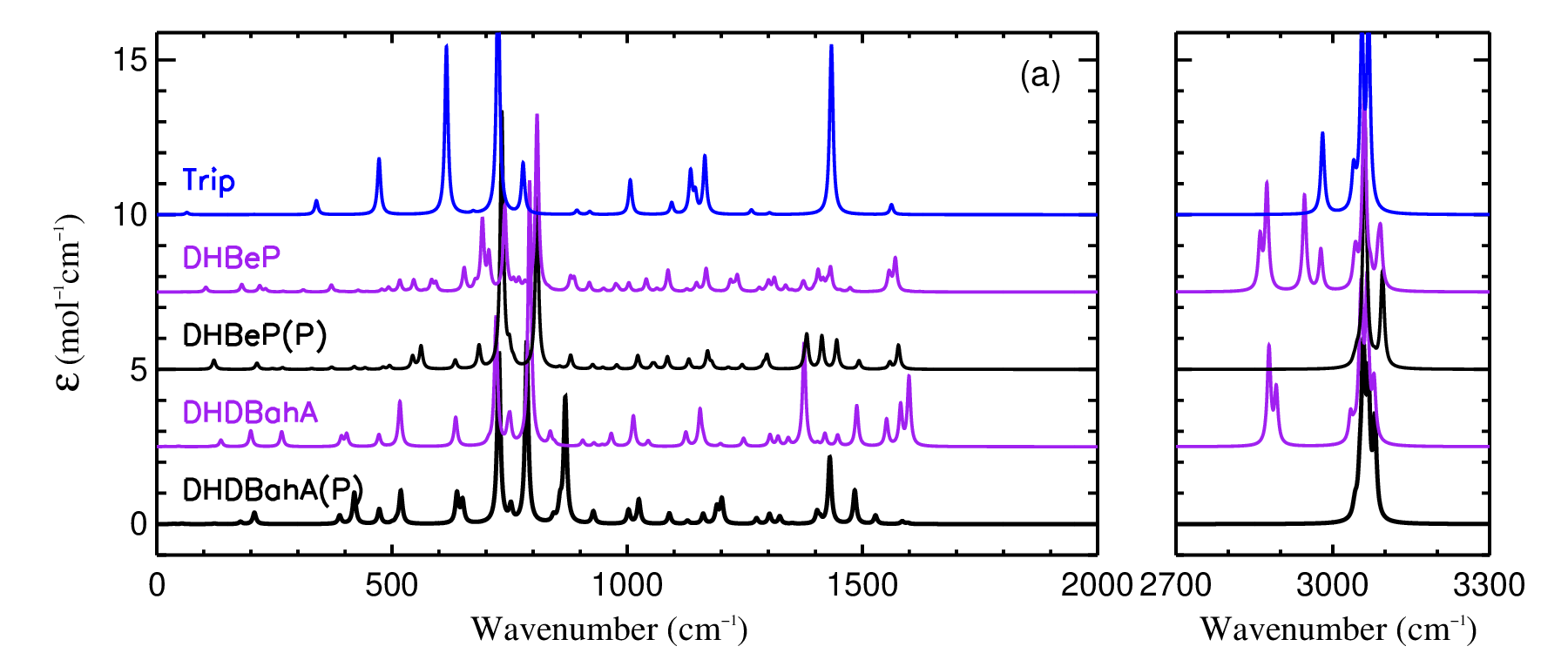}}\vspace{0.2cm}
\resizebox{16.8cm}{8cm}{\includegraphics[clip]{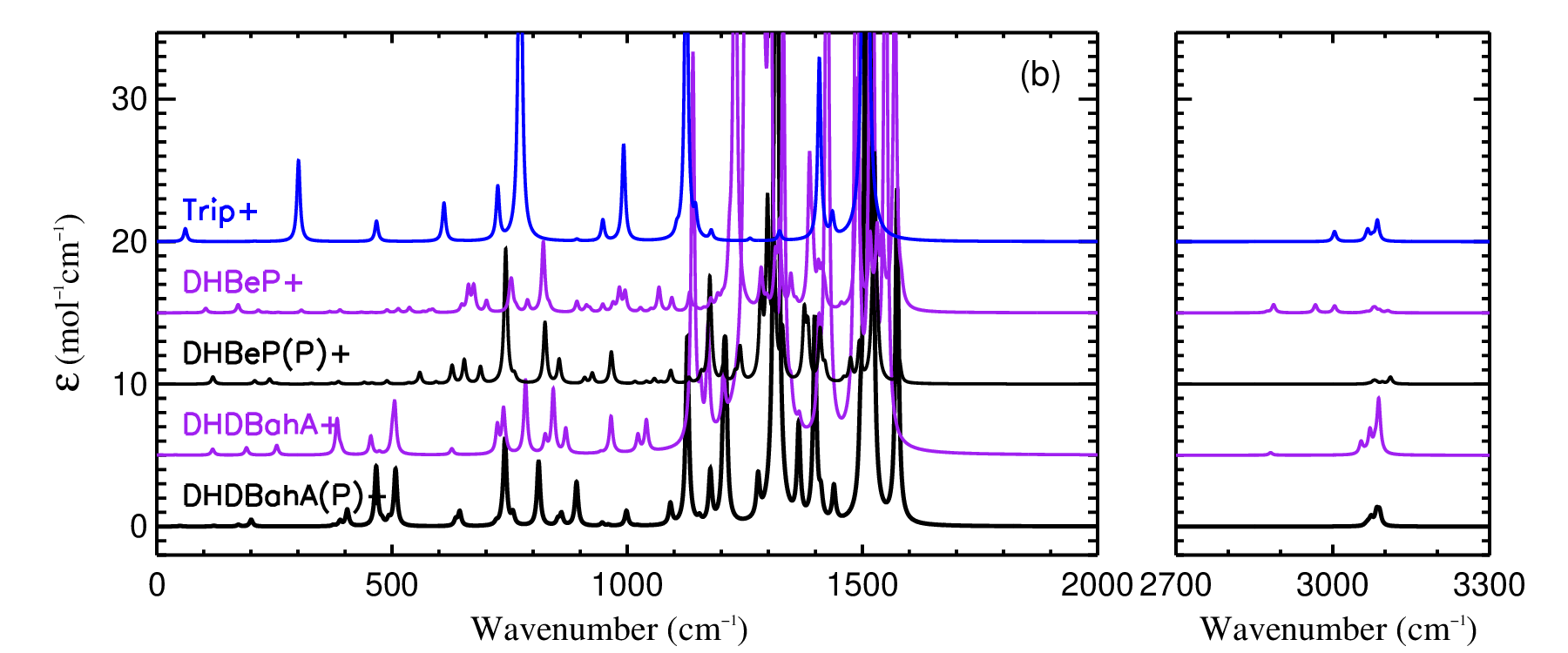}}\vspace{0.2cm}
\end{minipage}\hspace{-1.5cm}
\end{center}
\caption{\footnotesize
         Continued, but for  the ``Series H''
         molecules of Sandford et al.\ (2013).
         }
\end{figure*}

\clearpage

\begin{figure*}
\centering
{
\includegraphics[width=1.2\textwidth,angle=0]{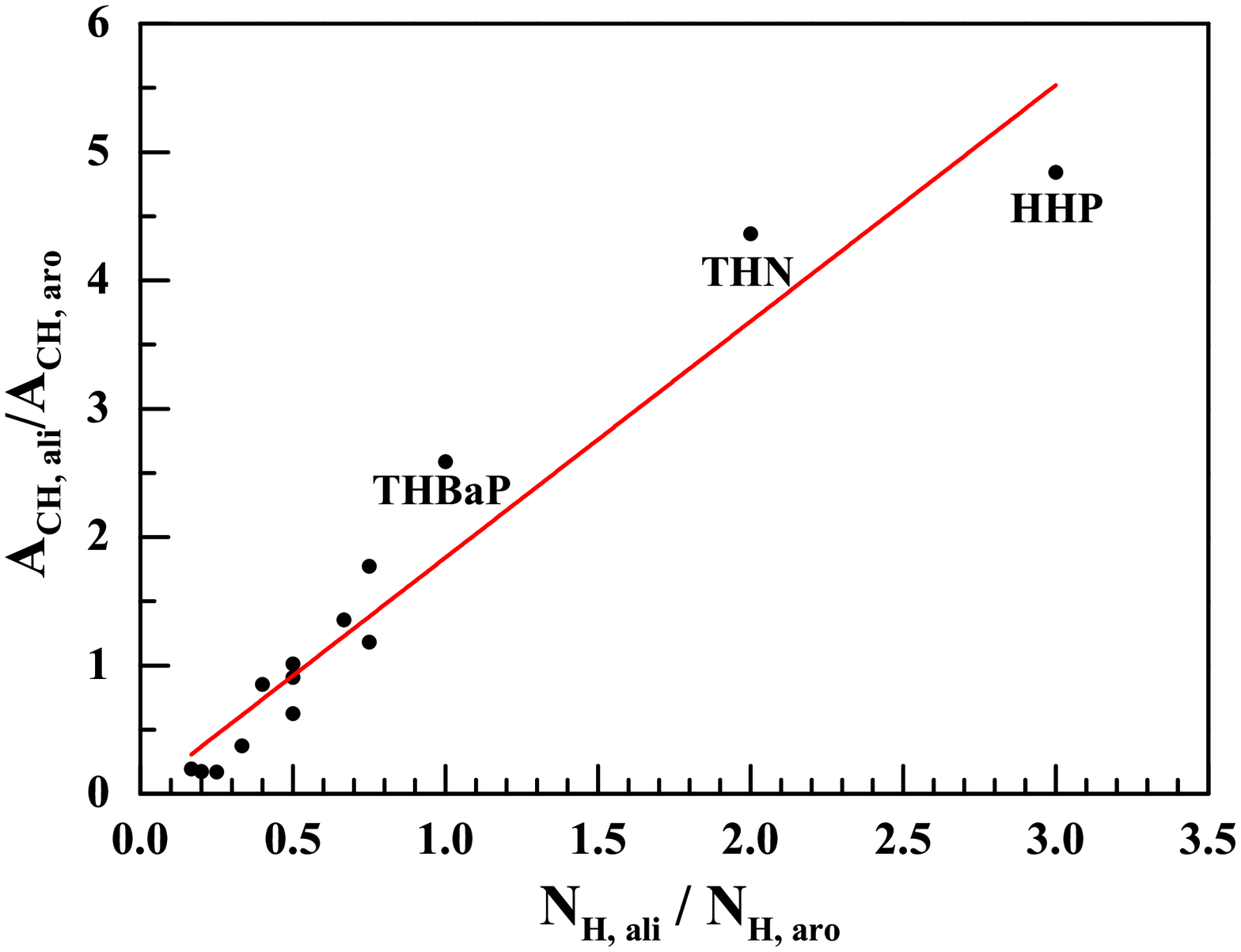}
}
\caption{\footnotesize
         \label{fig:NHaliOverNHaro}
         The ratios of the strengths of the 3.4$\mum$
         aliphatic C--H stretch ($A_{\rm CH,ali}$)
         to that of the 3.3$\mum$ aromatic C--H stretch
         ($A_{\rm CH,aro}$)
         versus the ratios of the number of
         aliphatic C--H bonds  ($N_{\rm H,ali}$)
         to the number of aromatic C--H bonds
         ($N_{\rm H,aro}$) for the Sandford et al.\ (2013) molecules.
         The red solid line, with a slope of $\simali$1.85,
         is the least-square fit to the data.
         The slope would increase to $\simali$1.98
          if the derivaties of benzene, naphthalene,
          and perylene are also included. 
         }
\end{figure*}

\begin{figure*}
\begin{center}
\begin{minipage}[t]{1.0\textwidth}
\resizebox{16.8cm}{8cm}{\includegraphics[clip]{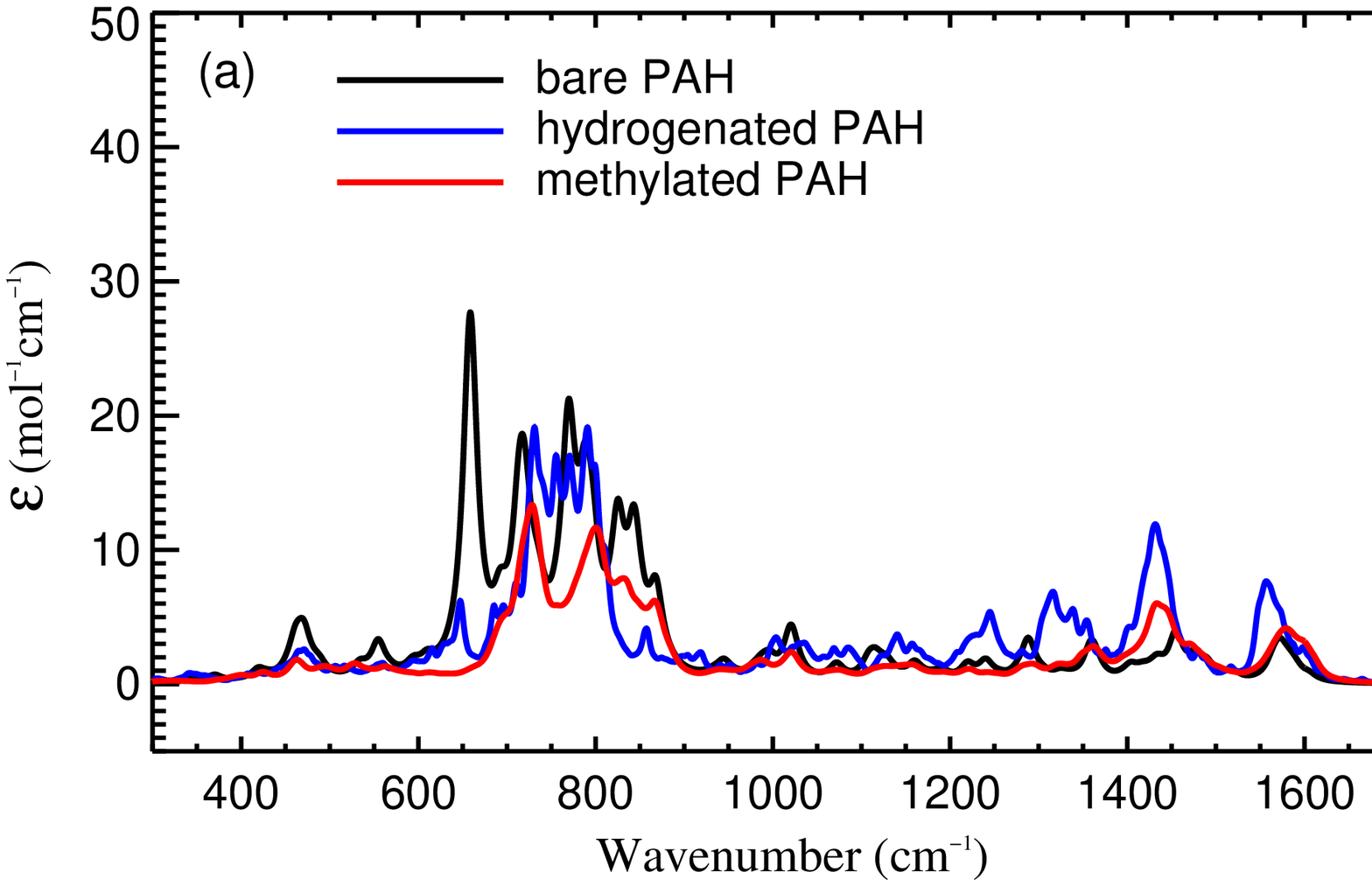}}\vspace{0.2cm}
\resizebox{16.8cm}{8cm}{\includegraphics[clip]{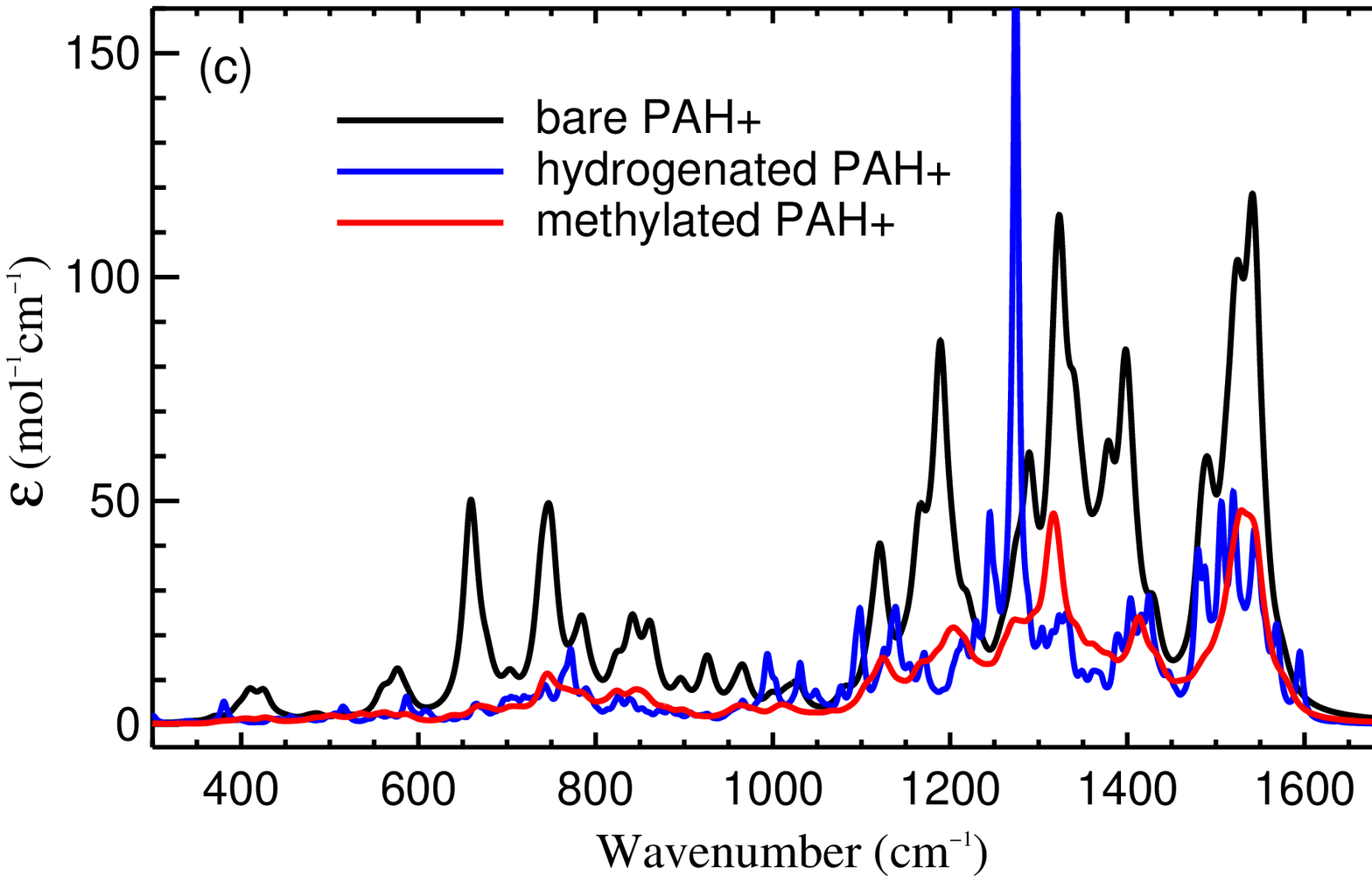}}\vspace{0.2cm}
\end{minipage}\hspace{-1.5cm}
\end{center}
\caption{\footnotesize
         \label{fig:AverageSpec_Compare}
         Comparison of the mean spectra of
         hydrogenated PAHs with methyl PAHs
         and their bare parental PAHs.
         The upper panels are for neutrals
         and the lower panels are for ions.
         The mean spectra of hydrogenated PAHs
         are derived by averaging the computed spectra,
         on a per unit aliphatic C--H bond basis,
         over all the hydrogenated species shown in
         Figures~\ref{fig:HBenzene_structure}--\ref{fig:HPAH_Sandford_structure}.
         The mean spectra of methylated PAHs
         are averaged, on a per C atom basis,
         over all the molecules
         listed in Figure~2 of Yang et al.\ (2013).
         The mean spectra of bare PAHs
         are obtained by averaging over benzene,
         naphthalene, anthracene, phenanthrene,
         pyrene, perylene and coronene,
         on a per C atom basis.
         For clarity, the mean spectra of methyl PAHs
         and bare PAHs are scaled by a factor of
         2 and 40, respectively.
         }
\end{figure*}

\begin{figure*}
\centering
{
\includegraphics[width=1.0\textwidth,angle=0]{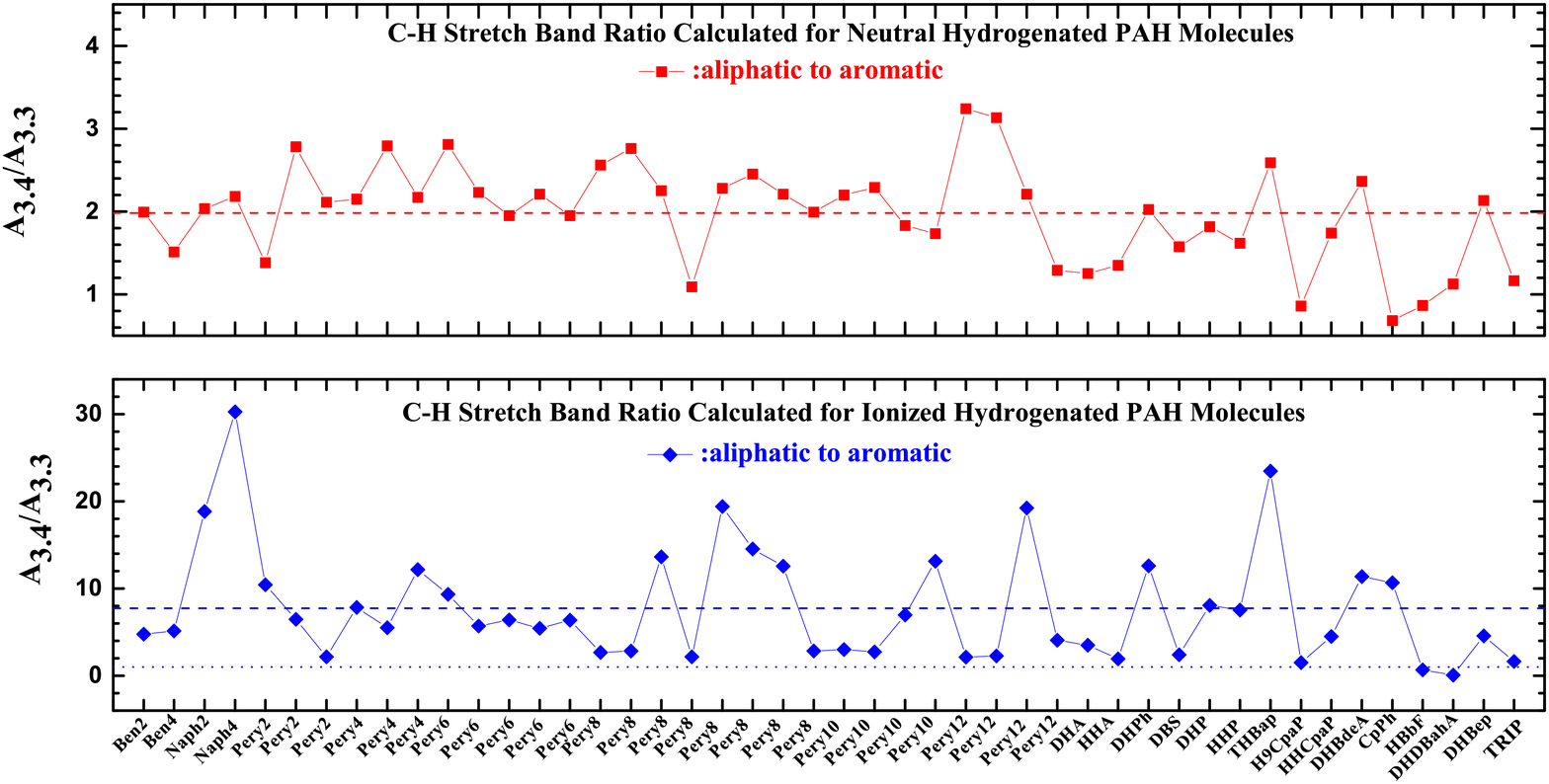}
}
\caption{\footnotesize
         \label{fig:A33A34_all}
         Band-strength ratios ($\Aratio$) computed at level
         {\rm B3LYP/6-311+G$^{\ast\ast}$}
         for the hydrogenated PAH molecules shown in
         Figures~\ref{fig:HBenzene_structure}--\ref{fig:HPAH_Sandford_structure}.
         The dashed horizontal lines plot
         the mean value of
         $\langle\Aratio\rangle\approx1.98$ for the neutrals
         and $\approx7.73$ for the cations.
         The dotted horizontal line plots
         the low-end  value of $\Aratio\simgt1.0$
         for all the hydrogenated PAH ions.
         }
\end{figure*}

\begin{figure*}
\centering
{
\includegraphics[width=1.5\textwidth,angle=0]{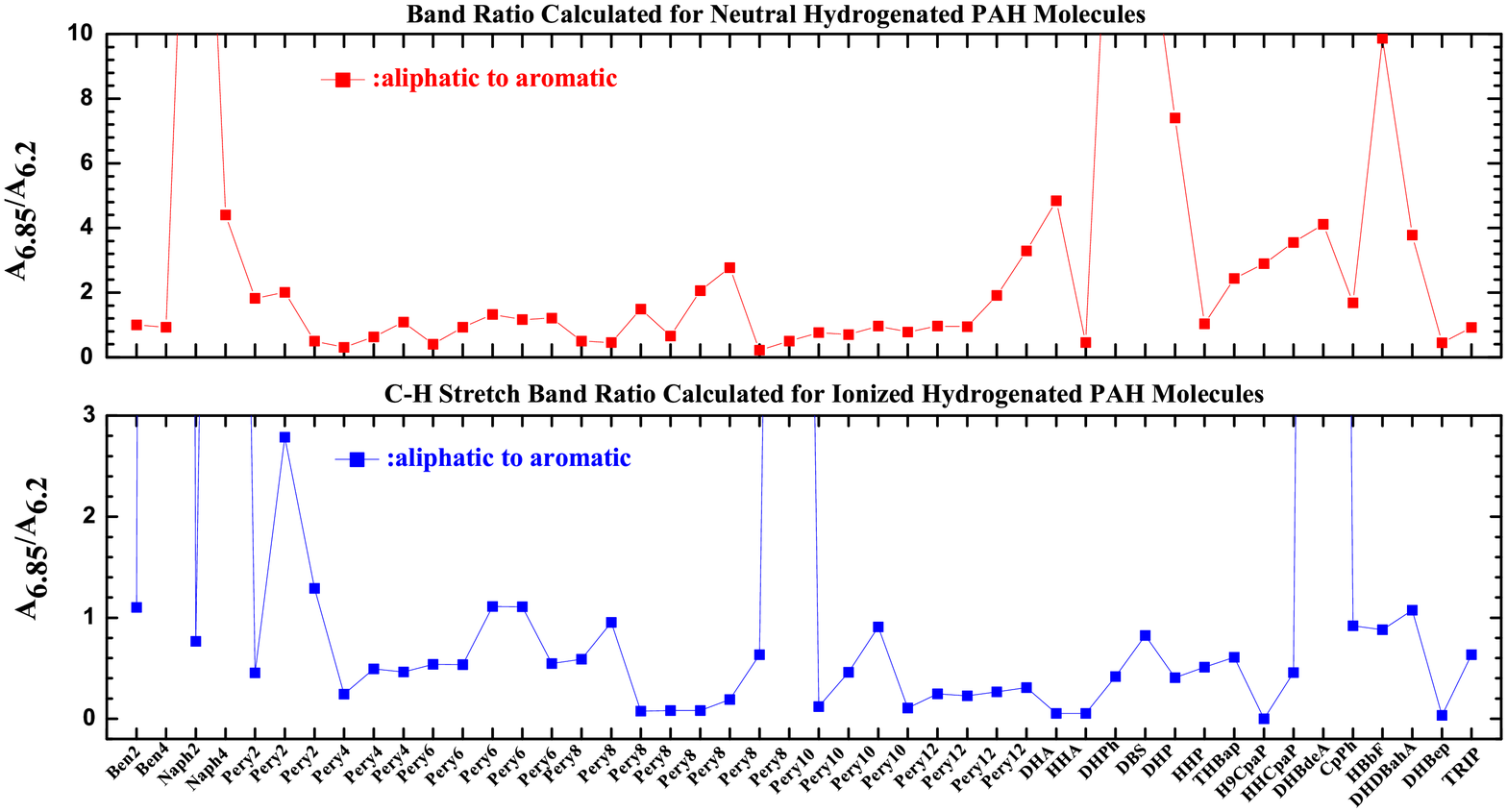}\vspace{-5cm}
}
\caption{\footnotesize
         \label{fig:A685A62_all}
         Same as Figure~\ref{fig:A33A34_all}
         but for $\Adfa/\Acc$.
         }
\end{figure*}

\begin{table*}
\footnotesize
\begin{center}
\caption[]{\footnotesize
           Computed Total Energies
           and Thermochemical Parameters
           for the Hydrogenated Benzene Molecules
           and Their Ions
           as Shown in Figure~\ref{fig:HBenzene_structure}
           at the {\rm B3LYP/6-311+G$^{\ast\ast}$} Level.
           }
\label{tab:E_ThermPara_Benzene}
\begin{tabular}{lccccccc}
\noalign{\smallskip} \hline \hline \noalign{\smallskip}
Compound	&	$\Etot$$^a$	
                &	VZPE$^b$	
                &	TE$^c$	
                &	$S$$^d$	
                &	$\nu_1$$^e$	
                &	$\nu_2$$^e$	
                &	$\mu$$^f$\\	
\noalign{\smallskip} \hline \noalign{\smallskip}
Ben$\_$2H	&	-233.483877 	&	76.42 	&	79.68 	&	70.95 	&	196.11 	&	300.94 	&	0.4992 	\\
Ben$\_$4H	&	-234.713155 	&	91.36 	&	94.82 	&	73.91 	&	165.69 	&	275.50 	&	0.3573 	\\
Ben$\_$6H	&	-235.944820 	&	106.34 	&	109.94 	&	74.68 	&	230.63 	&	230.90 	&	0.0000 	\\\hline
Ben$\_$2H+	&	-233.194323 	&	75.78 	&	79.23 	&	74.30 	&	95.51 	&	250.66 	&	0.0454 	\\
Ben$\_$4H+	&	-234.394150 	&	89.24 	&	92.97 	&	75.53 	&	165.44 	&	205.57 	&	2.5497 	\\
Ben$\_$6H+	&	-235.588818 	&	101.51 	&	105.63 	&	78.71 	&	198.63 	&	261.93 	&	0.0000 	\\
\hline
\noalign{\smallskip} \noalign{\smallskip}
\end{tabular}
\begin{description}
\item[$^{a}$] Total energies in atomic units.
\item[$^{b}$] Vibrational zero-point energies (VZPE)
                  in $\kcal\mol^{-1}$.
\item[$^{c}$] Thermal energies (TE) in $\kcal\mol^{-1}$.
\item[$^{d}$] Molecular entropies ($S$) in $\cals\mol^{-1}\K^{-1}$.

\item[$^{e}$] The lowest vibrational modes $\nu_1$ and $\nu_2$
                  in $\cm^{-1}$.
\item[$^{f}$] Dipole moment in Debye.
\end{description}
\end{center}
\end{table*}

\begin{table*}
\footnotesize
\begin{center}
\caption[]{\footnotesize
           IR Intensity (km\,mol$^{-1}$) of
           the 3.4 and 6.85$\mum$ Aliphatic C--H Bands,
           the 3.3$\mum$ Aromatic C--H Stretch Band,
           and the 6.2$\mum$ Aromatic C--C Stretch Band
           Computed at the {\rm B3LYP/6-311+G$^{\ast\ast}$} Level
           for All the Hydrogenated Benzenes
           as Shown in Figure~\ref{fig:HBenzene_structure}.
           The $\Aaro$, $\Aali$ and $\Adfa$ 
           Band Strengths Are on a per C--H Bond Basis,
           While the $\Acc$ Band Strength is 
           on a per C Atom Basis.
           Also Tabulated Are the Band-Strength Ratios
           $A_{3.4}/A_{3.3}$ and $A_{6.85}/A_{6.2}$.
           }
\label{tab:AValueARatio_Benzene}
\begin{tabular}{lcccccc}
\noalign{\smallskip} \hline \hline \noalign{\smallskip}
Compound	    &	$A_{3.4}$
                &	$A_{6.85}$
                &	$A_{3.3}$
                &   $A_{6.2}$
                &   $A_{3.4}/A_{3.3}$
                &   $A_{6.85}/A_{6.2}$
\\ \noalign{\smallskip} \hline \noalign{\smallskip}
Ben$\_$2H	&	32.92 	&	0.49 	&	16.54 	&	0.49 	&	1.99 	&	1.00 	\\
Ben$\_$4H	&	38.46 	&	2.29 	&	25.52 	&	2.48 	&	1.51 	&	0.93 	\\
Ben$\_$6H	&	45.39 	&	2.33 	&	$-$	&	$-$	&	$-$	&	$-$	\\
Average	        &	38.92 	&	1.70 	&	21.03 	&	1.48 	&	1.75 	&	0.96 	\\ \hline
Ben$\_$2H+	&	14.24 	&	30.08 	&	2.99 	&	27.29 	&	4.77 	&	1.10 	\\
Ben$\_$4H+	&	24.90 	&	3.03 	&	4.85 	&	0.02 	&	5.13 	&	139.10 	\\
Ben$\_$6H+	&	40.75 	&	2.20 	&	$-$	&	$-$	&	$-$	&	$-$	\\
Average	        &	26.63 	&	11.77 	&	3.92 	&	13.65 	&	4.95 	&	70.10 	\\ \hline
\noalign{\smallskip} \noalign{\smallskip}
\end{tabular}
\end{center}
\end{table*}

\begin{table*}
\footnotesize
\begin{center}
\caption[]{\footnotesize
            Same as Table~\ref{tab:E_ThermPara_Benzene}
            but for the Hydrogenated Derivatives of Naphthalene
            as Shown in Figure~\ref{fig:HNaph_structure}.
            }
\label{tab:E_ThermPara_Naph}
\begin{tabular}{lccccccc}
\noalign{\smallskip} \hline \hline \noalign{\smallskip}
Compound	&	$\Etot$
                &	VZPE
                &	TE
                &	$S$
                &	$\nu_1$
                &	$\nu_2$
                &	$\mu$\\	
\noalign{\smallskip} \hline \noalign{\smallskip}
DHN (Naph$\_$2H)	&	-387.180337 	&	106.21 	&	110.92 	&	85.00 	&	134.45 	&	152.34 	&	0.6693 	\\
THN (Naph$\_$4H)	&	-388.405052 	&	121.04 	&	126.01 	&	85.70 	&	96.16 	&	143.04 	&	0.7829 	\\
OHN (Naph$\_$8H)	&	-390.810255 	&	149.73 	&	155.35 	&	88.65 	&	91.21 	&	141.58 	&	0.0000 	\\
c-PHN (Naph$\_$10Ha)	&	-392.031003 	&	165.09 	&	170.64 	&	88.69 	&	138.52 	&	154.01 	&	0.0269 	\\
t-PHN (Naph$\_$10Hb)	&	-392.036103 	&	164.83 	&	170.42 	&	89.19 	&	128.35 	&	138.68 	&	0.0000 	\\\hline
DHN+	    &	-386.898228 	&	105.83 	&	110.70 	&	87.72 	&	113.20 	&	135.78 	&	1.1938 	\\
THN+	    &	-388.101882 	&	119.58 	&	124.85 	&	89.52 	&	88.20 	&	91.18 	&	2.0163 	\\
OHN+	    &	-390.524810 	&	148.44 	&	154.25 	&	91.85 	&	81.26 	&	81.49 	&	0.0000 	\\
c-PHN+	&	-391.691884 	&	160.73 	&	166.81 	&	94.67 	&	126.77 	&	141.78 	&	2.2291 	\\
t-PHN+	&	-391.706269 	&	161.77 	&	167.83 	&	93.38 	&	109.62 	&	143.02 	&	0.0000 	\\
\hline
\noalign{\smallskip} \noalign{\smallskip}
\end{tabular}
\end{center}
\end{table*}

\begin{table*}
\footnotesize
\begin{center}
\caption[]{\footnotesize
            Same as Table~\ref{tab:AValueARatio_Benzene}
            but for the Hydrogenated Derivatives of Naphthalene
            as Shown in Figure~\ref{fig:HNaph_structure}.
           }
\label{tab:AValueARatio_Naph}
\begin{tabular}{lcccccc}
\noalign{\smallskip} \hline \hline \noalign{\smallskip}
Compound	    &	$A_{3.4}$
                &	$A_{6.85}$
                &	$A_{3.3}$
                &   $A_{6.2}$
                &   $A_{3.4}/A_{3.3}$
                &   $A_{6.85}/A_{6.2}$
\\ \noalign{\smallskip} \hline \noalign{\smallskip}
DHN (Naph$\_$2H)	    &	31.05 	&	4.16	&	15.28 	&	0.17	&	2.03 	&	24.85 	\\
THN (Naph$\_$4H)	    &	35.85 	&	3.00	&	16.43 	&	0.68	&	2.18 	&	4.40 	\\
OHN (Naph$\_$8H)	    &	44.40 	&	1.61	&	$-$	&	0.00	&	$-$	&	$-$	\\
c-PHN (Naph$\_$10Ha)	&	44.65 	&	2.29	&	$-$	&	$-$	&	$-$	&	$-$	\\
t-PHN (Naph$\_$10Hb)	&	43.61 	&	1.91	&	$-$	&	$-$	&	$-$	&	$-$	\\
Average	            &	39.91 	&	2.59 	&	15.85 	&	0.28 	&	2.11 	&	14.62 	\\ \hline
DHN+	            &	7.28	&	28.67 	&	0.39	&	37.49	&	18.82	&	0.76 	\\
THN+	            &	17.87	&	2.32 	&	0.59	&	0.12	&	30.28	&	19.69 	\\
OHN+	            &	12.61	&	1.91 	&	$-$	&	0.00	&	$-$	&	$-$	\\
c-PHN+	            &	28.65	&	4.08 	&	$-$	&	$-$	&	$-$	&	$-$	\\
t-PHN+	            &	21.81	&	1.99 	&	$-$	&	$-$	&	$-$	&	$-$	\\
Average	            &	17.64 	&	7.79 	&	0.49 	&	12.54 	&	24.55 	&	10.23 	\\ \hline
\noalign{\smallskip} \noalign{\smallskip}
\end{tabular}
\end{center}
\end{table*}

\begin{table*}
\tiny
\begin{center}
\caption[]{\footnotesize
            Same as Table~\ref{tab:E_ThermPara_Benzene}
            but for the Hydrogenated Derivatives of Perylene
            as Shown in Figure~\ref{fig:HPery_structure}.
            }
\label{tab:E_ThermPara_Pery_Neutral}
\begin{tabular}{lccccccc}
\noalign{\smallskip} \hline \hline \noalign{\smallskip}
Compound	&	$\Etot$	
                &	VZPE
                &	TE
                &	$S$
                &	$\nu_1$
                &	$\nu_2$
                &	$\mu$\\	
\noalign{\smallskip} \hline \noalign{\smallskip}
Pery$\_$2H$\_$RamII	&	-770.753921 	&	172.62 	&	181.11 	&	114.95 	&	37.33 	&	64.53 	&	0.6515 	\\
Pery$\_$2H$\_$RamIII	&	-770.772978 	&	172.78 	&	181.24 	&	113.61 	&	50.92 	&	92.01 	&	0.8480 	\\
Pery$\_$2H$\_$RamIV	&	-770.744407 	&	172.18 	&	180.77 	&	114.57 	&	49.57 	&	84.50 	&	0.7150 	\\
Pery$\_$4H	&	-771.953401 	&	187.06 	&	195.83 	&	115.86 	&	47.24 	&	86.27 	&	1.0315 	\\
Pery$\_$4H$\_$RG1	&	-771.951377 	&	186.87 	&	195.70 	&	116.69 	&	34.57 	&	85.78 	&	1.3194 	\\
Pery$\_$4H$\_$RG2	&	-771.986296 	&	187.76 	&	196.38 	&	114.93 	&	51.61 	&	77.12 	&	0.9694 	\\
Pery$\_$6H	&	-773.178734 	&	201.90 	&	210.92 	&	117.59 	&	40.81 	&	84.47 	&	1.3198 	\\
Pery$\_$6H$\_$Yal	&	-773.205900 	&	202.12 	&	211.18 	&	118.20 	&	50.54 	&	60.42 	&	1.4594 	\\
Pery$\_$6H1$\_$Yal	&	-773.205072 	&	202.04 	&	211.17 	&	119.50 	&	42.50 	&	52.96 	&	1.4314 	\\
Pery$\_$6H$\_$ZinkeR	&	-773.196404 	&	202.07 	&	211.12 	&	117.66 	&	57.26 	&	73.40 	&	0.0460 	\\
Pery$\_$6H1$\_$Zinke	&	-773.196165 	&	202.09 	&	211.15 	&	117.85 	&	56.58 	&	67.78 	&	0.0000 	\\
Pery$\_$8H	&	-774.333885 	&	215.26 	&	224.77 	&	120.87 	&	39.69 	&	78.16 	&	1.4590 	\\
Pery$\_$8Hb	&	-774.335840 	&	215.32 	&	224.79 	&	119.99 	&	53.13 	&	81.01 	&	1.4564 	\\
Pery$\_$8H$\_$Yal	&	-774.404986 	&	216.86 	&	226.01 	&	118.14 	&	62.98 	&	69.23 	&	1.3854 	\\
Pery$\_$8Hb$\_$Yal	&	-774.392233 	&	216.94 	&	226.17 	&	119.13 	&	52.82 	&	80.29 	&	1.4419 	\\
Pery$\_$8H$\_$Zinke	&	-774.398164 	&	216.87 	&	225.97 	&	117.79 	&	57.52 	&	76.52 	&	0.5307 	\\
Pery$\_$8Hb$\_$Zinke	&	-774.385387 	&	216.62 	&	226.01 	&	123.33 	&	15.58 	&	56.86 	&	0.0000 	\\
Pery$\_$8H$\_$RG1	&	-774.384872 	&	216.71 	&	225.97 	&	119.82 	&	28.89 	&	86.27 	&	1.4488 	\\
Pery$\_$8H$\_$RG2	&	-774.387559 	&	216.72 	&	225.82 	&	117.94 	&	38.78 	&	84.51 	&	1.1210 	\\
Pery$\_$10H	&	-775.541432 	&	230.08 	&	239.83 	&	122.70 	&	42.09 	&	73.37 	&	0.9738 	\\
Pery$\_$10Hb	&	-775.539497 	&	230.09 	&	239.77 	&	121.83 	&	45.13 	&	78.15 	&	1.0945 	\\
Pery$\_$10H$\_$RG1	&	-775.612714 	&	231.25 	&	240.85 	&	121.84 	&	49.44 	&	62.02 	&	1.0798 	\\
Pery$\_$10H$\_$RG2	&	-775.563323 	&	231.13 	&	240.68 	&	121.99 	&	31.45 	&	68.40 	&	1.3365 	\\
Pery$\_$12H	&	-776.766232 	&	244.94 	&	254.92 	&	124.01 	&	43.15 	&	71.18 	&	0.3473 	\\
Pery$\_$12Hb	&	-776.768985 	&	245.00 	&	254.95 	&	123.31 	&	51.44 	&	78.04 	&	0.2981 	\\
Pery$\_$12H$\_$RG1	&	-776.800583 	&	245.45 	&	255.36 	&	123.69 	&	48.54 	&	67.71 	&	0.5676 	\\
Pery$\_$12H$\_$RG2	&	-776.799299 	&	245.29 	&	255.38 	&	125.67 	&	45.82 	&	64.18 	&	0.5083 	\\
Pery$\_$14H$\_$Yal	&	-778.023443 	&	260.28 	&	270.43 	&	125.47 	&	45.81 	&	69.16 	&	0.0643 	\\
\hline
HC$\_$Pery$\_$2H$\_$RamII+	&	-770.496974 	&	172.38 	&	180.96 	&	116.38 	&	43.61 	&	76.94 	&	2.1098 	\\
HC$\_$Pery$\_$2H$\_$RamIII+	&	-770.526761 	&	172.81 	&	181.40 	&	116.28 	&	44.18 	&	85.96 	&	1.4860 	\\
HC$\_$Pery$\_$2H$\_$RamIV+	&	-770.522678 	&	172.73 	&	181.35 	&	116.50 	&	45.79 	&	79.38 	&	1.3025 	\\
HC$\_$Pery$\_$4H+	&	-771.723440 	&	187.41 	&	196.25 	&	117.96 	&	44.23 	&	80.46 	&	0.9571 	\\
HC$\_$Pery$\_$4H$\_$RG1+	&	-771.716524 	&	187.11 	&	196.01 	&	118.63 	&	36.45 	&	81.25 	&	0.7951 	\\
HC$\_$Pery$\_$4H$\_$RG2+	&	-771.728963 	&	187.68 	&	196.49 	&	118.58 	&	33.29 	&	66.61 	&	2.1650 	\\
HC$\_$Pery$\_$6H+	&	-772.940502 	&	202.02 	&	211.15 	&	120.29 	&	32.91 	&	77.56 	&	1.6254 	\\
HC$\_$Pery$\_$6H$\_$Yal+	&	-772.951020 	&	201.91 	&	211.10 	&	119.31 	&	42.03 	&	65.22 	&	0.2005 	\\
HC$\_$Pery$\_$6H1$\_$Yal+	&	-772.949412 	&	201.75 	&	211.09 	&	124.84 	&	13.04 	&	56.54 	&	0.3613 	\\
HC$\_$Pery$\_$6H$\_$ZinkeR+	&	-772.956553 	&	202.26 	&	211.42 	&	120.29 	&	52.95 	&	64.70 	&	0.3530 	\\
HC$\_$Pery$\_$6H1$\_$Zinke+	&	-772.956228 	&	202.28 	&	211.45 	&	120.67 	&	49.67 	&	58.93 	&	0.0000 	\\
HC$\_$Pery$\_$8H+	&	-774.122849 	&	215.74 	&	225.31 	&	123.40 	&	31.23 	&	71.19 	&	1.8918 	\\
HC$\_$Pery$\_$8Hb+	&	-774.123811 	&	215.78 	&	225.32 	&	122.39 	&	44.13 	&	75.32 	&	1.8700 	\\
HC$\_$Pery$\_$8H$\_$Yal+	&	-774.136447 	&	216.67 	&	226.01 	&	121.08 	&	59.84 	&	61.18 	&	2.2739 	\\
HC$\_$Pery$\_$8Hb$\_$Yal+	&	-774.125730 	&	216.70 	&	226.11 	&	121.97 	&	50.31 	&	72.40 	&	1.8594 	\\
HC$\_$Pery$\_$8H$\_$Zinke+	&	-774.110956 	&	214.71 	&	224.21 	&	122.23 	&	47.63 	&	68.01 	&	0.8075 	\\
HC$\_$Pery$\_$8Hb$\_$Zinke+	&	-774.105278 	&	214.94 	&	224.50 	&	123.41 	&	40.66 	&	65.24 	&	1.1675 	\\
HC$\_$Pery$\_$8H$\_$RG1+	&	-774.144315 	&	216.82 	&	226.08 	&	120.61 	&	47.50 	&	73.98 	&	2.0791 	\\
HC$\_$Pery$\_$8H$\_$RG2+	&	-774.131549 	&	216.60 	&	225.84 	&	120.41 	&	37.24 	&	79.73 	&	3.6216 	\\
HC$\_$Pery$\_$10H+	&	-775.323173 	&	230.35 	&	240.17 	&	125.04 	&	35.95 	&	71.65 	&	2.4960 	\\
HC$\_$Pery$\_$10Hb+	&	-775.322751 	&	230.44 	&	240.36 	&	126.68 	&	29.92 	&	62.85 	&	2.4753 	\\
HC$\_$Pery$\_$10H$\_$RG1+	&	-775.359377 	&	230.97 	&	240.76 	&	125.13 	&	39.86 	&	60.39 	&	2.3740 	\\
HC$\_$Pery$\_$10H$\_$RG2+	&	-775.325197 	&	231.10 	&	240.59 	&	121.93 	&	53.18 	&	65.83 	&	3.4206 	\\
HC$\_$Pery$\_$12H+	&	-776.541119 	&	245.01 	&	255.12 	&	127.38 	&	28.86 	&	67.25 	&	2.3983 	\\
HC$\_$Pery$\_$12Hb+	&	-776.542472 	&	245.10 	&	255.15 	&	126.20 	&	36.41 	&	72.09 	&	2.4220 	\\
HC$\_$Pery$\_$12H$\_$RG1+	&	-776.544666 	&	245.15 	&	255.25 	&	127.08 	&	38.21 	&	63.75 	&	2.8541 	\\
HC$\_$Pery$\_$12H$\_$RG2+	&	-776.549805 	&	244.92 	&	255.13 	&	128.35 	&	42.55 	&	49.79 	&	3.7425 	\\
HC$\_$Pery$\_$14H$\_$Yal+	&	-777.754163 	&	258.90 	&	269.47 	&	131.13 	&	28.64 	&	63.37 	&	0.0418 	\\
\hline
\noalign{\smallskip} \noalign{\smallskip}
\end{tabular}
\end{center}
\end{table*}

\begin{table*}
\tiny
\begin{center}
\caption[]{\footnotesize
               Same as Table~\ref{tab:AValueARatio_Benzene}
               but for the Hydrogenated Derivatives of Perylene
               as Shown in Figure~\ref{fig:HPery_structure}.
               }
\label{tab:AValueARatio_NeuPery}
\begin{tabular}{lcccccc}
\noalign{\smallskip} \hline \hline \noalign{\smallskip}
Compound	    &	$A_{3.4}$
                &	$A_{6.85}$
                &	$A_{3.3}$
                &   $A_{6.2}$
                &   $A_{3.4}/A_{3.3}$
                &   $A_{6.85}/A_{6.2}$
\\ \noalign{\smallskip} \hline \noalign{\smallskip}
Pery$\_$2H$\_$RamII	    &	20.13 	&	4.83 	&	14.61 	&	2.65	&	1.38 	&	1.82 	\\
Pery$\_$2H$\_$RamIII	&	36.05 	&	3.54 	&	12.95 	&	1.76	&	2.78 	&	2.01 	\\
Pery$\_$2H$\_$RamIV	    &	30.63 	&	1.35 	&	14.54 	&	2.70	&	2.11 	&	0.50 	\\\hline
Pery$\_$4H	            &	34.47 	&	0.87 	&	16.06 	&	2.89	&	2.15 	&	0.30 	\\
Pery$\_$4H$\_$RG1	    &	38.10 	&	1.47 	&	13.65 	&	2.33	&	2.79 	&	0.63 	\\
Pery$\_$4H$\_$RG2	    &	30.53 	&	2.96 	&	14.10 	&	2.73	&	2.17 	&	1.09 	\\\hline
Pery$\_$6H	            &	39.43 	&	1.59 	&	14.02 	&	3.94	&	2.81 	&	0.40 	\\
Pery$\_$6H$\_$Yal	    &	34.57 	&	2.02 	&	15.67 	&	2.17	&	2.21 	&	0.93 	\\
Pery$\_$6H1$\_$Yal	    &	34.86 	&	2.92 	&	15.63 	&	2.20	&	2.23 	&	1.32 	\\
Pery$\_$6H$\_$ZinkeR	&	33.73 	&	3.20 	&	17.30 	&	2.75	&	1.95 	&	1.16 	\\
Pery$\_$6H1$\_$Zinke	&	33.82 	&	2.72 	&	17.33 	&	2.24	&	1.95 	&	1.21 	\\\hline
Pery$\_$8H	            &	41.49 	&	1.72 	&	16.19 	&	3.48	&	2.56 	&	0.50 	\\
Pery$\_$8Hb	            &	40.25 	&	1.54 	&	14.58 	&	3.41	&	2.76 	&	0.45 	\\
Pery$\_$8H$\_$Yal	    &	33.95 	&	2.96 	&	15.33 	&	1.98	&	2.21 	&	1.49 	\\
Pery$\_$8Hb$\_$Yal	    &	34.12 	&	1.48 	&	15.18 	&	2.26	&	2.25 	&	0.66 	\\
Pery$\_$8H$\_$Zinke	    &	32.59 	&	3.81 	&	16.35 	&	1.85	&	1.99 	&	2.06 	\\
Pery$\_$8Hb$\_$Zinke	&	27.87 	&	6.15 	&	25.54 	&	2.22	&	1.09 	&	2.77 	\\
Pery$\_$8H$\_$RG1	    &	35.75 	&	1.28 	&	15.65 	&	5.95	&	2.28 	&	0.22 	\\
Pery$\_$8H$\_$RG2	    &	36.46 	&	1.38 	&	14.85 	&	2.77	&	2.45 	&	0.50 	\\\hline
Pery$\_$10H	            &	40.28 	&	1.59 	&	18.30 	&	2.07	&	2.20 	&	0.77 	\\
Pery$\_$10Hb	        &	39.28 	&	1.74 	&	17.16 	&	2.48	&	2.29 	&	0.70 	\\
Pery$\_$10H$\_$RG1	    &	38.74 	&	1.92 	&	21.21 	&	1.99	&	1.83 	&	0.97 	\\
Pery$\_$10H$\_$RG2	    &	32.73 	&	2.19 	&	18.97 	&	2.82	&	1.73 	&	0.78 	\\\hline
Pery$\_$12H	            &	42.09 	&	1.56 	&	12.97 	&	1.62	&	3.24 	&	0.96 	\\
Pery$\_$12Hb	        &	42.14 	&	1.49 	&	13.47 	&	1.58	&	3.13 	&	0.95 	\\
Pery$\_$12H$\_$RG1	    &	39.62 	&	1.84 	&	17.91 	&	0.96	&	2.21 	&	1.91 	\\
Pery$\_$12H$\_$RG2	    &	39.01 	&	1.81 	&	30.13 	&	0.55	&	1.29 	&	3.29 	\\\hline
Average	                &	35.66 	&	2.29 	&	16.65 	&	2.46 	&	2.22 	&	1.12 	\\
\hline\hline
Pery$\_$2H$\_$RamII+    &	10.83 	&	9.99 	&	1.04 	&	21.98 	&	10.45 	&	0.45 	\\
Pery$\_$2H$\_$RamIII+	&	7.94 	&	38.26 	&	1.23 	&	13.72 	&	6.46 	&	2.79 	\\
Pery$\_$2H$\_$RamIV+    &	3.96 	&	19.57 	&	1.82 	&	15.18 	&	2.17 	&	1.29 	\\\hline
Pery$\_$4H+	            &	13.85 	&	5.04 	&	1.77 	&	20.77 	&	7.82 	&	0.24 	\\
Pery$\_$4H$\_$RG1+    	&	9.80 	&	14.52 	&	1.78 	&	29.49 	&	5.51 	&	0.49 	\\
Pery$\_$4H$\_$RG2+   	&	13.78 	&	11.54 	&	1.13 	&	24.94 	&	12.16 	&	0.46 	\\\hline
Pery$\_$6H+	            &	15.15 	&	8.85 	&	1.63 	&	16.42 	&	9.32 	&	0.54 	\\
Pery$\_$6H2$\_$Yal+	    &	10.07 	&	11.91 	&	1.86 	&	22.29 	&	5.42 	&	0.53 	\\
Pery$\_$6H1$\_$Yal+    	&	10.85 	&	28.83 	&	1.90 	&	25.93 	&	5.71 	&	1.11 	\\
Pery$\_$6H$\_$Zinke+   	&	10.39 	&	13.39 	&	1.63 	&	12.07 	&	6.37 	&	1.11 	\\
Pery$\_$6H1$\_$Zinke+	&	10.55 	&	6.96 	&	1.65 	&	12.74 	&	6.40 	&	0.55 	\\\hline
Pery$\_$8H+	            &	13.54 	&	7.78 	&	5.09 	&	13.24 	&	2.66 	&	0.59 	\\
Pery$\_$8Hb+         	&	12.97 	&	12.90 	&	4.60 	&	13.53 	&	2.82 	&	0.95 	\\
Pery$\_$8H$\_$Yal+    	&	13.02 	&	2.93 	&	1.04 	&	39.28 	&	12.57 	&	0.07 	\\
Pery$\_$8Hb$\_$Yal+	    &	13.02 	&	2.66 	&	0.95 	&	33.44 	&	13.65 	&	0.08 	\\
Pery$\_$8H$\_$Zinke+    &	13.37 	&	9.34 	&	4.70 	&	117.20 	&	2.85 	&	0.08 	\\
Pery$\_$8Hb$\_$Zinke+	&	10.67 	&	19.65 	&	4.90 	&	103.34 	&	2.18 	&	0.19 	\\
Pery$\_$8H$\_$RG1+	    &	14.52 	&	19.37 	&	0.75 	&	30.54 	&	19.39 	&	0.63 	\\
Pery$\_$8H$\_$RG2+	    &	18.17 	&	50.40 	&	1.25 	&	2.32 	&	14.55 	&	21.71 	\\\hline
Pery$\_$10H+	        &	14.75 	&	2.73 	&	4.93 	&	22.99 	&	2.99 	&	0.12 	\\
Pery$\_$10Hb+	        &	14.14 	&	8.50 	&	5.18 	&	18.49 	&	2.73 	&	0.46 	\\
Pery$\_$10H$\_$RG1+	    &	15.32 	&	7.00 	&	2.20 	&	7.70 	&	6.96 	&	0.91 	\\
Pery$\_$10H$\_$RG2+	    &	15.79 	&	4.48 	&	1.20 	&	42.06 	&	13.14 	&	0.11 	\\\hline
Pery$\_$12H+	            &	14.76 	&	7.58 	&	6.86 	&	30.68 	&	2.15 	&	0.25 	\\
Pery$\_$12Hb+         	&	14.85 	&	7.70 	&	6.53 	&	33.82 	&	2.28 	&	0.23 	\\
Pery$\_$12H$\_$RG1+	    &	15.78 	&	9.57 	&	0.82 	&	36.09 	&	19.24 	&	0.27 	\\
Pery$\_$12H$\_$RG2+	    &	14.99 	&	12.95 	&	3.69 	&	41.99 	&	4.07 	&	0.31 	\\\hline
Average	                &	12.85 	&	13.13 	&	2.67 	&	29.71 	&	7.48 	&	1.35 	\\\hline
\noalign{\smallskip} \noalign{\smallskip}
\end{tabular}
\end{center}
\end{table*}



\begin{table*}
\footnotesize
\begin{center}
\caption[]{\footnotesize
            Same as Table~\ref{tab:E_ThermPara_Benzene}
            but for the Sandford et al.\ (2013) Molecules
            as Shown in Figure~\ref{fig:HPAH_Sandford_structure}
            at the {\rm B3LYP/6-311+G$^{\ast\ast}$} Level.
           }
\label{tab:E_ThermPara_LongChain}
\begin{tabular}{lccccccc}
\noalign{\smallskip} \hline \hline \noalign{\smallskip}
Compound	&	$\Etot$	
                &	VZPE
                &	TE
                &	$S$
                &	$\nu_1$
                &	$\nu_2$
                &	$\mu$\\	
\noalign{\smallskip} \hline \noalign{\smallskip}

DHA	&	-540.868165 	&	135.60 	&	141.90 	&	96.93 	&	53.31 	&	122.15 	&	0.3557 	\\
HHA	&	-543.219349 	&	162.56 	&	170.10 	&	108.79 	&	36.87 	&	70.27 	&	0.0023 	\\
DHPh	&	-540.871768 	&	135.93 	&	142.18 	&	96.05 	&	86.59 	&	103.76 	&	0.5922 	\\
DBS	&	-580.183622 	&	153.93 	&	160.91 	&	103.72 	&	42.47 	&	93.64 	&	0.5200 	\\
tPHF	&	-508.771366 	&	205.07 	&	212.43 	&	105.17 	&	60.46 	&	104.25 	&	0.0993 	\\
DHP	&	-617.117922 	&	143.84 	&	150.51 	&	98.63 	&	86.75 	&	134.11 	&	0.6216 	\\
HHP	&	-619.530713 	&	172.79 	&	180.11 	&	104.72 	&	84.13 	&	97.29 	&	0.0000 	\\
THBaP	&	-772.007230 	&	187.51 	&	196.24 	&	115.50 	&	54.77 	&	78.13 	&	1.1531 	\\
H9CpaP	&	-731.455224 	&	154.18 	&	161.96 	&	108.25 	&	67.45 	&	113.41 	&	0.8871 	\\
HHCpaP	&	-731.455224 	&	154.18 	&	161.96 	&	108.25 	&	67.45 	&	113.41 	&	0.8871 	\\
DDHTP	&	-700.577894 	&	237.65 	&	247.33 	&	123.09 	&	50.06 	&	76.72 	&	0.0340 	\\
PHC 	&	-936.545478 	&	349.56 	&	361.11 	&	132.27 	&	64.00 	&	64.14 	&	0.0000 	\\
DHBdeA	&	-656.439215 	&	161.77 	&	169.27 	&	106.81 	&	67.38 	&	85.13 	&	0.9977 	\\
CpPh	&	-577.776337 	&	125.39 	&	131.42 	&	95.40 	&	94.07 	&	184.13 	&	0.6588 	\\
HBbF	&	-655.222412 	&	146.60 	&	153.90 	&	105.83 	&	60.67 	&	95.01 	&	0.6798 	\\
DHDBahA	&	-848.218890 	&	194.02 	&	203.80 	&	126.06 	&	19.21 	&	48.09 	&	0.0000 	\\
DHBep	&	-770.778358 	&	172.70 	&	181.23 	&	114.08 	&	61.35 	&	81.20 	&	0.7271 	\\
TRIP	&	-770.756412 	&	172.81 	&	181.11 	&	113.02 	&	66.45 	&	66.53 	&	0.0001 	\\\hline

DHA+        &	-540.572936 	&	133.52 	&	140.23 	&	102.32 	&	28.03 	&	93.78 	&	0.2662 	\\
HHA+        &	-542.936608 	&	160.22 	&	167.84 	&	109.54 	&	52.80 	&	83.74 	&	0.1517 	\\
DHPh+   	&	-540.593203 	&	135.71 	&	142.15 	&	99.01 	&	72.67 	&	95.49 	&	0.1409 	\\
DBS+	    &	-579.894009 	&	153.31 	&	160.50 	&	107.12 	&	34.65 	&	80.78 	&	0.3793 	\\
tPHF+    	&	-508.448630 	&	202.54 	&	210.39 	&	109.82 	&	53.47 	&	95.12 	&	3.2675 	\\
DHP+	    &	-616.844866 	&	143.39 	&	150.23 	&	101.20 	&	83.70 	&	124.12 	&	0.8562 	\\
HHP+	    &	-619.272752 	&	172.53 	&	180.02 	&	107.46 	&	81.67 	&	85.99 	&	0.0000 	\\
THBaP+	&	-771.755072 	&	187.51 	&	196.35 	&	117.79 	&	54.83 	&	69.96 	&	2.1796 	\\
H9CpaP+	&	-731.199725 	&	154.07 	&	161.94 	&	110.34 	&	68.24 	&	105.24 	&	1.2416 	\\
HHCpaP+	&	-734.823483 	&	197.89 	&	206.82 	&	118.72 	&	45.45 	&	81.58 	&	0.6217 	\\
DDHTP+	&	-700.312268 	&	236.08 	&	246.05 	&	124.76 	&	52.59 	&	67.68 	&	0.1916 	\\
PHC+ 	&	-936.252216 	&	346.64 	&	358.57 	&	135.86 	&	61.64 	&	64.77 	&	0.0000 	\\
DHBdeA+	&	-656.173080 	&	161.47 	&	169.14 	&	109.63 	&	58.14 	&	80.12 	&	0.5296 	\\
CpPh+	&	-577.500270 	&	124.81 	&	130.97 	&	97.57 	&	96.68 	&	175.49 	&	0.1925 	\\
HBbF+	&	-654.956057 	&	146.56 	&	153.97 	&	108.03 	&	57.63 	&	93.88 	&	0.8366 	\\
DHDBahA+	&	-847.951115 	&	193.15 	&	203.14 	&	130.79 	&	8.46 	&	41.40 	&	0.0000 	\\
DHBep+	&	-770.529159 	&	172.76 	&	181.39 	&	116.62 	&	56.46 	&	66.49 	&	0.9530 	\\
TRIP+	&	-770.480563 	&	172.58 	&	181.08 	&	116.10 	&	63.07 	&	63.16 	&	0.0000 	\\
\hline
\noalign{\smallskip} \noalign{\smallskip}
\end{tabular}
\end{center}
\end{table*}

\begin{table*}
\tiny
\begin{center}
\caption[]{\footnotesize
            Same as Table~\ref{tab:AValueARatio_Benzene}
            but for the Sandford et al.\ (2013) Molecules
            as Shown in Figure~\ref{fig:HPAH_Sandford_structure}.
            }
\label{tab:AValueARatio_NeuSandford}
\begin{tabular}{lcccccc}
\noalign{\smallskip} \hline \hline \noalign{\smallskip}
Compound	    &	$A_{3.4}$
                &	$A_{6.85}$
                &	$A_{3.3}$
                &   $A_{6.2}$
                &   $A_{3.4}/A_{3.3}$
                &   $A_{6.85}/A_{6.2}$
\\ \noalign{\smallskip} \hline \noalign{\smallskip}
DHA	    &	18.81 	&	2.01	&	15.06 	&	0.42 	&	1.25 	&	4.84 	\\
HHA	    &	40.15 	&	1.08	&	29.78 	&	2.37 	&	1.35 	&	0.46 	\\
DHPh	&	28.20 	&	6.11	&	13.95 	&	0.31 	&	2.02 	&	19.70 	\\
DBS	    &	24.06 	&	5.36	&	15.28 	&	0.38 	&	1.57 	&	14.17 	\\
tPHF 	&	39.26 	&	1.32	&	$-$	&	$-$	&	$-$	&	$-$	\\
DHP	    &	28.04 	&	3.37	&	15.44 	&	0.46 	&	1.82 	&	7.41 	\\
HHP	    &	36.01 	&	2.19	&	22.31 	&	2.12 	&	1.61 	&	1.03 	\\
THBaP	&	37.64 	&	4.21	&	14.55 	&	1.73 	&	2.59 	&	2.44 	\\
H9CpaP	&	11.85 	&	3.95	&	13.80 	&	1.37 	&	0.86 	&	2.89 	\\
HHCpaP	&	34.34 	&	3.29	&	19.76 	&	0.92 	&	1.74 	&	3.56 	\\
DDHTP   &	39.28 	&	2.15	&	$-$	&	0.52 	&	$-$	&	4.11 	\\
PHC  	&	47.18 	&	1.81	&	$-$	&	$-$	&	$-$	&	$-$	\\
DHBdeA	&	33.85 	&	3.18	&	14.31 	&	1.88 	&	2.36 	&	1.69 	\\
CpPh	&	10.55 	&	4.69	&	15.50 	&	0.48 	&	0.68 	&	9.87 	\\
HBbF	&	12.12 	&	7.23	&	13.99 	&	1.92 	&	0.87 	&	3.78 	\\
DHDBahA	&	15.35 	&	1.23	&	13.65 	&	2.75 	&	1.12 	&	0.45 	\\
DHBep	&	29.64 	&	1.61	&	13.90 	&	1.74 	&	2.13 	&	0.93 	\\
TRIP	&	16.35 	&	$-$	&	14.05 	&	0.23 	&	1.16 	&	$-$	\\\hline
Average	&	27.93 	&	3.22 	&	16.35 	&	1.22 	&	1.54 	&	5.15 	\\\hline
\hline
DHA+	    &	11.44	&	4.44 	&	3.28	&	86.14 	&	3.49	&	0.05 	\\
HHA+	    &	34.76	&	13.82 	&	18.03	&	261.89 	&	1.93	&	0.05 	\\
DHPh+	&	4.24	&	12.78 	&	0.34	&	30.70 	&	12.60	&	0.42 	\\
DBS+  	&	4.44	&	55.68 	&	1.85	&	67.51 	&	2.40	&	0.82 	\\
tPHF+	&	13.87	&	3.62 	&	$-$	&	$-$	&	$-$	&	$-$	\\
DHP+	    &	5.70	&	7.94 	&	0.71	&	19.55 	&	8.06	&	0.41 	\\
HHP+	    &	9.22	&	10.69 	&	1.22	&	20.96 	&	7.55	&	0.51 	\\
THBaP+	&	17.19	&	7.71 	&	0.73	&	12.71 	&	23.47	&	0.61 	\\
H9CpaP+	&	0.99	&	0.00 	&	0.66	&	11.21 	&	1.51	&	0.00 	\\
HHCpaP+	&	9.75	&	16.45 	&	2.16	&	36.04 	&	4.50	&	0.46 	\\
DDHTP+	&	39.28	&	12.32 	&	$-$	&	0.36 	&	$-$	&	34.72 	\\
PHC+	    &	31.47	&	2.20 	&	$-$	&	$-$	&	$-$	&	$-$	\\
DHBdeA+	&	10.13	&	18.24 	&	0.89	&	19.80 	&	11.37	&	0.92 	\\
CpPh+	&	3.62	&	15.03 	&	0.34	&	17.07 	&	10.67	&	0.88 	\\
HBbF+	&	0.91	&	58.14 	&	1.34	&	54.13 	&	0.68	&	1.07 	\\
DHDBahA+	&	0.58	&	2.70 	&	7.19	&	80.36 	&	0.08	&	0.03 	\\
DHBep+	&	5.31	&	14.18 	&	1.16	&	22.42 	&	4.58	&	0.63 	\\
TRIP+	&	4.40	&	$-$	&	2.68	&	0.00 	&	1.64	&	$-$	\\\hline
Average	&	11.52 	&	15.06 	&	2.84 	&	46.30 	&	6.30 	&	2.77 	\\\hline
\noalign{\smallskip} \noalign{\smallskip}
\end{tabular}
%
\end{center}
\end{table*}


\begin{table*}
\begin{center}
\caption[]{\footnotesize
           {\it Mean} IR Intensities (km\,mol$^{-1}$) of
           the 3.4 and 6.85$\mum$ Aliphatic C--H Bands,
           the 3.3$\mum$ Aromatic C--H Stretch Band,
           and the 6.2$\mum$ Aromatic C--C Stretch Band
           Computed at the {\rm B3LYP/6-311+G$^{\ast\ast}$} Level
           for All the Hydrogenated Species Shown in 
           Figures~\ref{fig:HBenzene_structure}--\ref{fig:HPAH_Sandford_structure},
           Including Hydrogenated Benzenes, 
           Hydrogenated Naphthalenes, Hydrogenated Perylenes, 
           and the Hydrogenated Molecules of Sandford et al.\ (2013).
           The $\Aaro$, $\Aali$ and $\Adfa$ 
           Band Strengths Are on a per C--H Bond Basis,
           While the $\Acc$ Band Strength is 
           on a per C Atom Basis.
           Also Tabulated Are the Band-Strength Ratios
           $A_{3.4}/A_{3.3}$ and $A_{6.85}/A_{6.2}$.
            }
\label{tab:AValueARatio_All}
\begin{tabular}{lcccccc}
\noalign{\smallskip} \hline \hline \noalign{\smallskip}
Compound	    &	$A_{3.4}$
                &	$A_{6.85}$
                &	$A_{3.3}$
                &   $A_{6.2}$
                &   $A_{3.4}/A_{3.3}$
                &   $A_{6.85}/A_{6.2}$
\\ \noalign{\smallskip} \hline \noalign{\smallskip}
Neutrals &	33.62 	&	2.59 	&	16.71 	&	1.87 &	1.98 	&	3.02$^{a}$ \\
Cations &	13.63 	&	13.17 	&	2.69 	&	33.50 &	7.73 	&	5.19$^{b}$\\
\hline	 				
\noalign{\smallskip} \noalign{\smallskip}
\end{tabular}
\begin{description}
\item[$^{a}$] $\langle\Adfa/\Acc\rangle\approx1.53\pm0.56$
                     if we exclude those molecules with
                     extreme $\Adfa/\Acc$ ratios.
\item[$^{b}$] $\langle\Adfa/\Acc\rangle\approx1.23\pm0.50$
                     if we exclude those molecules with
                     extreme $\Adfa/\Acc$ ratios.
\end{description}
\end{center}
\end{table*}



\begin{thebibliography}{30}
\expandafter\ifx\csname natexlab\endcsname\relax\def\natexlab#1{#1}\fi
%


\bibitem[]{}Acke, B., Bouwman, J., \& Juh\'{a}sz, A.\
            2010, ApJ, 718, 558

\bibitem[]{}Allamandola, L.J., Tielens, A.G.G.M., \& Barker, J.R.\
            1985, ApJ, 290, L25

\bibitem[]{}Allamandola, L.J., Tielens, A.G.G.M., \& Barker, J.R.\
            1989, ApJS, 71, 733

\bibitem[]{}Allamandola, L.J., Hudgins, D.M., \& Sandford, S.A.\
            1999, ApJ, 511, 115

\bibitem[]{}Andrews, H., Candian, A., \& Tielens, A.G.G.M.\
            2016, A\&A, 595, 23

\bibitem[]{}Bakes, E.L.O., \& Tielens, A.G.G.M. \
            1994, ApJ, 427, 822

\bibitem[]{}Barker, J.~R., Allamandola, L.~J., 
                 \& Tielens, A.~G.~G.~M.\ 1987, 
                 ApJL, 315, L61

\bibitem[]{}Bauschlicher, C.~W., Jr. \
            1998, ApJ, 509, L125

\bibitem[]{}Bernstein, M.P., Sandford, S.A., \& Allamandola, L.J.\
            1996, ApJ, 472, L127

\bibitem[]{}Borowski, P.\
            2012, J. Phys. Chem. A, 116, 3866

\bibitem[]{}Boschman, L., Reitsma, G., Cazaux, S., et al.\
            2012, ApJ, 761, L33

\bibitem[]{}Brenner, J., \& Barker, J.~R.\ 1992, 
                 ApJL, 388, L39


\bibitem[]{}Cassam-Chena{\"i}, P., Pauzat, F., \& Ellinger, Y.\
            1994, AIPC, 312, 543

\bibitem[]{}Cazaux, S., Boschman, L., Rougeau, N.\
            2016, Scientific Reports, 6, 19835

\bibitem[]{}Chen, T., Luo, Y., \& Li, A.\ 
                  2019, A\&A, 632, A71

\bibitem[]{}Draine, B.T., \& Li, A.\ 2001, ApJ, 551, 807

\bibitem[]{}Frisch, M. J., Trucks, G. W., Schlegel, H. B.,
            et al.\ 2009, Gaussian 09, Revision B01,
            Gaussian, Inc., Wallingford CT

\bibitem[]{}Geballe, T.R., Lacy, J.H., Persson, S.E.,
            McGregor, P.J., \& Soifer, B.T.\
            1985, ApJ, 292, 500

\bibitem[]{}Geballe, T.~R., Tielens, A.~G.~G.~M.,
            Allamandola, L.~J., Moorhouse, A.,
            \& Brand, P.~W.~J.~L.\
            1989, ApJ, 341, 278

\bibitem[]{}Geballe, T.~R., Joblin, C., 
                  d'Hendecourt, L.~B., et al.\ 1994, 
                  ApJL, 434, L15

\bibitem[]{}Goto, M., Usuda, T., Takato, N., et al.\
            2003, ApJ, 589, 419

\bibitem[]{}Halasinski, T.~M., Salama, F.,
                  \& Allamandola, L.~J.\ 2005, ApJ, 628, 555

\bibitem[]{}Hammonds, M., Pathak, A., \& Sarre, P.~J.\
                 2009, Phys. Chem. Chem. Phys., 11, 4458

\bibitem[]{}Joblin, C., d'Hendecourt, L., L\'eger, A.,
                 \& Defourneau, D.\ 1994, A\&A, 281, 923

\bibitem[]{}Joblin, C., Tielens, A.G.G.M., Allamandola, L.J.,
            \& Geballe, T.R.\ 1996, ApJ, 458, 610

\bibitem[]{}Jensen, P. A.m Leccese, M., Simonsen, F. D. S., et al.\
            2019, MNRAS, 486, 5492

\bibitem[]{}Jourdain de Muizon, M., Geballe, T.R., d'Hendecourt, L.B.,
            \& Baas, F.\ 1986, ApJ, 306, L105

\bibitem[]{}Jourdain de Muizon, M., d'Hendecourt, L.B.,
            \& Geballe, T.R.\ 1990, A\&A, 235, 367

\bibitem[]{}Kl{\ae}rke, B., Toker, Y., Rahbek, D. B., Hornek{\ae}r, L., \& Andersen, L. H.\
            2013, A\&A, 549, 84

\bibitem[]{}Kondo, T., Kaneda, H., Oyabu, S., et al.\
            2012, ApJ, 751, L18

\bibitem[]{}L\'{e}ger, A., \& Puget, J.
            1984, A\&A, 137, L5

\bibitem[]{}Le Page, V., Snow, T. P., Bierbaum, V. M.\
            2009, ApJ, 704, 274


\bibitem[]{}Li, A., \& Draine, B.T.\ 2012, ApJ, 760, L35

\bibitem[]{}Materese, C. K., Bregman, J. D., \& Sandford, S. A.\
            2017, ApJ, 850, 165


\bibitem[]{}Maltseva, E., Petrignani, A.,
                 Candian, A., et al.\ 2016, ApJ, 831, 58

\bibitem[]{}Maltseva, E., Mackie, C. J., Candian, A., et al.\
                 2018, A\&A, 610, 65

\bibitem[]{}Matsuura, M., Bernard-Salas, J., 
                  Lloyd Evans, T., et al.\ 2014, 
                  MNRAS, 439, 1472

\bibitem[]{}Nagata, T., Tokunaga, A. T., Sellgren, K., et al.\
            1988, ApJ, 326, 157


\bibitem[]{}Pauzat, F., \& Ellinger, Y.\ 2001,
                  MNRAS, 324, 355



\bibitem[]{}Peeters, E., Allamandola, L.J., Hudgins, D.M.,
                  Hony, S., \& Tielens, A.G.G.M.\ 2004,
                  in Astrophysics of Dust (ASP Conf. Ser. 309),
                  ed. A.N. Witt, G.C. Clayton, \& B.T. Draine
                  (San Francisco, CA: ASP), 141


\bibitem[]{}Quiti{\'a}n-Lara, H.~M., Fantuzzi, F.,
                  Nascimento, M.~A.~C., et al.\ 2018,
                  ApJ, 854, 61


\bibitem[]{}Rasmussen, J. A., Henkelman, G., \& Hammer, B.\
            2011, JChPh, 134, 164703

\bibitem[]{}Rauls, E., \& Hornek{\ae}r, L.\
            2008, ApJ, 679, 531

\bibitem[]{}Ricks, A.~M., Douberly, G.~E.,
                 \& Duncan, M.~A.\ 2009, ApJ, 702, 301

\bibitem[]{}Sandford, S.A.\ 1991, ApJ, 376, 599

\bibitem[]{}Sandford, S. A., Allamandola, L. J., Tielens, A. G. G. M., et al.\
            1991, ApJ, 371, 601

\bibitem[]{}Sandford, S. A., Bernstein, M. P., \& Materese, C. K.\
            2013, ApJS, 205, 8


\bibitem[]{}Schutte, W.~A., Tielens, A.~G.~G.~M., 
                  \& Allamandola, L.~J.\ 1993, ApJ 415, 397

\bibitem[]{}Seok, J. Y., \& Li, A.\ 2017, ApJ, 835, 291

\bibitem[]{}Sloan, G.C., Bregman, J.D., Geballe, T.R.,
            Allamandola, L.J., \& Woodward, C.E.\
            1997, ApJ, 474, 735

\bibitem[]{}Sloan, G. C., Lagadec, E., Zijlstra, A. A., et al.\
            2014, ApJ, 791, 28


\bibitem[]{}Smith, T.~L., Clayton, G.~C.,
                  \& Valencic, L.\ 2004, AJ, 128, 357

\bibitem[]{}Steglich, M., J{\"a}ger, C.,
                  Huisken, F., et al.\ 2013, ApJS, 208, 26

\bibitem[]{}Thrower, J. D., Friis, E. E., Skov, A. L., et al.\
                  2014, Phys. Chem. Chem. Phys., 16, 3381

\bibitem[]{}Thrower, J. D., J{\o}rgensen, B., Friis, E. E., et al.\
            2012, ApJ, 752, 3

\bibitem[]{}Tielens, A. G. G. M. \ 2008, ARA\&A, 46, 289

\bibitem[]{}Wagner, D.~R., Kim, H., \& Saykally, R.~J.\
                  2000, ApJ, 545, 854

\bibitem[]{}Weingartner, J. C., \& Draine, B. T. \
            2001, ApJS, 134, 263

\bibitem[]{}Wolf, M., Kiefer, H. V., Langeland, J., et al. \
            2016, ApJ, 832, 24

\bibitem[]{}Yamagishi, M., Kaneda, H., Ishihara, D., et al.\
            2012, A\&A, 541, A10

\bibitem[]{}Yang, X.~J., Glaser, R., Li, A.,
            \& Zhong, J.~X.\
            2013, ApJ, 776, 110

\bibitem[]{}Yang, X.~J., Glaser, R., Li, A.,
            \& Zhong, J.~X.\
            2016a, MNRAS, 462, 1551

\bibitem[]{}Yang, X.~J., Li, A., Glaser, R.,
            \& Zhong, J.~X.\
            2016b, ApJ, 825, 22

\bibitem[]{}Yang, X.~J., Glaser, R., Li, A.,
                  \& Zhong, J.~X.\
                  2017a, New Astron. Rev., 77, 1

\bibitem[]{}Yang, X.~J., Li, A., Glaser, R.,
            \& Zhong, J.~X.\
            2017b, ApJ, 837, 171







%
\end{thebibliography}
\end{document}